\documentclass[10pt,prd,singlecolumn,aps,
nofootinbib,
noshowkeys,
noshowpacs,
superscriptaddress,
floatfix
]{revtex4-2}
\usepackage{amsmath,amsfonts,amsthm,amssymb,bm}
\usepackage[dvips]{graphics,graphicx}
\usepackage[usenames,dvipsnames]{color}
\definecolor{darkblue}{RGB}{0,0,196}
\definecolor{darkgreen}{RGB}{0,120,0}
\usepackage[colorlinks=true,linktocpage=true,linkcolor=darkblue,citecolor=red,urlcolor=darkblue]{hyperref}
\usepackage{cancel}
\usepackage{bbold}
\usepackage{multirow}
\usepackage{longtable}
\usepackage{color}
\usepackage[normalem]{ulem}
\usepackage{hyperref}
\usepackage{bigints}
\usepackage{xparse}
\usepackage{physics}
\usepackage{verbatim}
\usepackage{minibox}
\usepackage{comment}
\usepackage{appendix}
\usepackage{slashed}
\usepackage{marginnote}
\usepackage{graphicx}
\usepackage[nice]{nicefrac}
\usepackage{amsmath}
\usepackage{cancel}
\usepackage[normalem]{ulem}
\usepackage[colorinlistoftodos]{todonotes}
\usepackage{hepunits}
\usepackage{stackengine,scalerel}
\newcommand\hstar[1]{\ThisStyle{\ensurestackMath{%
  \setbox0=\hbox{$\SavedStyle#1$}%
  \stackengine{0pt}{\copy0}{\kern.2\ht0\smash{\SavedStyle\star}}{O}{c}{F}{T}{S}}}}

\definecolor {darkgreen}{rgb}{0.2,0.7,0.2}

\begin{document}


\title{Spin dynamics and polarization in relativistic systems: recent developments}

\author{Sourav Dey \footnote{sourav.dey@niser.ac.in}}

\address{School of Physical Sciences, National Institute of Science Education and Research, An OCC of
Homi Bhabha National Institute, Jatni-752050, India}
\address{Department of Physics,  Birla Institute of Technology and Science Pilani, Pilani Campus, Pilani,  Rajasthan-333031, India}

\author{Arpan Das \footnote{arpan.das@pilani.bits-pilani.ac.in}}

\address{Department of Physics,  Birla Institute of Technology and Science Pilani, Pilani Campus, Pilani,  Rajasthan-333031, India}

\author{Hiranmaya Mishra \footnote{hiranmaya@niser.ac.in}}
\address{School of Physical Sciences, National Institute of Science Education and Research, An OCC of
Homi Bhabha National Institute, Jatni-752050, India}

\address{Institute of Physics, Sachivalaya Marg, Bhubaneswar-751005, India}

\author{Amaresh Jaiswal \footnote{a.jaiswal@niser.ac.in}}
\address{School of Physical Sciences, National Institute of Science Education and Research, An OCC of
Homi Bhabha National Institute, Jatni-752050, India}


\begin{abstract}
We review recent theoretical and experimental developments in spin dynamics and polarization phenomena in relativistic systems, with a particular focus on heavy-ion collisions. The large angular momentum and magnetic field generated in non-central collisions induce vorticity in the quark–gluon plasma, leading to observable spin polarization of emitted hadrons. We discuss the theoretical foundations of spin polarization arising from spin–vorticity coupling, including formulations based on relativistic hydrodynamics, kinetic theory, and quantum statistical approaches such as the Zubarev density operator. A central theme of the review is the role of pseudo-gauge freedom and its implications for defining energy–momentum and spin tensors, which can influence theoretical predictions of polarization observables. We further examine different formulations of spin hydrodynamics, emphasizing the impact of gradient expansions, spin chemical potential, and entropy-current analysis on the structure of the theory and associated transport coefficients. In addition, we discuss the recent developments in heavy flavor spin dynamics within the framework of rotational Brownian motion, where spin degrees of freedom undergo stochastic evolution due to interactions with the medium. This framework provides a complementary perspective on spin relaxation and diffusion by incorporating the effects of strong initial magnetic fields and establishes connections between spin polarization and the initial geometry through the definition of polarization harmonics. This review provides a comprehensive overview of relativistic spin hydrodynamics as well as non-equilibrium spin dynamics, and outlines future directions toward a consistent and predictive description of spin phenomena in strongly interacting matter.
\end{abstract}

\maketitle
\section{Introduction}
\label{intro}

It has been speculated that the thermalized medium produced in non-central relativistic heavy-ion collisions can carry large angular momentum, on the order of $10^{3\sim5}\hbar$~\cite{Becattini:2007sr}. Within the hydrodynamic framework, this angular momentum may manifest through vorticity, which is defined in terms of the gradients in the hydrodynamic variables, e.g., fluid flow ($u^{\mu}$), temperature ($T$), chemical potential ($\mu_B$), etc. Unlike the measurement of the flow anisotropies, e.g., elliptic flow, triangular flow, etc., which can be experimentally probed through the correlation studies of the momentum spectra of the emitted particles, the vortical structure of the strongly interacting fluid remained largely unexplored. This remained the case until the observation of hadron spin polarization in non-central heavy-ion collisions, which marked the advent of spin physics in the study of relativistic heavy-ion systems~\cite{STAR:2007ccu}. We emphasize that the orbital angular momentum generated in non-central heavy-ion collisions is not an observable. Instead, its effects are inferred through the coupling between the fluid’s mechanical angular momentum and the spin polarization of emitted particles. Although the observation of spin polarization in heavy ion collisions is a recent phenomenon, the first observation of the interplay of spin-vorticity dates back over a century, which is commonly known as the Barnett effect-- spontaneous magnetization of non-magnetized metal induced by rotation~\cite{barnett1915}. Its fluid analogue, which arises from the interaction between the bulk vorticity of the fluid and quantum spin polarization, was first identified by Takahashi et al. in 2016. In their study, they showed that the coupling between spin and vorticity leads to the formation of a polarization gradient, which can be experimentally detected through the inverse spin Hall effect~\cite{takahashi2016}. 
We emphasize that in Barnett's and Takahashi's experiments, macroscopic rotation was a controlled parameter, and the spin polarization could be measured directly. However, high-energy nuclear collisions involve event-by-event fluctuations in both the magnitude and direction of angular momentum. As a result, extracting meaningful spin polarization signals in heavy ion physics is rather non-trivial as compared to condensed matter systems. 

The total angular momentum ($\vec{J}$) in a heavy-ion collision can be written in terms of the impact parameter vector ($\vec{b}$), and the beam momentum in the center-of-momentum frame ($\vec{p}_{\text{beam}}$),  $\vec{J} = \vec{b} \times \vec{p}_{\text{beam}}$, 
where $\vec{b}$ is the transverse vector connecting the centers of the target and beam nuclei, and $\vec{p}_{\text{beam}}$ is the beam momentum in the center-of-momentum (c.o.m.) frame. The magnitude $|\vec{b}|$ is estimated from the total charged particle multiplicity emitted perpendicular to the beam axis, while its direction $\widehat{{b}}$ is inferred from the sideward deflection of particles emitted near the beam direction. The QGP (Quark Gluon Plasma) fluid exhibits complex flow patterns, with local vorticity fluctuating across the system. However, the event-averaged vorticity aligns with the total angular momentum $\vec{J}$, making spin alignment along $\widehat{{J}}$ known as “global” polarization. Once the direction of $\widehat{{J}}$ is determined, the challenge lies in measuring spin polarization along it. If the QGP possesses nonzero vorticity, then due to spin-vorticity coupling, emitted particles should exhibit spin alignment with $\widehat{J}$. Among the produced particles, only a few undergo parity-violating weak decays that reveal their spin via anisotropies in the momentum distribution of the decay products. Of these, only a limited subset is abundantly produced for a meaningful statistical analysis.

The best candidate for measuring spin polarization is the $\Lambda$-hyperon, which undergoes weak decay via $\Lambda \rightarrow p + \pi^{-}$. As a parity-violating weak process, this decay preferentially emits the daughter proton along the direction of the $\Lambda$'s polarization, following \cite{STAR:2007ccu}
\begin{equation}
    \frac{dN}{d\cos\theta^*}=\frac{1}{2}\Big( 1 + \alpha_{\Lambda}\vec{P}_{\Lambda}\cdot \widehat{p}_{p}^{~*}\Big)\label{Eq1}
\end{equation}
where $\theta^{*}$ is the angle between the polarization vector $\vec{P}_{\Lambda}$ and the daughter proton momentum $\vec{p}_{p}^{~*}$ in the hyperon rest frame. The decay parameter $\alpha_{\Lambda}$ indicates the strength of the interaction and has the value $\sim0.732$. The so-called global polarization $\vec{P}_{\Lambda}$ is obtained by integrating over all possible momenta of the emitted $\Lambda$-hyperons. By symmetry, it must be proportional to the system’s total angular momentum $\vec{J}$. 
\begin{figure}[t]  
    \centering
    \includegraphics[width=0.65\textwidth]{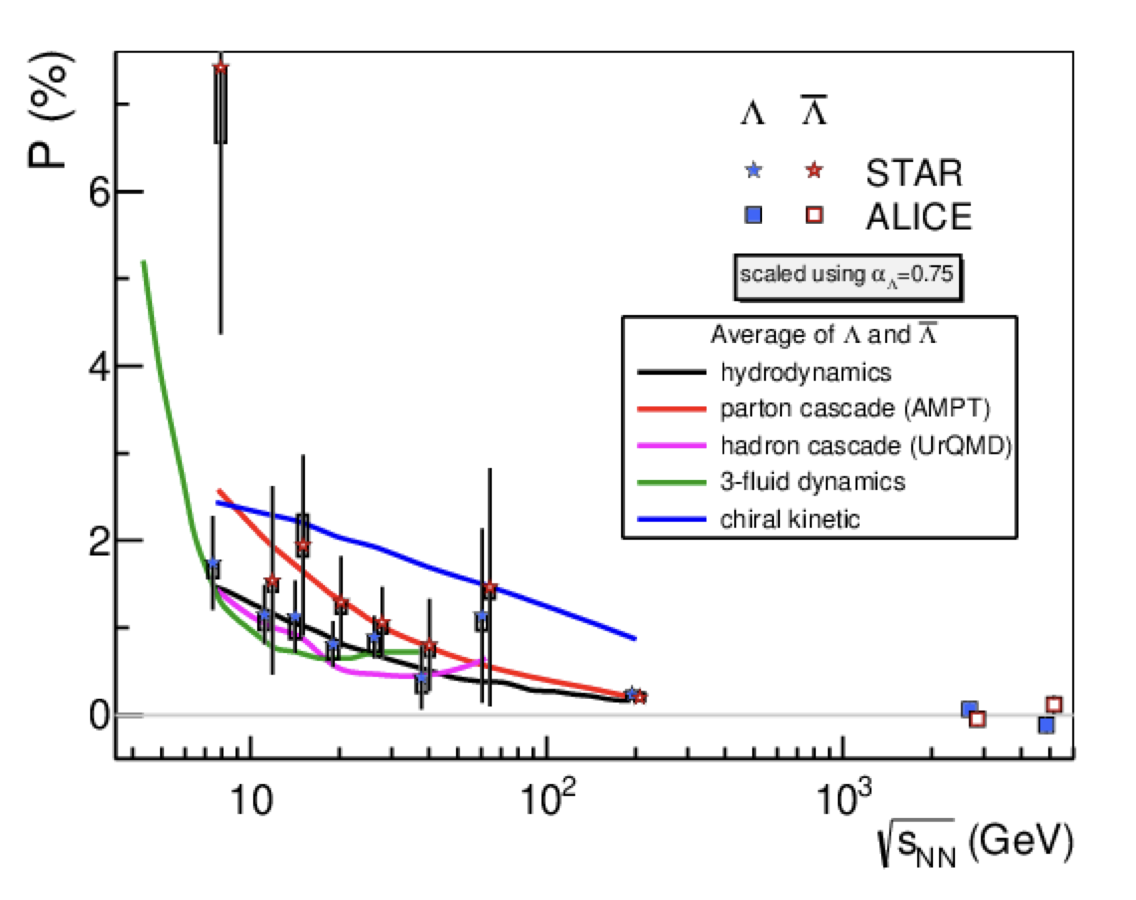}  
    \caption{Comparison of the global $\Lambda$ polarization measurements reported by STAR \cite{STAR:2007ccu,STAR:2017ckg,STAR:2018gyt} and ALICE \cite{ALICE:2019onw} with theoretical predictions obtained from various hydrodynamic and transport model simulations \cite{Karpenko:2016jyx,Li:2017slc,Vitiuk:2019rfv,Sun:2017xhx}. Plot adapted from Ref.~\cite{Becattini:2020ngo}.}
    \label{fig:SpinD}
\end{figure}
Experimentally, the polarization is inferred by measuring the average projection of the daughter proton momentum along a chosen direction. Thus, one calculates by setting $\vec{P}_{\Lambda}=P_{\Lambda}\widehat{J}$ \cite{Becattini:2007sr,STAR:2007ccu,Becattini:2017gcx,STAR:2017ckg,STAR:2018gyt,ALICE:2019onw}
\begin{align}
    \left\langle \widehat{{p}}_p^{~*} \cdot \widehat{J} \right\rangle_{\Lambda,e} 
= \left\langle \int d\cos\theta^*\frac{dN}{d\cos\theta^*} 
\, \widehat{p}_p^{~*} \cdot \widehat{{J}} \right\rangle_{e}
= \frac{\alpha_\Lambda}{3} \, \vec{P}_\Lambda \cdot \widehat{{J}}.
\end{align}
Here, $\langle\cdots\rangle_{\Lambda,e}$ indicates an average taken over all $\Lambda$ particles and over all events.
\begin{align}
    \vec{P}_\Lambda\cdot \widehat{{J}}
= \frac{3}{\alpha_\Lambda} \left\langle \widehat{{p}}_p^{~*} \cdot \widehat{{J}} \right\rangle_{\Lambda,e} 
= \frac{3}{\alpha_\Lambda} \left\langle \cos \theta^{*} \right\rangle_{\Lambda,e}
\end{align}
Here, $\theta^{*}$ is the angle between the proton momentum in the hyperon rest frame and the angular momentum of the system. The measurement of global $\Lambda$ polarization by the STAR collaboration became well-known as compelling evidence that “the fluid produced in heavy-ion collisions is by far the most vortical system ever observed.” The basis for estimating the vorticity $\omega$ in Ref.~\cite{STAR:2017ckg} relies on the relation
\begin{align}
    \omega = \frac{1}{T} \left( P_{\Lambda} + P_{\bar{\Lambda}} \right),
\end{align}
where $T$ is the temperature and $P_{\bar{\Lambda}}$ is the global polarization of anti-$\Lambda$ hyperons. A similar analysis was carried out by the ALICE experiment at the LHC. Combined with the measurements from STAR, these results represent the world’s dataset on global $\Lambda$ polarization, as illustrated in Fig.~\ref{fig:SpinD}.
Figure~\ref{fig:SpinD} presents measurements of global $\Lambda$ polarization as a function of collision energy. The data show a clear decreasing trend in polarization with increasing beam energy. Notably, both hyperons and anti-hyperons are polarized in the same direction, suggesting that the effect arises from the fluid's vorticity rather than electromagnetic interactions. Moreover, the degree of polarization is different for $\Lambda$, and $\bar{\Lambda}$, such a difference could originate due to the interaction with the magnetic field. 
Hyperons have a non-vanishing magnetic moment. The interaction of the magnetic moment with the magnetic field may result in different magnitudes of spin polarization. However, to come to such a conclusion conclusively one needs to study the polarization phenomena in  low energy scattering experiments. Note that $\Lambda$, and $\bar{\Lambda}$ may originate from different regions of the fireball, having different flow gradients. This can also give rise to a difference in the spin polarization of 
$\Lambda$, and $\bar{\Lambda}$. In contrast to global polarization, the so-called local (momentum-differential) polarization can exhibit components orthogonal to the system’s total angular momentum. If, instead of using the global angular momentum direction $\widehat{{J}}$, one chooses the beam axis $\widehat{{p}}_{\rm beam}$ as the reference, the longitudinal polarization is defined as
\begin{align}
P_{\Lambda}^{z} \equiv \vec{P}_{\Lambda} \cdot \widehat{{p}}^{~*}_{\rm beam}
= \frac{1}{\alpha_\Lambda} \frac{\langle \cos\theta_p^* \rangle_{\Lambda,\,e}}{\langle \cos^2\theta_p^* \rangle_{\Lambda,\,e}},
\end{align}
where the average is taken over $\Lambda$ hyperons within a fixed momentum bin, and $\theta_p^*$ is the angle between the proton momentum and $\vec{p}_{\rm beam}$ in the hyperon rest frame.

On the theory side, using the Zubarev density operator method, the average spin polarization ($S^{\mu}(p)$) of a massive (having mass $m$) spin-half particle (in this case $\Lambda$ or $\bar{\Lambda}$ particles) can be obtained as~\cite{Becattini:2020ngo}
\begin{align}
    S^{\mu}(p)=-\frac{1}{8m}\epsilon^{\mu\rho\sigma\tau}p_{\tau}\frac{\int d\Sigma_{\lambda}p^{\lambda}\varpi_{\rho\sigma}f_{FD}(1-f_{FD})}{\int d\Sigma_{\lambda}p^{\lambda}f_{FD}}.
    \label{equ2review}
\end{align}
Here $\beta^{\mu}\equiv u^{\mu}/T$, $T$ is the temperature, and $u^{\mu}$ is the flow velocity.  $\varpi_{\rho\sigma}=-(\partial_{\rho}\beta_{\sigma}-\partial_{\sigma}\beta_{\rho})/2$ is the thermal vorticity, which is dimensionless in the Natural units. The four-momentum of the particle is $p^{\mu}$, $f_{FD}$ is the Fermi-Dirac distribution function, and $d\Sigma_{\lambda}$ is the infinitesimal element of the space-like 3-dimensional hypersurface. Normally, in the context of heavy ion physics, this surface is chosen as the freezeout hypersurface to estimate the polarization of the hyperon at the freezeout.  $\left(S^{y}(p)\right)$ and $\left(S^{z}(p)\right)$ represent global spin polarization and local (longitudinal) spin polarization, respectively. Here $z$-axis represents the beam direction, and $x$- axis is along the impact factor direction. $y$-axis is the direction of the global angular momentum. Various theoretical models, e.g., relativistic hydrodynamic models~\cite{DelZanna:2013eua,Karpenko:2013wva,Ivanov:2019ern}),parton and hadron cascade models~\cite{Li:2017slc,Vitiuk:2019rfv}, etc., have been used to compute global $\left(S^{y}(p)\right)$ and longitudinal $\left(S^{z}(p)\right)$ spin polarization, using the spin-thermal vorticity coupling as given in Eq.~\eqref{equ2review}. These model calculations can explain the $\sqrt{s}_{NN}$ (center of mass energy) dependence of the average global spin polarization $\left(S^{y}(p)\right)$ data, which indicates that with increasing value of $\sqrt{s}_{NN}$ the global spin polarization decreases (shown in Fig.~\eqref{fig:SpinD}). However, the model based on spin-thermal vorticity coupling fails to explain the azimuthal angle ($\phi_p$) dependence of longitudinal polarization (local polarization), i.e., $S^{z}(\phi_p)$. Here, the azimuthal angle $\phi_p$ is defined as $\tan\phi_p=p_y/p_x$, and the azimuthal angle dependence can be obtained by taking an average over the transverse momentum (transverse to the beam direction).
Building theoretical models that can explain both global and local polarization data consistently poses an exciting theoretical challenge, which has stimulated much work in the field. New approaches involving thermal shear-induced spin polarization~\cite{Liu:2020dxg,Yi:2021ryh,Sun:2021nsg,Florkowski:2021xvy},
chiral kinetic theory (CKT) based model~\cite{Hidaka:2017auj}, etc., have been developed to address the longitudinal spin polarization problem. A detailed discussion on spin polarization in heavy ion collisions can be found in Refs.~\cite{Becattini:2020ngo,Becattini:2022zvf,Gao:2020vbh, Becattini:2020sww,Wang:2017jpl}. The spin-hydrodynamic framework is also a promising alternative approach takes into the hydrodynamical evolution of spin of microscopic constituents, has triggered a lot of investigation~\cite{Florkowski:2018ahw,Florkowski:2017dyn,Florkowski:2017ruc,Florkowski:2019qdp,Florkowski:2019voj,Hattori:2019lfp,Fukushima:2020ucl,Li:2020eon,She:2021lhe,Montenegro:2017lvf,Montenegro:2017rbu,Florkowski:2017ruc,Florkowski:2018myy,Bhadury:2020puc,Shi:2020qrx,Weickgenannt:2022zxs,Weickgenannt:2019dks,Weickgenannt:2020aaf,Speranza:2021bxf,Wang:2021ngp,Gallegos:2021bzp,Wang:2021ngp,Gallegos:2021bzp}. Moreover, relativistic kinetic theory approach~\cite{Florkowski:2017ruc,Florkowski:2017dyn,Hidaka:2017auj,Florkowski:2018myy,Weickgenannt:2019dks,Bhadury:2020puc,Weickgenannt:2020aaf,Shi:2020htn,Speranza:2020ilk,Bhadury:2020cop,Singh:2020rht,Bhadury:2021oat,Peng:2021ago,Sheng:2021kfc,Sheng:2022ssd,Hu:2021pwh,Hu:2022lpi,Fang:2022ttm,Wang:2022yli,Bhadury:2024ckc}, Zubarev's quantum statistical density operator approach~\cite{Becattini:2007nd,Becattini:2009wh,Becattini:2012pp,Becattini:2012tc,Becattini:2018duy,Hu:2021lnx}, effective Lagrangian approach~\cite{Montenegro:2017rbu,Montenegro:2017lvf,Montenegro:2018bcf,Montenegro:2020paq}, etc. have been considered to study the spin physics in a vortical QCD (Quantum Chromo Dynamics) medium.

This review is organized as follows. In Sec.~\ref{intro}, we introduced the basic concepts of spin polarization in relativistic heavy-ion collisions and summarized the key experimental observations, including global and local polarization of hadrons. Section~\ref{sec2} is devoted to the theoretical foundations of spin hydrodynamics, with a particular emphasis on pseudo-gauge freedom and its consequences for the definition of energy–momentum and spin tensors, as well as their impact on polarization observables. In Sec.~\ref{sec3}, we present different formulations of spin hydrodynamics, discussing entropy-current analysis, gradient expansion schemes, and the role of spin chemical potential, along with the associated transport coefficients. Section~\ref{sec4} focuses on microscopic approaches, including kinetic theory and quantum statistical methods, which provide a connection between underlying particle dynamics and macroscopic spin phenomena. Finally, in Sec.~\ref{sec5}, we discuss recent developments in heavy-flavor spin dynamics within the framework of rotational Brownian motion, highlighting the stochastic evolution of spin degrees of freedom, spin relaxation and diffusion mechanisms, and the connection between spin polarization, initial magnetic fields, and the geometry of the system. Finally, Sec.~\ref{sec6} provides a summary of the topics discussed in this review and an outlook on open problems and future directions in the study of relativistic spin dynamics.

In this manuscript the we use the following notations. $X^{\{\mu}Y^{\nu\}}=(X^{\mu}Y^{\nu}+X^{\nu}Y^{\mu})/2$ and $X^{[\mu}Y^{\nu]}=(X^{\mu}Y^{\nu}-X^{\nu}Y^{\mu})/2$ represents the symmetric and anti-symmetric combinations, respectively. 
$X^{\langle\mu\rangle}\equiv \Delta^{\mu\nu}X_{\nu}$ represents the projection of a four vector $X^{\mu}$ orthogonal to $u^{\mu}$. $X^{\langle\mu}Y^{\nu\rangle}\equiv \Delta^{\mu\nu}_{\alpha\beta}X^{\alpha}Y^{\beta}\equiv \frac{1}{2}\left(\Delta^{\mu}_{~\alpha}\Delta^{\nu}_{~\beta}+\Delta^{\mu}_{~\beta}\Delta^{\nu}_{~\alpha}-\frac{2}{3}\Delta^{\mu\nu}\Delta_{\alpha\beta}\right)X^{\alpha}Y^{\beta}$, represents the 
traceless and symmetric projection operator orthogonal to $u^{\mu}$.
Similarly, $X^{\langle[\mu}Y^{\nu]\rangle}\equiv \Delta^{[\mu\nu]}_{[\alpha\beta]}X^{\alpha}Y^{\beta}\equiv \frac{1}{2}\left(\Delta^{\mu}_{~\alpha}\Delta^{\nu}_{~\beta}-\Delta^{\mu}_{~\beta}\Delta^{\nu}_{~\alpha}\right)X^{\alpha}Y^{\beta}$ represents the anti-symmetric projection operator orthogonal to the flow vector. The partial derivative operator ($\partial_{\mu}$) can be decomposed in the following way, $\partial_{\mu}=u_{\mu}D+\nabla_{\mu}$, where $D\equiv u^{\mu}\partial_{\mu}$ is the comoving derivative, and  $\nabla_{\mu}\equiv\Delta_{\mu}^{~\alpha}\partial_{\alpha}$ is the derivative orthogonal to $u^{\mu}$. $\sigma_{\mu\nu}\equiv\frac{1}{2}(\nabla_{\mu} u_{\nu}+\nabla_{\nu} u_{\mu})-\frac{1}{3}\theta\Delta_{\mu\nu}$, is the symmetric traceless combination of the derivative of the fluid flow.

\section{Spin hydrodynamics, pseudo gauge choice and its effect on the spin observable}
\label{sec2}

The standard hydrodynamic framework (spinless fluid) is based on the conservation of macroscopic currents $T^{\mu\nu}$, and $J^{\mu}$. 
\begin{align}
& \partial_{\mu}T^{\mu\nu}=0, ~~~~\partial_{\mu}J^{\mu}=0.\label{equ1ver1}
\end{align}
Here  $T^{\mu\nu}$ is the energy-momentum tensor, $J^{\mu}$ is the current associated with a symmetry, e.g., net baryon number current in QCD, etc. In the absence of dissipation $T^{\mu\nu}$, and  $J^{\mu}$ can be expressed as, 
\begin{align}
T^{\mu\nu}=\varepsilon u^{\mu}u^{\nu}-P \Delta^{\mu\nu}, ~~~~J^{\mu}=nu^{\mu}.
\end{align}
$u^{\mu}$ is the normalized fluid four-velocity, i.e., $u^{\mu}u_{\mu}= 1$. The projector  $\Delta^{\mu\nu}\equiv g^{\mu\nu}-u^{\mu}u^{\nu}$ is orthogonal to $u^{\mu}$, i.e., $\Delta^{\mu\nu}u_{\mu}=0=\Delta^{\mu\nu}u_{\nu}$. $\varepsilon (T,\mu)$, $P (T,\mu)$, and $n (T,\mu)$ represent energy density, pressure, and number density. $\varepsilon$, $P$, and $n$ are not independent and the are related via the Equation-of-State (EoS). $T$ is the temperature, $\mu$ is the chemical potential associated with the conserved charge. The conservation of  $T^{\mu\nu}$ and $J^{\mu}$ give rise to five equations, and in the absence of dissipation, these five equations give the evolution of hydrodynamic variables $T$, $\mu$, and $u^{\mu}$. For the spinful fluid along with the conservation of energy-momentum tensor, and conserved current, we also take into account the conservation of the total angular momentum tensor $J^{\lambda\mu\nu}$, i.e, $\partial_{\lambda}J^{\lambda\mu\nu}=0$. The additional conservation equation is essential because neither the orbital angular momentum tensor $L^{\lambda\alpha\beta}=x^{\alpha}T^{\lambda\beta}-x^{\beta}T^{\lambda\alpha}$, nor the spin tensor $S^{\lambda\alpha\beta}$ is separately conserved. This is crucial for allowing the conversion of spin and orbital angular momentum, also known as the spin-orbit interaction. Note that $J^{\lambda\alpha\beta}$, and $L^{\lambda\alpha\beta}$ are anti-symmetric in the last two indices, hence the conservation of $J^{\lambda\alpha\beta}$ boils down to six equations. These six equations represent the evolution of six components of spin chemical potential, which is anti-symmetric in the last two indices~\cite{Florkowski:2018fap}. Therefore, we can summarize the governing equation of the spin hydrodynamic framework as, 
\begin{align}
& \partial_{\mu}T^{\mu\nu}=0, ~~~~\partial_{\mu}J^{\mu}=0,~~~~\partial_{\lambda}J^{\lambda\mu\nu}=0.\label{equ5review}
\end{align}

It is well known that the conserved energy-momentum tensor does not have a unique form. Consider the conserved energy-momentum tensor $T^{\mu\nu}$, i.e. $\partial_\mu T^{\mu\nu}=0$. We can always construct an equivalent new energy-momentum tensor $T^{\prime \,\mu\nu}= T^{\mu\nu} + \partial_\lambda \Phi^{\nu\mu \lambda}$, where the $\Phi^{\nu\mu \lambda}$ is anti-symmetric in the last two indices. Note that $T^{\mu\nu}$, and $T^{\prime \,\mu\nu}$ both are conserved, hence this redefinition of $T^{\mu\nu}$ does not affect the conservation equations~\cite{Chen:2018cts,HEHL197655,Speranza:2020ilk}. Such different choices of conserved energy-momentum tensor are also known as pseudo-gauge choices, and the tensor $\Phi^{\mu\nu\lambda}$ is defined as the pseudo-gauge parameter. The pseudo-gauge transformation of only the energy-momentum tensor will certainly affect the conservation of the total angular momentum tensor. Therefore, to keep the conservation of the total angular momentum tensor intact, we need to make a suitable redefinition of the spin tensor as well, e.g., if we start with $T^{\mu\nu}$, and $S^{\lambda\mu\nu}$, then the pseudo-gauge transformed energy-momentum tensor $T^{\prime\mu\nu}=T^{\mu\nu}+\frac{1}{2}\partial_{\lambda}\left(\Phi^{\lambda\mu\nu}-\Phi^{\mu\lambda\nu}-\Phi^{\nu\lambda\mu}\right)$, and spin tensor $S^{\prime\lambda\mu\nu}=S^{\lambda\mu\nu}-\Phi^{\lambda\mu\nu}$ will not affect the conservation equations, i.e., $\partial_{\mu}T^{\prime\mu\nu}=0$, and $\partial_{\lambda}J^{\prime\lambda\mu\nu}=0$, where, $J^{\prime\lambda\mu\nu}=x^{\mu}T^{\prime\lambda\nu}-x^{\nu}T^{\prime\lambda\mu}+S^{\prime \lambda\mu\nu}$. For massive spin-half particles, different pseudo-gauge choices of energy-momentum tensor and spin tensor have been discussed in the literature, e.g., Belinfante-Rosenfeld (BR) pseudo-gauge~\cite{BELINFANTE1939887,BELINFANTE1940449,Rosenfeld1940}, the de Groot-van Leeuwen-van  Weert  (GLW) pseudo-gauge~\cite{DeGroot:1980dk}, the Hilgevoord-Wouthuysen (HW) pseudo-gauge~\cite{HILGEVOORD19631,HILGEVOORD19651002}, etc.

Although at the level of conservation equations, different pseudo-gauge choices are equivalent, the manifestation of this equivalence is not always evident, which brings a certain level of pseudo-gauge dependence. Quantum effects inherently indicates pseudo-gauge dependent results~\cite{Buzzegoli:2021wlg,Das:2021aar}. On the other hand, classical approaches do not show pseudo-gauge dependence and different pseudo-gauge choices are equivalent. In Ref.~\cite{Dey:2023hft}, the authors argued that only the canonical spin tensor obtained using Noether’s theorem fulfills the $SO(3)$ algebra of angular momentum. One starts with the Lagrangian density of non-interacting massive spin-half particles, 
\begin{align}
\mathcal{L}_D(x)
= \frac{i}{2}\,\bar{\psi}(x)\gamma^\mu \!\stackrel{\leftrightarrow}{\partial}_\mu \psi(x)
- m\,\bar{\psi}(x)\psi(x)
\label{equ8review}
\end{align}
Here $\overleftrightarrow{\partial}_\mu
= \overrightarrow{\partial}_\mu - \overleftarrow{\partial}_\mu$, and $m$ is the mass of the particle. The canonical spin tensor can be shown to be, 
\begin{equation}
S^{\lambda\mu\nu}_{\rm can}
= -\frac{1}{2}\,\varepsilon^{\lambda\mu\nu\rho}\,
\bar{\psi}\gamma_\rho\gamma_5\psi.
\label{equ7review}
\end{equation}
Moreover, the spin operator can be defined as the integral of the spin tensor,
\begin{align}
 & S^{k}_{\rm can}(t)
= \,\epsilon^{kij} \int d^{3}x \, S^{0ij}_{\rm can}(t,\vec{x})
= \,\epsilon^{kij} S^{ij}_{\rm can}(t).\\
& S^{ij}_{\rm can}(t)
= \frac{1}{2}\,\varepsilon^{0ijk}
\int d^{3}x \,\bar{\psi}(t,\vec{x})\,\gamma^{k}\gamma_{5}\psi(t,\vec{x}),\\
& S^{k}_{\rm can}(t)
= \int d^{3}x \,\psi^{\dagger}(t,\vec{x})
\,\gamma_5\gamma_0\gamma^k\psi_{}(t,\vec{x})=\int d^{3}x \,\psi^{\dagger}(t,\vec{x})
\,\Sigma^k\psi_{}(t,\vec{x}).
\end{align}
Using the above expression of $S^{k}_{\rm can}$ and the equal-time anti-commutation relations for the Dirac field it can be shown that, $[S^{i}_{\rm can}, S^{j}_{\rm can}]=2i\epsilon^{ijk}S^{k}_{\rm can}$, which is expected angular momentum operator algebra for spin~\footnote{Here $S^{k}_{\rm can}$ should not be confused with the average spin polarization $S^{\mu}(p)$ as introduced in Eq.~\eqref{equ2review}.}. Instead of using the canonical spin tensor, as defined in Eq.~\eqref{equ7review}, we can also consider the GLW spin tensor, which is connected with the canonical spin tensor through a suitable pseudo-gauge transformation ($\Phi^{\lambda\alpha\beta}_{\mathrm{GLW}}$), 
\begin{align}
S^{k}_{\mathrm{GLW}}(t)
= \varepsilon^{kij}
\int d^{3}x \,
\Bigl(
S^{0ij}_{\rm can}(t,\vec{x})
- \Phi^{0ij}_{\mathrm{GLW}}(t,\vec{x})
\Bigr).  
\end{align}
However, it can be explicitly shown that GLW spin operators $S^{k}_{\mathrm{GLW}}$ are not consistent with the $SO(3)$ operator algebra, i.e., $[S^{i}_{\rm GLW}, S^{j}_{\rm GLW}]\neq 2i\epsilon^{ijk}S^{k}_{\rm GLW}$. A similar conclusion can be obtained for the spin operators in Hilgevoord-Wouthuysen (HW) pseudo-gauge~\cite{Dey:2023hft}. In the context of hyperon spin polarization as observed in the heavy-ion collision experiments, physical implications of pseudo-gauge transformation of the energy-momentum tensor and the spin tensor have been extensively discussed~\cite{HEHL197655,Speranza:2020ilk,Florkowski:2018fap,Leader:2013jra,Gallegos:2021bzp,Buzzegoli:2021wlg}.

Interestingly, it has been pointed out in Ref.~\cite{Buzzegoli:2021wlg} that the expression of the average spin polarization $S^{\mu}(p)$ as defined in Eq.~\eqref{equ2review} may itself depend on the different pseudo-gauges. This pseudo-gauge dependence in the expression of the average spin polarization ($S^{\mu}(p)$) originates from the pseudo-gauge dependence of the local equilibrium density operator. 
Following Zubarev's non-equilibrium statistical operator method, the local equilibrium density operator is~\cite{Becattini:2013fla,Florkowski:2018fap}, 
\begin{align}
\widehat{\rho}_{\mathrm{LEQ}} =
\frac{1}{\mathcal{Z}}
\exp\!\left[-\int_{\Sigma(\tau)} d\Sigma_\mu
\left(
\widehat{T}^{\mu\nu}\beta_\nu
-\frac{1}{2}\,\Omega_{\lambda\nu}\widehat{S}^{\mu\lambda\nu}
-\alpha\widehat{J}^{\mu}
\right)
\right].
\label{equ12review}
\end{align}
$\mathcal{Z}$ is the normalization such that $\text{Tr} \left(\widehat{\rho}_{\rm LEQ}\right)=1 $. Here $\alpha=\mu/T$, $\beta^{\mu}=u^{\mu}/T$, and $\Omega_{\lambda\nu}$ is the spin chemical potential. $\widehat{T}^{\mu\nu}$, $\widehat{S}^{\mu\lambda\nu}$, and $\widehat{J}^{\mu}$ are the energy-momentum tensor, spin tensor, and conserved current operator.
Note that $\widehat{\rho}_{\rm LEQ}$ generically depends on the choice of the hypersurface $\Sigma (\tau)$. In heavy ion collisions, the hypersurface is chosen as the freeze-out hypersurface, which depends on the proper time $\tau$. Therefore, the local equilibrium density operator $\widehat{\rho}_{\mathrm{LEQ}}$ is time-dependent. Only in the global equilibrium, the density operator is hypersurface independent, i.e, $\tau$ independent. The density operator would become hypersurface independent if the divergence of the integrand in Eq.~\eqref{equ12review} vanishes~\cite{Florkowski:2018fap}, i.e.
\begin{align}
& \partial_{\mu}\bigg(\widehat{T}^{\mu\nu}_{}\beta_{\nu}-\frac{1}{2}\Omega_{\lambda\nu}\widehat{S}^{\mu\lambda\nu}_{}-\alpha\widehat{J}^{\mu}\bigg)=0\nonumber\\
\implies & \widehat{T}^{\mu\nu}_{ (s)}\partial_{\{\mu}\beta_{\nu\}}+\widehat{T}^{\mu\nu}_{ (a)}\left(\Omega_{\mu\nu}+\partial_{[\mu}\beta_{\nu]}\right)-\frac{1}{2}(\partial_{\mu}\Omega_{\lambda\nu})\widehat{S}^{\mu\lambda\nu}_{\rm can}-\widehat{J}^{\mu}\partial_{\mu}\alpha=0.
\label{equ13review}
\end{align}
Here $\widehat{T}^{\mu\nu}_{ (s)}\equiv \widehat{T}^{\{\mu\nu\}} $ and $\widehat{T}^{\mu\nu}_{ (a)}\equiv \widehat{T}^{[\mu\nu]}$  are the symmetric and anti-symmetric parts of the energy-momentum tensor, respectively. We have kept both symmetric and anti-symmetric parts of the energy-momentum tensor, as in the canonical framework, Noether’s theorem does not give a symmetric energy-momentum tensor. Moreover, to obtain the second line of the above equation, we have used the conservation of $\widehat{T}^{\mu\nu}$, $\widehat{J}^{\mu}$ and $\widehat{J}^{\lambda\mu\nu}$, i.e., $\partial_{\mu}\widehat{T}^{\mu\nu}=0$, $\partial_{\mu}\widehat{J}^{\mu}=0$, and $\partial_{\lambda}\widehat{J}^{\lambda\mu\nu}=0$. Note that in the presence of the anti-symmetric part of the energy-momentum tensor, the conservation of the total angular momentum tensor, i.e., $\partial_{\lambda}J^{\lambda\mu\nu}=0$, gives rise to non-conservation of the spin tensor. Therefore, if $\widehat{T}^{\mu\nu}_{ (s)}$, $\widehat{T}^{\mu\nu}_{ (a)}$, $\widehat{S}^{\mu\lambda\nu}_{\rm}$, and $\widehat{J}^{\mu}$ are non-vanishing, then in global equilibrium~\cite{Becattini:2013fla,Florkowski:2018fap,Rindori:2020qqa},
\begin{align}
\partial_{\mu}\beta_{\nu}+\partial_{\nu}\beta_{\mu} =0;~~\Omega_{\mu\nu}=-(\partial_{\mu}\beta_{\nu}-\partial_{\mu}\beta_{\nu})/2=\varpi_{\mu\nu}=\text{constant}; ~~\partial_{\mu}\alpha=0. 
\label{equ14review}
\end{align}
Here $\varpi^{\mu\nu}$ represents thermal vorticity.
We can conclude that if $\widehat{T}^{\mu\nu}_{ (a)}\neq 0$, then spin chemical potential $\Omega^{\mu\nu}$ is completely determined  in terms of thermal vorticity.
However, this identification of spin chemical potential and thermal vorticity can be relaxed in local equilibrium.  

The expression of the momentum-dependent spin polarization of $\Lambda$ hyperons produced in heavy-ion collisions can be obtained from the density operator in the Belinfante-Rosenfeld (BR) pseudo-gauge, i.e., for the Belinfante energy-momentum tensor~\cite{BELINFANTE1939887,BELINFANTE1940449,Rosenfeld1940}, 
\begin{align}
\widehat{\rho}_{\mathrm{LEQ}}^B =
\frac{1}{\mathcal{Z}}
\exp\!\left[-\int_{\Sigma(\tau)} d\Sigma_\mu
\left(
\widehat{T}^{\mu\nu}_B\beta_\nu
-\alpha\widehat{J}^{\mu}
\right)
\right].
\label{equ15review}
\end{align}
In this pseudo-gauge, the energy-momentum tensor is symmetric, but the spin tensor is identically zero. Belinfante-Rosenfeld (BR) pseudo-gauge can be obtained from the energy-momentum tensor and spin tensor of the canonical framework, by choosing an appropriate pseudo-gauge parameter $\Phi^{\lambda\mu\nu}$~\cite{Becattini:2013fla,Florkowski:2018fap,Buzzegoli:2021wlg}. Once we know $\widehat{T}^{\mu\nu}_{B}$, a generic energy-momentum tensor ($\widehat{T}^{\mu\nu}_{\Phi}$) and spin tensor ($\widehat{S}^{\lambda\mu\nu}_{\Phi}$) can be obtained using the following pseudo-gauge transformation~\cite{Buzzegoli:2021wlg},
\begin{align}
& \widehat{T}^{\mu\nu}_{\Phi}
= \widehat{T}^{\mu\nu}_{B}
+ \frac{1}{2}\,\partial_{\lambda}
\left(
\widehat{\Phi}^{\lambda\mu\nu}
- \widehat{\Phi}^{\mu\lambda\nu}
- \widehat{\Phi}^{\nu\lambda\mu}
\right),\label{equ16review}\\
&\widehat{S}^{\lambda\mu\nu}_{\Phi}
= -\,\widehat{\Phi}^{\lambda\mu\nu}
+ \partial_{\rho}\widehat{Z}^{\mu\nu\lambda\rho}.
\label{equ17review}
\end{align}
For simplicity, we can set the pseudo-gauge parameter $\widehat{Z}^{\mu\nu\lambda\rho}=0$. Therefore, for a given pseudo-gauge choice, i.e, for a choice of $\widehat{T}^{\mu\nu}_{\Phi}$, and $\widehat{S}^{\lambda\mu\nu}_{\Phi}$, the local equilibrium density operator can be expressed as
\begin{align}
\widehat{\rho}_{\mathrm{LTE}}^{\Phi}
= \frac{1}{\mathcal{Z}}\,
\exp\!\left[
- \int d\Sigma_\mu
\left(
\widehat{T}^{\mu\nu}_{\Phi}\,\beta_\nu
- \frac{1}{2}\,\Omega_{\lambda\nu}\,
\widehat{S}^{\mu\lambda\nu}_{\Phi}
- \widehat{J}^{\mu}\,\alpha
\right)
\right].
\label{equ18review}
\end{align}
Naturally, for the Belinfante-Rosenfeld (BR) pseudo-gauge, Eq.~\eqref{equ18review} boils down to Eq.~\eqref{equ15review}. Moreover, using Eqs.~\eqref{equ16review}-\eqref{equ17review} back into Eq~\eqref{equ18review} one find~\cite{Buzzegoli:2021wlg}, 
\begin{equation}
\widehat{\rho}_{\mathrm{LTE}}^{\Phi}
= \frac{1}{\mathcal{Z}}\,
\exp\!\left\{
- \int d\Sigma_\mu
\left[
\widehat{T}^{\mu\nu}_{B}\,\beta_\nu
- \frac{1}{2}\,(\varpi_{\lambda\nu}-\Omega_{\lambda\nu})\,
\widehat{\Phi}^{\mu\lambda\nu}
- \xi_{\lambda\nu}\,\widehat{\Phi}^{\lambda\mu\nu}
- \widehat{J}^{\mu}\,\alpha
\right]
\right\}.
\label{equ19review}
\end{equation}
Here, $\varpi_{\lambda\nu}$ is the thermal vorticity, and $\xi_{\mu\nu}=(\partial_{\mu}\beta_{\nu}+\partial_{\nu}\beta_{\mu})/2$ is the thermal shear~\cite{Becattini:2013fla,Florkowski:2018fap,Buzzegoli:2021wlg}.
Eq.~\eqref{equ19review} indicates that different pseudo-gauge choices, i.e., for different choices of $\widehat{\Phi}^{\lambda\mu\nu}$, the local density operator $\widehat{\rho}_{\mathrm{LTE}}^{\Phi}$ does not remain invariant. However, in global global equilibrium, where $\varpi_{\mu\nu}=\Omega_{\mu\nu}$ and $\xi_{\mu\nu}=0$, the  $\widehat{\Phi}^{\lambda\mu\nu}$ dependence of the density operator drops~\cite{Becattini:2013fla,Buzzegoli:2021wlg}. The pseudo-gauge dependence of the density operator ($\widehat{\rho}_{\mathrm{LTE}}^{\Phi}$) can also translate into the average spin polarization of hyperons, i.e., the theoretical prediction of the average spin polarization of hyperons depends on the choice of the energy-momentum tensor and the spin tensor~\cite{Buzzegoli:2021wlg}.

Given a density operator in local equilibrium, we can find the expression for the average spin polarization~\cite{Buzzegoli:2021wlg}, 
\begin{align}
S^{\mu}(p)
= \frac{1}{2}\,
\frac{\displaystyle \int_{\Sigma} d\Sigma_{\lambda} p^{\lambda} \,
\operatorname{Tr}\!\left[\gamma^{\mu}\gamma^{5} W^{+}(x,p)\right]}
{\displaystyle \int_{\Sigma} d\Sigma_{\lambda} p^{\lambda} \,
\operatorname{Tr}\!\left[ W^{+}(x,p)\right]}, 
\end{align}
Here $W^{+}(x,p)$ is the particle contribution to the Wigner function ($W_{AB}(x,p)$) of a massive spin-half particle,
\begin{align}
W_{AB}(x,p) = \operatorname{Tr}\!\left(\widehat{\rho}^{\Phi}_{\rm LTE}\,\widehat{W}_{AB}(x,p)\right),
\label{equ21review}
\end{align}
where $A,\,B$ denote the spinorial indices, and $\widehat{W}$ denotes the
Wigner operator,
\begin{equation}
\widehat{W}_{AB}(x,p)
= \int \frac{d^{4}y}{(2\pi)^{4}}\, e^{-i p\cdot y}\,
:\,\bar{\psi}_{B}\!\left(x+\frac{y}{2}\right)
\psi_{A}\!\left(x-\frac{y}{2}\right)\,:\, ,
\label{eq:Wigner_operator}
\end{equation}
and the symbol $:\ :$ denotes the normal ordering. As the local density operator ($\widehat{\rho}^{\Phi}$) carries the pseudo-gauge dependence, naturally the average spin polarization $S^{\mu}(p)$ can also becomes pseudo-gauge dependent, e.g., it can be shown explicitly that the spin polarization in the  Belinfante pseudo-gauge (keeping terms only up to first order in thermodynamic gradient) is~\cite{Buzzegoli:2021wlg}
\begin{align}
S^{\mu}_{B}(p)
= -\frac{1}{8m}\,
\epsilon^{\mu\rho\sigma\tau} p_{\tau}\,
\frac{\displaystyle \int_{\Sigma} d\Sigma_{\lambda} p^{\lambda} \;
f_{FD}(1-f_{FD})\,\varpi_{\rho\sigma}}
{\displaystyle \int_{\Sigma} d\Sigma_{\lambda} p^{\lambda} \; f_{FD}}
-\,\frac{1}{4m}\,
\epsilon^{\mu\alpha\sigma\tau}
\frac{p_{\tau} p^{\rho}}{\varepsilon_{p}}\,
\frac{\displaystyle \int_{\Sigma} d\Sigma_{\lambda} p^{\lambda} \;
f_{FD}(1-f_{FD})\,\widehat{t}_{\alpha}\xi_{\rho\sigma}}
{\displaystyle \int_{\Sigma} d\Sigma_{\lambda} p^{\lambda} \; f_{FD}}. 
\label{equ23review}
\end{align}
Here $\varepsilon_p=\sqrt{p^2+m^2}$ is the single particle energy, and $\widehat{t}_{\lambda}$ is a time-like four vector. On the other hand, for the canonical choice, the expression of the average spin polarization becomes~\cite{Buzzegoli:2021wlg}, 
\begin{align}
S^{\mu}_{C}(p)\simeq S^{\mu}_{B}(p)+\frac{\epsilon^{\lambda\rho\sigma\tau}\,\widehat{t}_{\lambda}
\left(p^{\mu} p_{\tau}-g^{\mu}{}_{\tau} m^{2}\right)}
{8m\,\varepsilon_{p}}\;
\frac{\displaystyle \int_{\Sigma} d\Sigma_{\alpha} p^{\alpha} \;
f_{FD}(1-f_{FD})\,
\left(\varpi_{\rho\sigma}-\Omega_{\rho\sigma}\right)}
{\displaystyle \int_{\Sigma} d\Sigma_{\alpha} p^{\alpha} \; f_{FD}}
\label{equ24review}
\end{align}
From Eqs.~\eqref{equ23review}, and \eqref{equ24review}, we conclude that the spin polarization obtained using the local equilibrium density operator can be pseudo-gauge dependent. Moreover, in the global equilibrium, where $\xi_{\mu\nu}=0$, and $\varpi_{\mu\nu}=\Omega_{\mu\nu}$, expression of spin polarization becomes pseudo-gauge choice independent. Therefore, the choice of a pseudo-gauge can affect the theoretical predictions for the spin polarization measured in heavy-ion collisions. A natural question would be to ask whether it is possible to find a pseudo-gauge invariant form of the local equilibrium density operator. This question has been addressed in Ref.~\cite{Becattini:2025twu}, where the authors obtained a pseudo-gauge invariant form of the local equilibrium density operator, 
\begin{align}
\widehat{\rho}_{\mathrm{LTE}}
= \frac{1}{\mathcal{Z}}\,
\exp\!\left[
-\int_{\Sigma} d\Sigma_{\mu}
\left(
\widehat{T}^{\mu\nu}\,\beta_{\nu}
- \frac{1}{2}\,\varpi_{\lambda\nu}\,\widehat{S}^{\mu\lambda\nu}
- \xi_{\lambda\nu}\,\widehat{S}^{\lambda\mu\nu}
\right)
\right].
\end{align}
The argument to obtain such a pseudo-gauge invariant form of the density operator is based on the assumption that the argument within the exponential should be linear in the energy-momentum tensor and the spin tensor, i.e., 
\begin{align}
\widehat{\rho}_{\mathrm{LTE}}
= \frac{1}{\mathcal{Z}}\,
\exp\!\left[
-\int_{\Sigma} d\Sigma_{\mu}
\left(
\widehat{T}^{\mu\nu} X_{\nu}
+ Y_{\lambda\nu}\,\widehat{S}^{\mu\lambda\nu}
+ Z_{\lambda\nu}\,\widehat{S}^{\lambda\mu\nu}
\right)
\right].
\label{equ26review}
\end{align}
Here $Y_{\lambda\nu}=-Y_{\nu\lambda}$, and $Z_{\lambda\nu}=Z_{\nu\lambda}$. If we want that the argument within the exponential in $\widehat{\rho}_{\mathrm{LTE}}$ to be invariant under pseudo-gauge transformation, then it can be argued that, $Y_{\lambda\nu}= \frac{1}{2}\,\partial_{[\lambda} X_{\nu]}, Z_{\lambda\nu}= -\,\partial_{\{\lambda} X_{\nu\}}$. Moreover, if we demand that $\widehat{\rho}_{\mathrm{LTE}}$ as given in Eq.~\eqref{equ26review} to reproduce the correct density operator in global equilibrium, then we can identify $X_{\mu}\equiv\beta_{\mu}$. Therefore one immediately find, $Y_{\lambda\nu}=-\frac{1}{2}\varpi_{\lambda\nu}$, and $Z_{\lambda\nu}=-\xi_{\lambda\nu}$. Hence, Eq.~\eqref{equ26review} boils down to a pseudo-gauge invariant local equilibrium density operator. We emphasize that the local equilibrium density operator can be used to obtain a hydrodynamic framework~\cite{Huang:2011dc,She:2025qri,She:2024rnx,Tiwari:2024trl}. Therefore, the pseudo-gauge invariant form of the local equilibrium density operator can have non-trivial implications in the development of the spin hydrodynamic framework.

\section{Frameworks of spin hydrodynamics}
\label{sec3}

\subsection{Spin hydrodynamic framework: entropy current analysis}
\label{subsec3a}
In the previous section, we encountered the non-trivial aspect of the pseudo-gauge dependence in spin hydrodynamics. In this section, we elaborate on another non-trivial feature associated with the spin hydrodynamic frameworks. This is associated with the hydrodynamic gradient ordering of spin hydrodynamic variables. Standard hydrodynamic theory (spinless fluid) can be written as gradient ordering of hydrodynamic variables~\cite{Kovtun:2019hdm} 
\begin{align}
T^{\mu\nu}=\mathcal{O}(1)+\mathcal{O}(\partial)+ \mathcal{O}(\partial^2)+\mathcal{O}(\partial^3)+....\\
J^{\mu}=\mathcal{O}(1)+\mathcal{O}(\partial)+ \mathcal{O}(\partial^2)+\mathcal{O}(\partial^3)+....
\end{align}
Here the k-th order derivative of hydrodynamic variables, i.e, derivatives of $T, \mu$, and  $u^{\mu}$, is represented by $\mathcal{O}(\partial^k)$ terms. 
For example $\partial^2 T$, $\partial^2 u$, $\partial T \partial u$, $(\partial T)^2$, $(\partial u)^2$, etc., are $\mathcal{O}(\partial^2)$ terms. 
Hydrodynamic gradient ordering implicitly assumes that $T, \mu, u^{\mu}$ are leading order or $\mathcal{O}(1)$ terms. Equilibrium thermodynamic quantities, i.e, energy density, pressure, and number density, are also $\mathcal{O}(1)$ terms in the gradient expansion, and these thermodynamic quantities can be expressed in terms of $T$, $\mu$.  Similar to the standard hydrodynamic frameworks in spin hydrodynamics, we should be able to write the spin tensor as a  gradient expansion, 
\begin{align}
S^{\lambda\mu\nu}=\mathcal{O}(1)+\mathcal{O}(\partial)+ \mathcal{O}(\partial^2)+\mathcal{O}(\partial^3)+....
\end{align}
Moreover in spin hydrodynamics the hydrodynamic variables are $T$, $\mu$, $u^{\mu}$, and spin chemical potential ($\Omega^{\mu\nu}$). Therefore, we have to specify the gradient ordering of the spin chemical potential to properly define the gradient expansion of macroscopic currents, $T^{\mu\nu}, J^{\mu}, S^{\mu\alpha\beta}$ in spin hydrodynamics. When it comes to the gradient ordering of $\Omega^{\mu\nu}$, different possibilities are allowed. In some studies, spin chemical potential
has been considered as a $\mathcal{O}(\partial)$ term to obtain a spin hydrodynamics framework~\cite{Hattori:2019lfp,Biswas:2023qsw,Biswas:2022bht,Daher:2022xon,Daher:2022wzf}. Moreover, in other studies, a different gradient ordering of spin chemical potential has been considered, e.g., in Refs.~\cite{She:2021lhe,Daher:2022wzf,Dey:2024cwo} spin hydrodynamic framework has been consider where spin chemical potential is $\mathcal{O}(1)$ term. The rationale behind such a different ordering scheme can be understood from Eq.~\eqref{equ13review}. From Eq.~\eqref{equ13review} we find that in global equilibrium, if $\widehat{T}^{\mu\nu}_{ (a)}\neq 0$, then spin chemical potential is completely determined by the thermal vorticity. Since thermal vorticity ($\varpi_{\mu\nu}$) is $\mathcal{O}(\partial)$ in the hydrodynamic gradient expansion, it is natural to consider that spin chemical potential is also $\mathcal{O}(\partial)$. For a symmetric energy-momentum tensor, where  $\widehat{T}^{\mu\nu}_{(a)}=0$, the spin chemical potential and thermal vorticity may not be related to each other. In that case, one can consider the spin chemical potential as a  $\mathcal{O}(1)$ term.  

Different gradient ordering of spin chemical potential gives rise to a different spin hydrodynamic framework.  To demonstrate this, we consider the \textit{phenomenological} framework where the constitutive relations for macroscopic currents are ~\cite{Weyssenhoff:1947iua,Florkowski:2018fap,Florkowski:2017ruc,Hattori:2019lfp,Biswas:2023qsw,Biswas:2022bht,Daher:2022xon,Daher:2022wzf},
\begin{align}
    & T^{\mu\nu}_{} = T^{\mu\nu}_{(0)}+T^{\mu\nu}_{(1)}, ~~T^{\mu\nu}_{(0)}=\varepsilon u^{\mu}u^{\nu}-P\Delta^{\mu\nu}, \label{equ6ver1}\\
    & J^{\mu}_{} = J^{\mu}_{(0)}+J^{\mu}_{(1)}, ~~J^{\mu}_{(0)}=nu^{\mu},\label{equ7ver1}\\
    & S^{\mu\alpha\beta}_{}=S^{\mu\alpha\beta}_{(0)}+S^{\mu\alpha\beta}_{(1)},~~S^{\mu\alpha\beta}_{(0)}=u^{\mu}S^{\alpha\beta}.  \label{equ8ver1}
\end{align}  
Here $T^{\mu\nu}_{(0)}$, $J^{\mu}_{(0)}$, and $S^{\mu\alpha\beta}_{(0)}$ represent $\mathcal{O}(1)$ terms in the hydrodynamic gradient expansion. $T^{\mu\nu}_{(1)}$, $J^{\mu}_{(1)}$, and $S^{\mu\alpha\beta}_{(1)}$ represent first order derivative corrections, i.e, $\mathcal{O}(\partial)$ terms, which satisfy the following conditions, $T^{\mu\nu}_{(1)}u_{\mu}u_{\nu}=0$, $J^{\mu}_{(1)}u_{\mu}=0$, and $S^{\mu\alpha\beta}_{(1)}u_{\mu}=0$ respectively. The spin tensor  $ S^{\mu\alpha\beta}_{}$ is anti-symmetric in the last two indices. $S^{\mu\nu}$ is the spin density tensor which is anti-symmetric, i.e. $S^{\mu\nu}=-S^{\nu\mu}$. In spin hydrodynamic frameworks energy density ($\varepsilon$), pressure ($P$), number density ($n$), and spin density ($S^{\alpha\beta}$) are  
 $\mathcal{O}(1)$ terms which can be expressed in terms of spin hydrodynamic variables, i.e., temperature ($T$), chemical potential ($\mu$), and spin chemical potential. The generalized local thermodynamic relations in the presence of a spin tensor can be written as~\cite{Hattori:2019lfp,Fukushima:2020ucl},
\begin{align} 
& \varepsilon+P =Ts+\mu n+\omega_{\alpha\beta}S^{\alpha\beta},\label{equ9ver1}\\
& d\varepsilon =Tds+\mu dn+\omega_{\alpha\beta}dS^{\alpha\beta},\label{equ10ver1}\\ 
& dP=sdT+n d\mu+S^{\alpha\beta}d\omega_{\alpha\beta}.
\label{equ11ver1}
\end{align}
$s$ is the entropy density, and $\omega^{\alpha\beta}$ is the spin chemical potential. Here we introduced the notation $\omega^{\alpha\beta}$ to denote spin chemical potential. Previously, we used $\Omega^{\alpha\beta}$ as the spin chemical potential in the density operator. $\Omega^{\alpha\beta}$ and $\omega^{\alpha\beta}$ differ in dimension. In the Natural unit $\Omega^{\alpha\beta}$ is dimensionless, but $\omega^{\alpha\beta}$ has mass dimension one. One can consider $\omega^{\alpha\beta}\sim T \Omega^{\alpha\beta}$. Since $T\sim \mathcal{O}(1)$ the derivative ordering of $\omega^{\alpha\beta}$, and $\Omega^{\alpha\beta}$   are same. In spin hydrodynamics, all thermodynamic quantities appearing in the above thermodynamic relations are functions of $T,\mu,\omega^{\alpha\beta}$, i.e., $\varepsilon (T,\mu,\omega^{\alpha\beta})$, $P (T,\mu,\omega^{\alpha\beta})$, $s (T,\mu,\omega^{\alpha\beta})$, and $S^{\alpha\beta} (T,\mu,\omega^{\alpha\beta})$. The above thermodynamic relations also imply, 
\begin{align}
s=\left.\frac{\partial P}{\partial T}\right|_{\mu,\omega^{\alpha\beta}},~~ n=\left.\frac{\partial P}{\partial \mu}\right|_{T,\omega^{\alpha\beta}}, ~~S^{\alpha\beta}=\left.\frac{\partial P}{\partial \omega_{\alpha\beta}}\right|_{T,\mu}
\label{equ36review}
\end{align}
Therefore, once $P (T,\mu,\omega^{\alpha\beta})$ is defined, using Eqs.~\eqref{equ9ver1}-\eqref{equ36review}, we can find, entropy density ($s$), number density ($n$), spin density ($S^{\alpha\beta}$), and energy density ($\varepsilon$). One may wonder what prompts us to consider Eq.~\eqref{equ9ver1} as the generalized thermodynamic relation in spin hydrodynamics. The argument stems from an interesting result in spin kinetic theory, which we discuss in detail in a later section. The argument is the following: in spin kinetic theory, we can define the entropy current in terms of the single-particle distribution function. Moreover, in spin kinetic theory, the single-particle distribution function gets modified because of the presence of spin density and spin chemical potential~\cite{Bhadury:2020cop}. In such a theory, the phase space is enhanced to incorporate the spin chemical potential ($\Omega^{\alpha\beta}$) in the distribution function. If one uses such a generalized spin-dependent distribution function in the definition of the entropy current, then one will end up with an expression of equilibrium entropy current of the form $\mathcal{S}^{\mu}_{\rm eq}=T^{\mu\nu}_{\rm eq}\beta_{\nu}+P\beta^{\mu}-\alpha J^{\mu}_{\rm eq}-\Omega_{\alpha\beta}S^{\mu\alpha\beta}_{\rm eq}.$ This indicates the covariant form of generalized out-of-equilibrium entropy current~\cite{Hattori:2019lfp,Fukushima:2020ucl,Hu:2021lnx,Hu:2022azy,She:2021lhe}, 
\begin{align}
\mathcal{S}^{\mu}=T^{\mu\nu}\beta_{\nu}+P\beta^{\mu}-\alpha J^{\mu}-\beta\omega_{\alpha\beta}S^{\mu\alpha\beta}.
\label{equ12ver1}
\end{align}
Here $\beta^{\mu}=\beta u^{\mu}=u^{\mu}/T$, and $\alpha=\mu/T$. It is interesting to note that, if we remove  $\beta\omega_{\alpha\beta}S^{\mu\alpha\beta}$ term from the above equation, then Eq.~\eqref{equ12ver1} boils down to the expression of the non-equilibrium entropy current in standard dissipative hydrodynamics (spin-less fluid). Moreover, if we write $\mathcal{S}^{\mu}_{\rm eq}\equiv s u^{\mu}$, $T^{\mu\nu}_{\rm eq} \equiv T^{\mu\nu}_{(0)}$, $J^{\mu}_{\rm eq} \equiv J^{\mu}_{(0)}$, and $S^{\mu\alpha\beta}_{\rm eq}\equiv S^{\mu\alpha\beta}_{(0)} $, then the thermodynamic relation in spin hydrodynamics (Eq.~\eqref{equ9ver1}) can be easily obtained from  Eq.~\eqref{equ12ver1}.

If the energy-momentum tensor has anti-symmetric component, then it can be argued that (for a detailed derivation, Appendix A of Ref.~\cite{Dey:2024cwo})  
\begin{align}
\partial_{\mu}\mathcal{S}^{\mu}_{\rm eq}=2\beta\omega_{\alpha\beta}T^{[\alpha\beta]}_{(1)}.
\label{equ14ver1}
\end{align}
To obtain the above equation, one uses spin hydrodynamic equations, i.e., $\partial_{\mu}T^{\mu\nu}=0$, $\partial_{\mu}J^{\mu}=0$, and $\partial_{\mu}J^{\mu\alpha\beta}=0$. This is a very interesting result with far-reaching consequences in the framework of spin hydrodynamics. Note that for an asymmetric energy-momentum tensor, when we consider $\omega^{\alpha\beta}\sim \mathcal{O}(\partial)$, and $S^{\alpha\beta}\sim \mathcal{O}(1)$, the thermodynamic relations indicate that entropy density ($s$) and entropy current ($\mathcal{S}^{\mu}_{\rm eq}$) contains terms up to order  $\mathcal{O}(\partial)$. Naturally, $\partial_{\mu}\mathcal{S}^{\mu}_{\rm eq}$ contains terms up to  $\mathcal{O}(\partial^2)$, which is consistent with the derivative ordering on the right-hand side (RHS) of Eq.~\eqref{equ14ver1}. Therefore, at the level of hydrodynamic derivative ordering, Eq.~\eqref{equ14ver1} is consistent, but in this case the notion of local equilibrium is lost, i.e., $\partial_{\mu}\mathcal{S}^{\mu}_{\rm eq}\neq 0$. However, the global equilibrium still exists in this case~\cite{Hattori:2019lfp}. We emphasize that to obtain the above conclusion, we considered spin density to be $\mathcal{O}(1)$, despite the fact that $\omega^{\alpha\beta}\sim \mathcal{O}(\partial)$. $S^{\alpha\beta}$ should be proportional to $\omega^{\alpha\beta}$, which is known as the spin equation-of-State (EoS). Spin equation-of-state that connects an order $\mathcal{O}(1)$ term to an order $\mathcal{O}(\partial)$ is challenging. In Ref.~\cite{Biswas:2022bht}, the authors argued a possible way to obtain such a spin equation-of-state, namely, $S^{\mu\nu}\sim \omega^{\mu\nu}/\sqrt{\omega^{\alpha\beta}\omega_{\alpha\beta}}$. Now, if we consider the other gradient ordering of spin chemical potential, i.e., $\omega^{\alpha\beta}\sim \mathcal{O}(1)$, then it is easy to argue that $S^{\alpha\beta}\sim \omega^{\alpha\beta}\sim \mathcal{O}(1)$. In this case entropy density ($s$) and entropy current ($\mathcal{S}^{\mu}_{\rm eq}$) contains terms only of order  $\mathcal{O}(1)$. Moreover, in such a situation, the energy-momentum tensor is symmetric, i.e., $T^{[\alpha\beta]}_{(1)}=0$. Therefore, from Eq.~\eqref{equ14ver1} we can conclude that local equilibrium can be achieved, i.e., $\partial_{\mu}\mathcal{S}^{\mu}_{\rm eq}=0$ if we consider $ \omega^{\alpha\beta}\sim \mathcal{O}(1)$.
Moreover, when $\omega^{\alpha\beta}\sim S^{\alpha\beta}\sim \mathcal{O}(1)$, we can obtain the standard hydrodynamic theory by considering the limit $\omega^{\alpha\beta}\rightarrow 0 $.
However, in this case, we cannot express the spin chemical potential in terms of thermal vorticity in global equilibrium.  

Now, taking the divergence of the entropy current, one finds (for a detailed derivation see Ref.~\cite{Dey:2024cwo}),
\begin{align}
\partial_{\mu}\mathcal{S}^{\mu} & = (\partial_{\mu}\beta_{\nu})T^{\mu\nu}+\beta_{\nu}\partial_{\mu}T^{\mu\nu}+\beta^{\mu}\partial_{\mu}P+P\partial_{\mu}\beta^{\mu}\nonumber\\
& ~~~~~-J^{\mu}\partial_{\mu}\alpha-\alpha \partial_{\mu}J^{\mu}-S^{\mu\alpha\beta}\partial_{\mu}(\beta\omega_{\alpha\beta})-(\beta\omega_{\alpha\beta}) \partial_{\mu}S^{\mu\alpha\beta}\nonumber\\
& = T^{\{\mu\nu\}}_{(1)}\partial_{\{\mu}\beta_{\nu\}}+T^{[\mu\nu]}_{(1)}\partial_{[\mu}\beta_{\nu]}-J^{\mu}_{(1)}\partial_{\mu}\alpha-S^{\mu\alpha\beta}_{(1)}\partial_{\mu}(\beta\omega_{\alpha\beta})+2(\beta\omega_{\alpha\beta})T^{[\alpha\beta]}_{(1)}.
\label{equ39review}
\end{align}
To obtain the above equation, one uses the spin hydrodynamic equations Eq.~\eqref{equ5review}. If we consider that the spin chemical potential $\omega^{\alpha\beta}\sim \mathcal{O}(\partial)$, then all terms on the RHS, apart from $S^{\mu\alpha\beta}_{(1)}\partial_{\mu}(\beta\omega_{\alpha\beta})$, are second order in derivative. The term $S^{\mu\alpha\beta}_{(1)}\partial_{\mu}(\beta\omega_{\alpha\beta})$ is $\mathcal{O}(\partial^3)$
in the hydrodynamic gradient expansion. Therefore, in the Navier-Stokes limit (first order theory) where $\partial_{\mu}\mathcal{S}^{\mu}$ contains terms up to second order in derivative, one can remove such higher order terms. Furthermore, demanding entropy must be produced in a dissipative system, i.e., $\partial_{\mu}\mathcal{S}^{\mu} \geq 0$, one can find the constitutive relation for various dissipative currents that appear in $T^{\mu\nu}$, and $J^{\mu}$. But we can not find a constitutive relation for the dissipative part of the spin tensor $S^{\mu\alpha\beta}_{(1)}$, because this gives rise to higher-order terms. On the other hand, if we consider $\omega^{\alpha\beta}\sim \mathcal{O}(1)$, then we must also consider energy-momentum tensor is symmetric. In this case $\partial_{\mu}\mathcal{S}^{\mu}$ boils down to, $\partial_{\mu}\mathcal{S}^{\mu}=T^{\{\mu\nu\}}_{(1)}\partial_{\{\mu}\beta_{\nu\}}-J^{\mu}_{(1)}\partial_{\mu}\alpha-S^{\mu\alpha\beta}_{(1)}\partial_{\mu}(\beta\omega_{\alpha\beta})$, and using the condition for entropy production, we can find the constitutive relation for $T^{\{\mu\nu\}}_{(1)}$, $J^{\mu}_{(1)}$, and $S^{\mu\alpha\beta}_{(1)}$. From the above discussion, it is clear that different gradient orders of the spin chemical potential give rise to distinct spin hydrodynamic frameworks. 

Next, let us first look into the spin hydrodynamic framework with spin chemical potential $\omega^{\alpha\beta}\sim \mathcal{O}(\partial)$~\cite{Hattori:2019lfp,Daher:2022xon}. For simplicity, let us also consider $J^{\mu}=0$. The energy-momentum tensor has symmetric and anti-symmetric parts, i.e., 
\begin{align} 
& T^{\mu\nu}= \varepsilon u^{\mu}u^{\nu}-P\Delta^{\mu\nu}+ T^{\mu\nu}_{(1s)}+T^{\mu\nu}_{(1a)},\label{equ40review}\\
& T^{\mu\nu}_{(1s)} \equiv T^{\{\mu\nu\}}_{(1)}= h^{\mu}u^{\nu}+h^{\nu}u^{\mu}+\pi^{\mu\nu}+\Pi\Delta^{\mu\nu},\label{equ41review}\\ 
&T^{\mu\nu}_{(1a)} \equiv T^{[\mu\nu]}_{(1)}= q^{\mu}u^{\nu}-q^{\nu}u^{\mu}+\phi^{\mu\nu}. 
\label{equ42review}
\end{align}
Here, $h^{\alpha}$, $\pi^{\alpha\beta}$, $\Pi$, $q^{\alpha}$ and $\phi^{\alpha\beta}$ are different dissipative currents and are orthogonal to the fluid flow, i.e., $q^{\alpha} u_{\alpha}=0, h^{\alpha} u_{\alpha}=0, \pi^{\alpha\beta}u_{\alpha}=0, \phi^{\alpha\beta}u_{\alpha}=0, \pi^{\alpha\beta}=\pi^{\beta\alpha}, \phi^{\alpha\beta}=-\phi^{\beta\alpha}$, $\pi^{\alpha}_{~\alpha}=0$. All dissipative currents are $\mathcal{O}(\partial)$ in the hydrodynamic gradient expansion. Using Eqs.~\eqref{equ5review}, \eqref{equ39review}-\eqref{equ42review}, one can show the divergence of the entropy current is, 
\begin{align}
\partial_{\mu}\mathcal{S}^{\mu}= &-\beta h^{\mu}\left(\beta \nabla_{\mu}T-Du_{\mu}\right)+\beta\pi^{\mu\nu}\sigma_{\mu\nu}+\beta\Pi \theta\nonumber\\ & -\beta q^{\mu}\left(\beta \nabla_{\mu}T+Du_{\mu}-4 \omega_{\mu\nu}u^{\nu}\right) +\phi^{\mu\nu}\left(\beta\nabla_{[\mu}u_{\nu]}+2\beta \Delta^{\alpha}_{~\mu}\Delta^{\beta}_{~\nu}\omega_{\alpha\beta}\right). \label{equ43review}
\end{align}
Imposing the second law of thermodynamics, i.e. $\partial_{\mu}\mathcal{S}^{\mu}\geq 0$, one finds the constitutive relations of different dissipative currents~\cite{Hattori:2019lfp,Daher:2022xon}, 
\begin{align} 
& h^{\alpha}=-\kappa\left(Du^{\alpha}-\beta\nabla^{\alpha}T\right),\label{equ44review}\\ 
& \pi^{\alpha\beta}=2\eta\sigma^{\alpha\beta},\label{equ45review}\\
& \Pi = \zeta \theta,\label{equ46review}\\
& q^{\alpha}=\lambda \left(\beta\nabla^{\alpha}T+Du^{\alpha}-4\omega^{\alpha\beta}u_{\beta}\right),\label{equ47review}\\ 
& \phi^{\alpha\beta}=\gamma\left(\beta\nabla^{[\alpha}u^{\beta]}+2\beta \Delta^{\alpha}_{~\mu}\Delta^{\beta}_{~\nu}\omega^{\mu\nu}\right),\nonumber\\
& ~~~~~=2\widetilde{\gamma}\left(\nabla^{[\alpha}u^{\beta]}+2\Delta^{\alpha}_{~\mu}\Delta^{\beta}_{~\nu}\omega^{\mu\nu}\right),\label{equ48review}
\end{align}
Here, $\kappa$, $\eta$, $\zeta$, $\lambda, \gamma\equiv 2\widetilde{\gamma}/\beta$ are different transport coefficients satisfying the conditions: $\kappa\geq 0$, $\eta\geq 0$, $\zeta\geq 0$, $\lambda \geq 0, \gamma\geq 0$. Note the consistency of the hydrodynamic gradient expansion in Eqs.~\eqref{equ44review}-\eqref{equ48review},  both LHS (Left Hand Side) and RHS (Right Hand Side) are order $\mathcal{O}(\partial)$. This is the Navier-Stokes theory of dissipative spin hydrodynamics. In this case, we can not find the constitutive relation for the dissipative part of the spin tensor. Building upon this first-order spin hydrodynamics, one can construct a second-order dissipative spin hydrodynamics, which is analogous to the Israel-Stewart theory of second-order hydrodynamics~\cite {Biswas:2023qsw}.

Now, we look into the spin hydrodynamic framework with spin chemical potential $\omega^{\alpha\beta}\sim \mathcal{O}(1)$~\cite{She:2021lhe,Dey:2024cwo}. In this case, the energy-momentum tensor is symmetric, so the dissipative part of the energy-momentum tensor is given in Eq.~\eqref{equ41review}. Apart from the dissipative part of the energy-momentum tensor, we also need the dissipative part of the spin tensor.  To proceed further, we write a generic decomposition of $S^{\mu\alpha\beta}_{(1)}$ in terms of the irreducible tensors which are $\mathcal{O}(\partial)$ in the gradient expansion~\cite{Becattini:2011ev,Biswas:2023qsw,Daher:2022xon}
\begin{align} 
& S^{\mu\alpha\beta}_{(1)}=2u^{[\alpha}\Delta^{\mu\beta]}\Phi+2u^{[\alpha}\tau^{\mu\beta]}_{(s)}+2u^{[\alpha}\tau^{\mu\beta]}_{(a)}+\Theta^{\mu\alpha\beta}.
\label{equ49review}
\end{align}
The $\mathcal{O}(\partial)$ terms appearing in the spin tensor are $\Phi,\tau^{\mu\nu}_{(s)}, \tau^{\mu\nu}_{(a)} $, and $\Theta^{\mu\alpha\beta}$. These currents satisfy the following conditions: $u_{\mu} \tau_{(s)}^{\mu \beta}=0$; $\tau^{\mu}_{~~\mu(s)}=0$; $u_{\mu} \tau_{(a)}^{\mu \beta}=0$; $u_{\mu} \Theta^{\mu \alpha \beta}=0$; $u_{\alpha} \Theta^{\mu \alpha \beta}=0$;  $ \tau_{(s)}^{\mu \beta}=\tau_{(s)}^{\beta \mu}$; $\tau_{(a)}^{\mu \beta}=-\tau_{(a)}^{\beta\mu}$; $\Theta^{\mu \alpha \beta}=-\Theta^{\mu \beta \alpha}$. Using Eqs.~\eqref{equ5review}, \eqref{equ39review}-\eqref{equ41review}, and Eq.~\eqref{equ49review}, one can show the divergence of the entropy current is, 
\begin{align}
\partial_{\mu}\mathcal{S}^{\mu}= &-\beta h^{\mu}\left(\beta \nabla_{\mu}T-Du_{\mu}\right)+\beta\pi^{\mu\nu}\sigma_{\mu\nu}+\beta\Pi \theta \nonumber\\
& -2\Phi u^{\alpha}\nabla^{\beta}(\beta\omega_{\alpha\beta})-2\tau^{\mu\beta}_{(s)}u^{\alpha}\Delta^{\gamma\rho}_{\mu\beta}\nabla_{\gamma}(\beta\omega_{\alpha\rho})-2\tau^{\mu\beta}_{(a)}u^{\alpha}\Delta^{[\gamma\rho]}_{[\mu\beta]}\nabla_{\gamma}(\beta\omega_{\alpha\rho})\nonumber\\
& -\Theta_{\mu\alpha\beta}\Delta^{\alpha\delta}\Delta^{\beta\rho}\Delta^{\mu\gamma}\nabla_{\gamma}(\beta\omega_{\delta\rho}),
\label{equ50review}
\end{align}
Once again, imposing the second law of thermodynamics, i.e. $\partial_{\mu}\mathcal{S}^{\mu}\geq 0$, one finds the constitutive relations of different dissipative currents. The constitutive relation for $h^{\alpha}$, $\pi^{\alpha\beta}$, and $\Pi$ is same as given in Eqs.~\eqref{equ44review}-\eqref{equ46review}, respectively. The constitutive relations for the dissipative part of the spin tensor can be written as, 
\begin{align}
& \Phi=-2\chi_{1} u^{\alpha}\nabla^{\beta}(\beta\omega_{\alpha\beta}),
\label{equ51review}\\
& \tau^{\mu\beta}_{(s)}=-2\chi_{2}\Delta^{\mu\beta,\gamma\rho}\nabla_{\gamma}(\beta\omega_{\alpha\rho})u^{\alpha},\label{equ52review}\\
& \tau^{\mu\beta}_{(a)}=-2\chi_{3} \Delta^{[\mu\beta][\gamma\rho]}\nabla_{\gamma}(\beta\omega_{\alpha\rho})u^{\alpha},
  \label{equ53review}\\
& \Theta^{\mu\alpha\beta}= \chi_4 \Delta^{\delta\alpha}\Delta^{\rho\beta}\Delta^{\gamma\mu}\nabla_{\gamma}(\beta\omega_{\delta\rho})\label{equ54review}.
\end{align}
The spin transport coefficients $\chi_1\geq 0$, $\chi_2\geq 0$, $\chi_{2}\geq 0$, and $\chi_{4}\geq 0$. In Eqs.~\eqref{equ51review}-\eqref{equ54review}, one can observe the consistency of the hydrodynamic gradient ordering. Since in this case we consider $\omega^{\alpha\beta}\sim\mathcal{O}(1)$, the LHS and RHS are order $\mathcal{O}(\partial)$.
We emphasize that when one considers $\omega^{\alpha\beta}\sim\mathcal{O}(\partial)$, then the spin transport coefficients do not appear within the first-order spin hydrodynamic framework (Navier-Stokes theory)~\cite{Hattori:2019ahi,Fukushima:2020ucl,Daher:2022wzf,Daher:2022xon,Biswas:2022bht}. 

Some interesting features appear when we consider $J^{\mu}\neq 0$ in the entropy current. In this case, the divergence of the entropy current boils down to, $\partial_{\mu}\mathcal{S}^{\mu}=T^{\{\mu\nu\}}_{(1)}\partial_{\{\mu}\beta_{\nu\}}-J^{\mu}_{(1)}\partial_{\mu}\alpha-S^{\mu\alpha\beta}_{(1)}\partial_{\mu}(\beta\omega_{\alpha\beta})$. For the symmetric energy-momentum tensor, using the decomposition of $T^{\{\mu\nu\}}_{(1)}$, and $S^{\mu\alpha\beta}_{(1)}$ as given in Eqs.~\eqref{equ41review}, and Eq.~\eqref{equ49review}, and spin hydrodynamic equation of motion one can show that~\cite{Daher:2022xon,Biswas:2023qsw,Dey:2024cwo}, 
\begin{align}
\partial_{\mu}\mathcal{S}^{\mu}=~& h^{\mu}\frac{S^{\alpha\beta}}{\varepsilon+P}\nabla_{\mu}(\beta\omega_{\alpha\beta})-\mathcal{J}^{\mu}\nabla_{\mu}\alpha+\beta\pi^{\mu\nu}\sigma_{\mu\nu}+\beta\Pi \theta \nonumber\\
& -2\Phi u^{\alpha}\nabla^{\beta}(\beta\omega_{\alpha\beta})-2\tau^{\mu\beta}_{(s)}u^{\alpha}\Delta^{\gamma\rho}_{\mu\beta}\nabla_{\gamma}(\beta\omega_{\alpha\rho})-2\tau^{\mu\beta}_{(a)}u^{\alpha}\Delta^{[\gamma\rho]}_{[\mu\beta]}\nabla_{\gamma}(\beta\omega_{\alpha\rho})\nonumber\\
& -\Theta_{\mu\alpha\beta}\Delta^{\alpha\delta}\Delta^{\beta\rho}\Delta^{\mu\gamma}\nabla_{\gamma}(\beta\omega_{\delta\rho})
\label{equ55review}
\end{align}
Here $\mathcal{J}^{\mu} = J^{\mu}_{(1)}-\frac{n}{\varepsilon+P}h^{\mu}$. Imposing the condition of entropy production, i.e., $\partial_{\mu}\mathcal{S}^{\mu}\geq0$ for a dissipative system, we obtain the constitutive relations for various dissipative currents in terms of the first-order derivative of spin hydrodynamic variables,
\begin{align}
& \Pi=\zeta\theta,\label{equ56review}\\
& h^{\mu}=-\kappa_{11}\frac{S^{\alpha\beta}}{\varepsilon+P}\nabla^{\mu}(\beta\omega_{\alpha\beta})-\kappa_{12}\nabla^{\mu}\alpha,\label{equ57review}\\
& \pi^{\mu\nu}=2\eta\sigma^{\mu\nu},\label{equ58review}\\
& \mathcal{J}^{\mu}=\widetilde{\kappa}_{11}\nabla^{\mu}\alpha+\widetilde{\kappa}_{12}\frac{S^{\alpha\beta}}{\varepsilon+P}\nabla^{\mu}(\beta \omega_{\alpha\beta}),\label{equ59review}\\
& \Phi=-2\chi_{1} u^{\alpha}\nabla^{\beta}(\beta\omega_{\alpha\beta}),
\label{equ60review}\\
& \tau^{\mu\beta}_{(s)}=-2\chi_{2}\Delta^{\mu\beta,\gamma\rho}\nabla_{\gamma}(\beta\omega_{\alpha\rho})u^{\alpha},\label{equ61review}\\
& \tau^{\mu\beta}_{(a)}=-2\chi_{3} \Delta^{[\mu\beta][\gamma\rho]}\nabla_{\gamma}(\beta\omega_{\alpha\rho})u^{\alpha},
  \label{equ62review}\\
& \Theta^{\mu\alpha\beta}= \chi_4 \Delta^{\delta\alpha}\Delta^{\rho\beta}\Delta^{\gamma\mu}\nabla_{\gamma}(\beta\omega_{\delta\rho})\label{equ63review}.
\end{align}
As compared to Eqs.~\eqref{equ44review}-\eqref{equ46review}, \eqref{equ51review}-\eqref{equ54review} we find new transport coefficients in Eqs.~\eqref{equ56review}-\eqref{equ63review}, these are $\kappa_{11}$, $\kappa_{12}$, $\widetilde{\kappa}_{11}$, and $\widetilde{\kappa}_{12}$. These transport coefficients relates $h^{\mu}$, and $\mathcal{J}^{\mu}$ with the gradient of $\beta\omega^{\alpha\beta}$, and $\beta\mu$,  respectively. $\kappa_{11}$, and $\widetilde{\kappa}_{11}$ are diagonal transport coefficients in the $(h^{\mu},\mathcal{J}^{\mu})$ sector. On the other hand, $\kappa_{12}$, $\widetilde{\kappa}_{12}$ are the cross-conductivity transport coefficients~\cite{Dey:2024cwo}. The coefficients $\kappa_{12}$, and  $\widetilde{\kappa}_{12}$ are not independent. Onsager relation relates these transport coefficients, $\kappa_{12}=\widetilde{\kappa}_{12}$~\cite{Landau_Physical_kinetics}. Moreover $\partial_{\mu}\mathcal{S}^{\mu}\geq 0$ implies that, $\kappa_{11}\geq 0$, $\widetilde{\kappa}_{11}\geq 0$ and $\kappa_{12}^{2}-\kappa_{11}\widetilde{\kappa}_{11}\leq0$. The cross-transport coefficients have also been discussed in Ref.~\cite{Hu:2022azy}, where the author discusses the cross-diffusion between the vector currents appearing in the $T^{\mu\nu}_{(1s)}$, and $T^{\mu\nu}_{(1a)}$, i.e., $h^{\mu}$, and $q^{\mu}$, respectively. But the cross transport coefficients that appear in Eqs.~\eqref{equ57review}-\eqref{equ59review} represent cross-diffusion between the vector current of the energy-momentum tensor ($T^{\mu\nu}$) and the conserved current ($J^{\mu}$). Interestingly, terms of the form $\frac{S^{\alpha\beta}}{\varepsilon+P}\nabla_{\mu}(\beta\omega_{\alpha\beta})$ as given in Eqs.~\eqref{equ57review}, and \eqref{equ59review} do not appear in the spin hydrodynamic theory where one considers $\omega^{\alpha\beta}\sim \mathcal{O}(\partial)$. If we consider $\omega^{\alpha\beta}\sim \mathcal{O}(\partial)$, then $\frac{S^{\alpha\beta}}{\varepsilon+P}\nabla_{\mu}(\beta\omega_{\alpha\beta})\sim \mathcal{O}(\partial^2)$ which is a higher ordered gradient term as compared to $\nabla^{\mu}\alpha$. 

In the development of the spin hydrodynamic theories, entropy current (Eq.~\eqref{equ14review}), and generalized thermodynamic relations (Eqs.~\eqref{equ9ver1}-\eqref{equ36review}) play a crucial role. But the entropy current as defined in Eq.~\eqref{equ14review} is not unique. Similar to a pseudo-gauge transformation of the energy-momentum tensor and spin tensor, one can also transform the entropy current without changing the entropy production. Such a transformation of the entropy current is defined as the entropy gauge transformation~\cite{Becattini:2023ouz}.
If $\mathcal{S}^{\mu}$ is the original entropy current to start with, then a new entropy current $\mathcal{S}^{\prime\mu}\equiv \mathcal{S}^{\mu}+\partial_{\lambda}A^{\lambda\mu}$ does not affect the entropy production, i.e., $\partial_{\mu}\mathcal{S}^{\mu}=\partial_{\mu}\mathcal{S}^{\prime\mu}$, if $A^{\lambda\mu}=-A^{\mu\lambda}$. It has been argued in Ref.~\cite{Becattini:2023ouz} that generic thermodynamic relations, i.e., Eq.~\eqref{equ36review}, are not invariant under entropy gauge transformation, e.g., the thermodynamic relation $\left.\frac{\partial P}{\partial T}\right|_{\mu,\omega} = s$ is not invariant under the entropy gauge transformation. In fact, it can be argued that pressure transforms non trivially under the entropy gauge transformation, $P^{\prime}=P+ T u_{\mu}\partial_{\lambda}A^{\lambda\mu}$, hence $\left.\frac{\partial P^{\prime}}{\partial T}\right|_{\mu,\omega} = s^{\prime}+\left.u_{\mu}T\frac{\partial}{\partial T}\partial_{\lambda}A^{\lambda\mu}\right|_{\mu,\omega}$. Here $s^{\prime}\equiv u_{\mu}\mathcal{S}^{\prime\mu}$, and $s^{}\equiv u_{\mu}\mathcal{S}^{\mu}$. In Ref.~\cite{Becattini:2023ouz} authors argue that in general a pseudo-gauge transformation of $T^{\mu\nu}$ and $S^{\lambda\mu\nu}$ does not lead to an entropy-gauge transformation. Hence the divergence of the entropy current under a general pseudo-gauge transformation, may not remain invariant. Although if for certain specific cases, the entropy gauge transformation (Eq.~\eqref{equ12ver1}) is related to the pseudo-gauge transformation of $T^{\mu\nu}$, and $S^{\lambda\mu\nu}$, then it would be interesting to study the transformation of the local thermodynamic relations under such a pseudo-gauge transformation.

The stability and causality of different spin hydrodynamic frameworks has also been looked into in different literature~\cite{Daher:2022wzf,Xie:2023gbo,Lu:2026ceo}. These studies indicate that the first order spin hydrodynamic frameworks (Navier-Stokes limit) are unstable for both the cases where one considers $\omega^{\mu\nu}\sim\mathcal{O}(\partial)$, and $\omega^{\mu\nu}\sim\mathcal{O}(1)$. In Refs.~\cite{Daher:2022wzf,Xie:2023gbo} authors consider the stability studies for non polarized global equilibrium configuration, but in Ref.~\cite{Lu:2026ceo} authors studies the stability and causality with polarized global equilibrium configurations. The non causal and unstable nature of first order spin hydrodynamic framework requires a second order generalization of spin hydrodynamic frameworks. In Refs.~\cite{Xie:2023gbo,Lu:2026ceo} authors considered the Maxwell–Cattaneo forms of the minimal causal spin hydrodynamics. Such a second order spin hydrodynamic equation can be obtained using the entropy current analysis, as discussed in Ref.~\cite{Biswas:2023qsw}. In Ref.~\cite{Daher:2024bah}, the stability and causality properties of the second order spin hydrodynamic framework developed in Ref.~\cite{Biswas:2023qsw} has been discussed. It has been argued that the stability and causality properties can crucially depend on the spin equation-of-state.

\subsection{Kinetic theory framework for massive particles: Classical description of spin}
\label{subsec3b}
The starting point to obtain a spin-hydrodynamic framework within the framework of the classical kinetic theory approach is the spin-dependent equilibrium distribution function~\cite{Bhadury:2020cop,Florkowski:2018fap,Bhadury:2020puc} 
\begin{align}
f^{\pm}_{s,\mathrm{eq}}(x,p,s)& =\exp\left[-p^\mu\beta_\mu(x)\pm\alpha(x)\right]\,
\exp\!\left[\frac{1}{2}\,\bar{\omega}_{\mu\nu}(x)\,s^{\mu\nu}\right]\nonumber\\
& = f_{\rm eq}^{\pm}(x,p)\exp\left[\frac{1}{2} \bar{\omega}_{\mu\nu} (x) s^{\mu\nu} \right]
\label{equ64review}
\end{align}
Here $s^{\alpha\beta}$ represents the internal angular momentum of spin-half particles, which can be expressed in terms of the momentum ($p^{\alpha}$), and spin vector ($s^{\alpha}$)~\cite{Mathisson:1937zz,Weickgenannt:2019dks}, 
\begin{align}
s^{\alpha\beta} = \frac{1}{m} \epsilon^{\alpha\beta\gamma\delta} p_\gamma s_\delta,
\end{align}
Using the definition of $s^{\alpha\beta}$ it is straightforward to show that $s^{\alpha\beta}p_{\alpha}=0$. In the particle rest frame (PRF), the four-momentum of a particle is $p^\mu = (m,0,0,0)$, and $m$ is the mass of the particle. In the PRF the spin four-vector $s^\alpha = (0,\vec{s}_*)$, with $s_{*}^2\equiv\vec{s}_{*}\cdot \vec{s}_{*}=\frac{1}{2}(1+\frac{1}{2})$. The spin-independent distribution function $f_{\rm eq}^{\pm}(x,p)$ can be obtained from Eq.~\eqref{equ64review} by integrating the spin degree of freedom~\cite{Florkowski:2018fap}, 
\begin{align}
\int dS\, f^{\pm}_{s,\mathrm{eq}}(x,p,s)
= \int \frac{m}{\pi s_{*}}\,
d^{4}s\,
\delta\!\left(s \cdot s + s_{*}^2\right)\,
\delta(p \cdot s)\, f^{\pm}_{s,\mathrm{eq}}(x,p,s)=
f^{\pm}_{\mathrm{eq}}(x,p)
\label{equ66review}
\end{align}
In the natural unit both $s^{\mu\nu}$, and $\bar{\omega}^{\mu\nu}$ are dimensionless. $\bar{\omega}^{\mu\nu}$ can be identified with dimensionless spin chemical potential. To obtain Eq.~\eqref{equ66review} one assumes that spin potential $\bar{\omega}^{\mu\nu}$ is small. Note that $\bar{\omega}^{\mu\nu}$ is dimensionless, hence an order by order expansion of $\exp(\bar{\omega}_{\mu\nu}s^{\mu\nu}/2)$ can be performed. After expanding $\exp(\bar{\omega}_{\mu\nu}s^{\mu\nu}/2)$ up to linear order in  $\bar{\omega}_{\mu\nu}$, and using the following identities~\cite{Bhadury:2020cop}, 
\begin{align}
& \int dS=2,\label{equ67review}\\
& \int dS~s^{\alpha}=0,\label{equ68review}\\
& \int dS~s^{\alpha}s^{\beta}=-\frac{2}{3}s_{\star}^2\left(g^{\alpha\beta}-\frac{p^{\alpha}p^{\beta}}{m^2}\right).
\label{equ69review}
\end{align}
one can easily obtain Eq.~\eqref{equ66review}. The spin-dependent distribution function can also be used to obtain the energy-momentum tensor, the spin tensor, and the net number density~\cite{Bhadury:2020cop}, 
\begin{align}
& T^{\mu\nu}_{\rm eq}=\int \frac{d^3p}{(2\pi)^3 \varepsilon_p} \int dS~p^{\mu}p^{\nu}\left(f^{+}_{s,\mathrm{eq}}+f^{-}_{s,\mathrm{eq}}\right)=\int dP \int dS~p^{\mu}p^{\nu}\left(f^{+}_{s,\mathrm{eq}}+f^{-}_{s,\mathrm{eq}}\right),\label{equ70review}\\
& J^{\mu}_{\rm eq}=\int \frac{d^3p}{(2\pi)^3 \varepsilon_p} \int dS~p^{\mu}\left(f^{+}_{s,\mathrm{eq}}-f^{-}_{s,\mathrm{eq}}\right)=\int dP \int dS~p^{\mu}\left(f^{+}_{s,\mathrm{eq}}-f^{-}_{s,\mathrm{eq}}\right),\label{equ71review}\\
& S^{\lambda\mu\nu}_{\rm eq}= \int \frac{d^3p}{(2\pi)^3 \varepsilon_p} \int dS p^{\lambda}s^{\mu\nu}\left(f^{+}_{s,\mathrm{eq}}+f^{-}_{s,\mathrm{eq}}\right)=\int dP \int dS ~p^{\lambda}s^{\mu\nu}\left(f^{+}_{s,\mathrm{eq}}+f^{-}_{s,\mathrm{eq}}\right).\label{equ72review}
\end{align}
Once again, by expanding the spin-dependent distribution function up to linear order in the spin potential and performing the spin and momentum integrations, one can find closed closed-form expression of $T^{\mu\nu}_{\mathrm{eq}}$, $N^{\mu}_{\mathrm{eq}}$, and   $S^{\lambda\mu\nu}_{\mathrm{eq}}$. The macroscopic currents defined in Eqs.~\eqref{equ70review}-\eqref{equ72review} can be obtained from the generating function~\cite{Drogosz:2025iyr,Drogosz:2025ihp,Bhadury:2025boe} 
\begin{align}
\mathcal{G}_{}
= 2 \int dP \, \cosh \alpha \, e^{- p \cdot \beta}
\int dS \, \exp\!\left( \frac{1}{2}\, \bar{\omega}_{\mu\nu} s^{\mu\nu} \right)= 2 \int dP \, \cosh \alpha \, e^{- p \cdot \beta}
\int dS \, \exp\!\left( \frac{1}{2}\, \bar{\omega}: s \right). 
\end{align}
One can argue that $T^{\mu\nu}_{\mathrm{eq}}$, $N^{\mu}_{\mathrm{eq}}$, and $S^{\lambda\mu\nu}_{\mathrm{eq}}$ can be obtained from generating function ($\mathcal{G}$), 
\begin{align}
T^{\mu\nu}_{\mathrm{eq}}
= \frac{\partial^2 \mathcal{G}}{\partial \beta_\mu \, \partial \beta_\nu}, ~~J^\mu_{\mathrm{eq}}
= - \frac{\partial^2 \mathcal{G}}{\partial \beta_\mu \, \partial \alpha}, ~~S^{\lambda\mu\nu}_{\mathrm{eq}}
= - \frac{\partial^2 \mathcal{G}}{\partial \beta_\lambda \, \partial \bar{\omega}_{\mu\nu}}.
\label{equ74review}
\end{align}
In the small $\bar{\omega}_{\mu\nu}$ limit one can expand the exponential involving $\bar{\omega}_{\mu\nu}$, and keep terms up to linear order in $\bar{\omega}_{\mu\nu}$. In this way, one can get a closed-form expression of $T^{\mu\nu}_{\mathrm{eq}}$, $N^\mu_{\mathrm{eq}}$, and  $S^{\lambda\mu\nu}_{\mathrm{eq}}$. However, the generating functional $\mathcal{G}$ also allow one to 
calculate the macroscopic currents for all orders of $\bar{\omega}_{\mu\nu}$~\cite{Drogosz:2025iyr}, by first writing down $\mathcal{G}$ in the following way, 
\begin{align}
\mathcal{G}= 2 \int dP \, e^{-p \cdot \beta} \, \cosh \alpha
\int dS \sum_{n=0}^{\infty} \frac{(\bar{\omega} : s)^n}{n! \, 2^{n}} \, .
\end{align}
The above equation can be simplified using the following identity involving spin ($s^{\alpha}$)~\cite{Wagner:2024rbt},
\begin{align}
\int dS\, s^{\mu_1}\cdots s^{\mu_n}
=
\begin{cases}
\displaystyle
 \, \frac{2\,(-s_{\star}^2)^{n/2}}{n+1}\,
K^{(\mu_1\mu_2}K^{\mu_3\mu_4}\cdots K^{\mu_{n-1}\mu_n)} ,
& \text{if $n$ is even}, \\[8pt]
0,
& \text{if $n$ is odd}.
\end{cases}
\label{equ76review}
\end{align}
Here $K^{\mu\nu}=g^{\mu\nu}-\frac{p^{\mu}p^{\nu}}{m^2}$, and $T^{(\mu_1\mu_2...\mu_n)}\equiv \frac{1}{n!}\sum_{\sigma}T^{\sigma(\mu_1)\sigma(\mu_2)\cdots\sigma(\mu_n)}$, $\sigma$ represents  permutations of the indices. The identities given in Eqs.~\eqref{equ67review}-\eqref{equ69review} are special cases of Eq.~\eqref{equ76review}. Using Eq.~\eqref{equ76review} one can perform the integration to obtain the generating function $\mathcal{G}$ as a function of $\beta^{\mu},\alpha, \bar{\omega}_{\mu\nu}$ for all orders of $\bar{\omega}^{\mu\nu}$. Such a closed form expression of $\mathcal{G}$ and using Eqs.~\eqref{equ74review} one can find ideal spin hydrodynamics at all order in $\bar{\omega}^{\mu\nu}$~\cite{Drogosz:2025iyr}.

The dissipative corrections to the energy-momentum tensor, number current, and spin tensor can be written in terms of the out-of-equilibrium distribution function ($\delta f^{\pm}_{s}$), 
\begin{align}
& \delta T^{\mu\nu}=\int dP \int dS~p^{\mu}p^{\nu}\left(\delta f^{+}_{s}+\delta f^{-}_{s}\right),\label{equ77review}\\
& \delta J^{\mu}=\int dP \int dS~p^{\mu}\left(\delta f^{+}_{s}-\delta f^{-}_{s}\right),\label{equ78review}\\
& \delta S^{\lambda\mu\nu}_{}=\int dP \int dS~p^{\lambda}s^{\mu\nu}\left(\delta f^{+}_{s}+\delta f^{-}_{s}\right).\label{equ79review}
\end{align}
The out-of-equilibrium distribution function ($\delta f^{\pm}_{s}$) is not arbitrary, but it satisfies the spin kinetic equation,
\begin{align}
p^{\mu}\partial_{\mu}f^{\pm}_{s}(x,p,s)=\mathcal{C}[f^{\pm}_{s}(x,p,s)],
\label{equ80review}
\end{align}
where $\mathcal{C}[f^{\pm}_{s}(x,p,s)]$ is the collision term which simplifies to,
\begin{align}
C\!\left[f^{\pm}_{s}(x,p,s)\right]
= p^{\alpha}u_{\alpha} \,
\frac{f^{\pm}_{s,\rm{eq}}(x,p,s) - f^{\pm}_{s}(x,p,s)}{\tau} \, 
\label{equ81review}
\end{align}
in the relaxation time ($\tau$) approximation. In the Chapman-Enskog-like expansion, the single particle distribution function ($f^{\pm}_{s}(x,p,s)$) can be expanded about its equilibrium value ($f^{\pm}_{s,\rm{eq}}(x,p,s)$) in powers of space-time gradients $f^{\pm}_{s}(x,p,s)=f^{\pm}_{s,\rm{eq}}(x,p,s)+\delta f^{\pm}_{s}(x,p,s)$~\cite{Jaiswal:2013npa, Jaiswal:2013vta, Jaiswal:2016hex}. In this approximation, keeping only the first-order terms in space-time gradients, one finds
\begin{align}
p^{\mu}\,\partial_{\mu} f^{\pm}_{s,\mathrm{eq}}(x,p,s)
= -\, p \cdot u \,
\frac{\delta f^{\pm}_{s}(x,p,s)}{\tau} \, .
\label{equ82review}
\end{align}
Using Eq.~\eqref{equ64review} in Eq.~\eqref{equ82review} one can find $\delta f^{\pm}_{s}(x,p,s)$ in terms of the first order gradient of temperature, chemical potential, and spin chemical potential, i.e., $\partial_{\mu}\beta_{\lambda}$, $\partial_{\mu}\alpha$, and $\partial_{\mu}\bar{\omega}_{\alpha\beta}$~\cite{Bhadury:2020puc, Bhadury:2020cop}. Using such an expression of $\delta f^{\pm}_{s}(x,p,s)$ in Eqs.~\eqref{equ77review}-\eqref{equ79review} one obtains dissipative corrections $\delta T^{\mu\nu}$, $\delta J^{\mu} $, and $\delta S^{\lambda\mu\nu}$~\cite{Bhadury:2020puc, Bhadury:2020cop}. In the kinetic-theory approach, one determines the structure of dissipative currents and associated kinetic coefficients, including spin transport coefficients that characterize spin diffusion.  

\subsection{Quantum Kinetic theory framework}
\label{subsec3c}
Since spin is inherently a quantum phenomenon, its dynamics can be described using quantum kinetic theory. The key quantity relevant for the quantum kinetic theory is the Wigner function~\cite{DeGroot:1980dk,Vasak:1987um,Das:2022azr}
\begin{align}
W_{\alpha\beta}(x,k)=\int\frac{d^4y}{(2\pi\hbar)^4}~e^{-\frac{i}{\hbar}k\cdot y}\langle:\bar{\psi}_{\beta}\left(x+\frac{y}{2}\right)\psi_{\alpha}\left(x-\frac{y}{2}\right):\rangle. 
 ~~\label{eq:WigFunc}
\end{align}
$\psi$, and $\bar{\psi} \equiv \psi^{\dagger} \gamma^{0}$ denotes Dirac field operator and its adjoint, respectively.  The Dirac field operator satisfies the Dirac equation~\cite{DeGroot:1980dk}
\begin{align}
 \left( i\hbar \gamma^{\mu}\partial_{\mu}-m\right)\!\psi(x)=\hbar\rho(x)\,,
\label{eq:DirEq}
\end{align}
here $\rho(x)$ = $-(1/\hbar)\partial\mathcal{L}_I/\partial \bar{\psi}$.  $\mathcal{L}_{I}(x)$ denotes the interaction part of the Dirac Lagrangian. $m$ is the mass of the spin-half particle. Using the Dirac equation, one can obtain the transport equation for the Wigner function~\cite{DeGroot:1980dk},
\begin{align}
\left[\gamma^{\mu} k_{\mu}
+\frac{i\hbar}{2}\gamma^{\mu}\partial_{\mu}-m\right]W_{\alpha\beta}(x,k)=\hbar~\mathcal{C}_{\alpha\beta}.
 \label{equ85review2}
\end{align}
Here, $\mathcal{C}_{\alpha\beta}$ is the collision kernel defined in terms of the interaction term~\cite{DeGroot:1980dk}
\begin{align}
 \mathcal{C}_{\alpha\beta}\equiv \int \frac{d^4y}{(2\pi\hbar)^4}e^{-\frac{i}{\hbar}k \cdot y}\langle:\rho_{\alpha}\left(x-\frac{y}{2}\right)\bar{\psi}_{\beta}\left(x+\frac{y}{2}\right):\rangle.
 \label{collisionterm}
\end{align}
The Wigner function $W(x,k)$ can be decomposed in the Clifford basis 
\begin{align}
W(x,k)=\frac{1}{4}\left[\boldsymbol{1} \, \mathcal{F}(x,k)+i \, \gamma^5 \, \mathcal{P}(x,k)+\gamma^{\mu} \, \mathcal{V}_{\mu}(x,k)+\gamma^5 \, \gamma^{\mu} \, \mathcal{A}_{\mu}(x,k)+\Sigma^{\mu\nu} \, \mathcal{S}_{\mu\nu}(x,k)\right],
 \label{equ87review2}
\end{align}
here $\Sigma^{\mu\nu}\equiv (1/2) \sigma^{\mu\nu}\equiv (i/4)[\gamma^\mu,\gamma^\nu]$ is the Dirac spin operator. The expansion coefficients of the Wigner function $\mathcal{F}$, $\mathcal{P}$, $\mathcal{V}_{\mu}$, $\mathcal{A}_{\mu}$  and $\mathcal{S}_{\mu\nu}$ transform as a scalar, pseudo-scalar, vector, axial-vector, and tensor, respectively, under Lorentz transformations ~\cite{Vasak:1987um}. Using the expansion of the Wigner function as given in Eq.~\eqref{equ87review2} back into Eq.~\eqref{equ85review2}, one obtains kinetic equations for different components of the Wigner function~\cite{Itzykson:1980rh}. These equations are generically complex. After separating the real and imaginary parts of these kinetic equations, it yields two sets of equations involving $\mathcal{F},\mathcal{P},\mathcal{V}^{\mu}, \mathcal{A}^{\mu}$ and $\mathcal{S}^{\mu\nu}$. The real parts give rise to the following equations, 

\begin{align}
& k^{\mu}\mathcal{V}_{\mu}-m \, \mathcal{F}=\hbar \, \mathcal{D}_{\mathcal{F}}\,,\label{equ88review2}\\
& \frac{\hbar}{2} \, \partial^{\mu}\mathcal{A}_{\mu}+m \, \mathcal{P}=-\hbar \, \mathcal{D}_{\mathcal{P}}\,,\label{equ89review2}\\
& k_{\mu}\mathcal{F}-\frac{\hbar}{2} \, \partial^{\nu}\mathcal{S}_{\nu\mu}-m \, \mathcal{V}_{\mu}=\hbar \, \mathcal{D}_{\mathcal{V},\mu}\,,\label{equ90review2}\\
&  -\frac{\hbar}{2} \, \partial_{\mu}\mathcal{P}+k^{\beta}\hstar{\mathcal{S}}_{\mu\beta}
 +m \, \mathcal{A}_{\mu}=-\hbar \, \mathcal{D}_{\mathcal{A},\mu}\,,\label{equ91review2}\\
& \frac{\hbar}{2} \, \left(\partial_{\mu}\mathcal{V}_{\nu}-\partial_{\nu}\mathcal{V}_{\mu}\right)-\epsilon_{\mu\nu\alpha\beta}k^{\alpha}\mathcal{A}^{\beta}-m \, \mathcal{S}_{\mu\nu}=\hbar \, \mathcal{D}_{\mathcal{S},{\mu\nu}}\label{equ92review2}\,.
\end{align}
In the above, the dual form of the tensor $\mathcal{S}^{\mu\nu}$ is defined as
$\hstar{\mathcal{S}}^{\mu\nu} = \frac{1}{2}\epsilon^{\mu\nu\alpha\beta}\mathcal{S}_{\alpha\beta}$, and $\mathcal{D}_{\mathcal{X}}\equiv \Re\text{Tr}\big[\Gamma_{\mathcal{X}}\mathcal{C}\big]$. Here $\mathcal{X}\in \left\{ \mathcal{F}, \mathcal{P}, \mathcal{V}, \mathcal{A}, \mathcal{S}\right\}$, with $\Gamma_\mathcal{F}=\boldsymbol{1}$, $\Gamma_\mathcal{P}=-i\gamma_5$, $\Gamma_\mathcal{V}=\gamma^{\mu}$, $\Gamma_\mathcal{A}=\gamma^{\mu}\gamma_5$, and $\Gamma_\mathcal{S}=2\Sigma^{\mu\nu}$~\cite{Das:2022azr}. On the other hand, the imaginary parts of the kinetic equations boil down to the following equations, 
\begin{align}
& \hbar~\partial^{\mu}\mathcal{V}_{\mu}=2\hbar \, \mathcal{C}_{\mathcal{F}}\,,\label{equ93review2}\\
& k^{\mu}\mathcal{A}_{\mu}=\hbar \, \mathcal{C}_{\mathcal{P}}\,,\label{equ94review2}\\
& \frac{\hbar}{2}\partial_{\mu}\mathcal{F}+k^{\nu}\mathcal{S}_{\nu\mu}=\hbar \, \mathcal{C}_{\mathcal{V},\mu}\,,\label{equ95review2}\\   
& k_{\mu}\mathcal{P}+\frac{\hbar}{2}\partial^{\beta}
\hstar{\mathcal{S}}_{\mu\beta}=-\hbar \,\mathcal{C}_{\mathcal{A},\mu}\,,\label{equ96review2}\\
& k_{\mu}\mathcal{V}_{\nu}-k_{\nu}\mathcal{V}_{\mu}
+\frac{\hbar}{2}\epsilon_{\mu\nu\alpha\beta}\partial^{\alpha}\mathcal{A}^{\beta}=-\hbar \, \mathcal{C}_{\mathcal{S},\mu\nu}\,.\label{equ97review2}
\end{align}
In Eqs.~\eqref{equ93review2}-\eqref{equ97review2}, $\mathcal{C}_{\mathcal{X}}\equiv\Im\text{Tr}\big[\Gamma_{\mathcal{X}}\mathcal{C}\big]$. From the above equation, it is evident that different components of the Wigner function are coupled. However, in the massless limit ($m=0$), and in the absence of a collision term, i.e., $\mathcal{C}=0$, the vector ($\mathcal{V}^{\mu}$) and the axial vector  ($\mathcal{A}^{\mu}$) components  are  coupled only to each other, but not with other components of the Wigner function. In the collisionless limit and for the massless case, the equations involving $\mathcal{V}^{\mu}$ and $\mathcal{A}^{\mu}$ are
\begin{align}
& k^{\mu}\mathcal{V}_{\mu}=0, \qquad k^{\mu}\mathcal{A}_{\mu}=0. \label{equ98review2}\\
& \partial^{\mu}\mathcal{V}_{\mu}=0, \qquad \partial^{\mu}\mathcal{A}_{\mu}=0 \label{equ99review2}\\
& \frac{\hbar}{2} \, \left(\partial_{\mu}\mathcal{V}_{\nu}-\partial_{\nu}\mathcal{V}_{\mu}\right)=\epsilon_{\mu\nu\alpha\beta}k^{\alpha}\mathcal{A}^{\beta},\label{equ100review2}\\
& k_{\mu}\mathcal{V}_{\nu}-k_{\nu}\mathcal{V}_{\mu}
+\frac{\hbar}{2}\epsilon_{\mu\nu\alpha\beta}\partial^{\alpha}\mathcal{A}^{\beta}=0\label{equ101review2}. 
\end{align}
Eqs.~\eqref{equ98review2}-\eqref{equ101review2} are particularly interesting as they give rise to spin hydrodynamic equations within the framework of chiral kinetic theory. Note that the canonical energy-momentum tensor ($T^{\mu\nu}$), spin tensor ($S^{\lambda\mu\nu}$), vector current ($J^{\mu}$), axial vector current ($J^{\mu}_5$), can be expressed in terms of $\mathcal{V}^{\mu}$, and  $\mathcal{A}^{\mu}$~\cite{Shi:2020htn,Speranza:2020ilk,Shi:2020qrx}, 
\begin{align}
& T^{\mu\nu}=\int\frac{d^4k}{(2\pi)^4} k^{\nu}\mathcal{V}^{\mu}, \label{equ102review2}\\
& S^{\lambda\mu\nu} = \frac{\hbar}{2}\epsilon^{\sigma\lambda\mu\nu}\int\frac{d^4k}{(2\pi)^4}\mathcal{A}_{\sigma}, \label{equ103review2}\\
& J^{\mu} = \int\frac{d^4k}{(2\pi)^4} \mathcal{V}^{\mu}, \label{equ104review2}\\
& J^{\mu}_5 = \int\frac{d^4k}{(2\pi)^4} \mathcal{A}^{\mu}.\label{equ105review2}
\end{align}
Using Eqs.~\eqref{equ99review2}-\eqref{equ101review2} back into Eqs.~\eqref{equ102review2}-\eqref{equ105review2} one immediately finds, 
\begin{align}
& \partial_{\mu}T^{\mu\nu}=\int\frac{d^4k}{(2\pi)^4} k^{\nu}\partial_{\mu}\mathcal{V}^{\mu}=0.\label{equ106review2}\\
& \partial_{\mu}J^{\mu}=\int\frac{d^4k}{(2\pi)^4} \partial_{\mu}\mathcal{V}^{\mu}=0.\label{equ107review2}\\
& \partial_{\mu}J^{\mu}_{5}=\int\frac{d^4k}{(2\pi)^4} \partial_{\mu}\mathcal{A}^{\mu}=0.\label{equ108review2}\\
& \partial_{\lambda}S^{\lambda\mu\nu} = \int \frac{d^4k}{(2\pi)^4} \frac{\hbar}{2}\epsilon^{\sigma\lambda\mu\nu}\partial_{\lambda}\mathcal{A}_{\sigma}\nonumber\\
& \qquad ~~~= \int \frac{d^4k}{(2\pi)^4} \left(k^{\mu}\mathcal{V}^{\nu}-k^{\nu}\mathcal{V}^{\mu}\right)=T^{\nu\mu}-T^{\mu\nu}.\label{equ109review2}
\end{align}
Eqs.~\eqref{equ106review2}-\eqref{equ109review2} are the conservation of energy-momentum tensor, vector current, axial vector current, and the total angular momentum tensor, respectively. It is important to note that the canonical energy-momentum tensor is not symmetric, i.e., $T^{\mu\nu}\neq T^{\nu\mu}$, hence the spin tensor is not separately conserved, i.e, $\partial_{\lambda}S^{\lambda\mu\nu}\neq 0$. But the total angular momentum tensor $J^{\lambda\mu\nu}\equiv x^{\mu}T^{\lambda\nu}-x^{\nu}T^{\lambda\mu}+S^{\lambda\mu\nu}$ is conserved, i.e, $\partial_{\lambda}J^{\lambda\mu\nu}=0$. In the chiral limit Dirac field $\psi$ can be decomposed into left chiral ($\psi_L=(1-\gamma_5)\psi/2$) and right chiral parts ($\psi_R=(1+\gamma_5)\psi/2$). Naturally, the vector current ($\mathcal{V}^{\mu}$) and axial vector current ($\mathcal{A}^{\mu}$) can be used to write down the right-handed (RH) and left-handed
(LH) currents, $\mathcal{J}^{\mu}_{\pm}=(\mathcal{V}^{\mu}\pm\mathcal{A}^{\mu})/2$. Using Eqs.~\eqref{equ98review2}-\eqref{equ101review2} it can be shown that~\cite{Shi:2020qrx,Shi:2020htn,Huang:2018wdl,Hidaka:2016yjf,Chen:2015gta}, 
\begin{align}
& k_{\mu}\mathcal{J}^{\mu}_{\pm} = \frac{1}{2}\left(k_{\mu}\mathcal{V}^{\mu}\pm k_{\mu}\mathcal{A}^{\mu}\right)=0,\label{equ110review2}\\
& \partial_{\mu}\mathcal{J}^{\mu}_{\pm} = \frac{1}{2}\left(\partial_{\mu}\mathcal{V}^{\mu}\pm \partial_{\mu}\mathcal{A}^{\mu}\right)=0,\label{equ111review2} \\
& \frac{\hbar}{2}\epsilon_{\mu\nu\rho\sigma}\partial^{\rho}\mathcal{J}^{\sigma}_{\pm}=\frac{\hbar}{4}\epsilon_{\mu\nu\rho\sigma}\partial^{\rho}\mathcal{V}^{\sigma}\pm \frac{\hbar}{4}\epsilon_{\mu\nu\rho\sigma}\partial^{\rho}\mathcal{A}^{\sigma}\nonumber\\
& ~~~~~~~~~~~~~~~~~= -\frac{1}{2}\left(k_{\mu}\mathcal{A}_{\nu}-k_{\nu}\mathcal{A}_{\mu}\right)\mp\frac{1}{2}\left(k_{\mu}\mathcal{V}_{\nu}-k_{\nu}\mathcal{V}_{\mu}\right)\nonumber\\
& ~~~~~~~~~~~~~~~~~= \pm\left(k_{\nu}\mathcal{J}_{\pm \mu}-k_{\mu}\mathcal{J}_{\pm \nu}\right).\label{equ112review2}
\end{align}
In general, solving the quantum kinetic equations~\eqref{equ110review2}--\eqref{equ112review2} completely is difficult due to their complicated structure involving different components of the Wigner function. But using the semi-classical approach (in the power of $\hbar$) one can find a closed form expression of $\mathcal{J}^{\mu}_{\pm}$ as a power series expansion in $\hbar$, and the chiral kinetic equation for $\mathcal{J}^{\mu}_{\pm}$, which has also been discussed in literature extensively~\cite{Shi:2020qrx,Shi:2020htn,Huang:2018wdl,Hidaka:2016yjf,Chen:2015gta}. A causal and stable spin hydrodynamic framework based on the chiral kinetic theory framework has been discussed in Ref.~\cite{Shi:2020qrx,Shi:2020htn}.

We observed that significant simplification occurs in the chiral limit. But for the massive spin-half particles, different components of the Wigner function couple together as can be seen from Eqs.~\eqref{equ88review2}-\eqref{equ97review2}. In this case, one can also take the approach of  
semi-classical expansion to reduce the complexity of the equations by expanding Eqs.~\eqref{equ88review2}-\eqref{equ97review2} in different powers of $\hbar$, which leads to a set of independent equations. To achieve this semi-classical expansion, we expand various components of the Wigner function in the form of a series expansion in $\hbar$,
${\mathcal{X}}=\sum_{n} \hbar^n{\mathcal{X}}^{(n)}$. Here $\mathcal{X}$ represents different components of the Wigner function, i.e., $\mathcal{F}$, $\mathcal{P}$, $\mathcal{V}^{\mu}$, $\mathcal{A}^{\mu}$, and $\mathcal{S}^{\mu\nu}$. Similarly, the respective collision terms $\mathcal{C}_\mathcal{X}$ and $\mathcal{D}_\mathcal{X}$ can also be expanded in the powers of $\hbar$, i.e., 
$\mathcal{C}_\mathcal{X}=\sum_n\hbar^{n}{\mathcal{C}^{(n)}_\mathcal{X}}$ and $\mathcal{D}_\mathcal{X}=\sum_n\hbar^{n}{\mathcal{D}^{(n)}_\mathcal{X}}$. Inserting the expansions of $\mathcal{F}$, $\mathcal{P}$, $\mathcal{V}^{\mu}$, $\mathcal{A}^{\mu}$, $\mathcal{S}^{\mu\nu}$, $\mathcal{C}_\mathcal{X}$, and $\mathcal{D}_\mathcal{X}$ back into Eqs.~\eqref{equ88review2}-\eqref{equ97review2}, we get simplified kinetic equations at different orders of $\hbar$.  

In the zeroth order in $\hbar$ (which is also known as the leading order), Eqs.~\eqref{equ88review2}-\eqref{equ92review2} become,
\begin{align}
& k^{\mu}\mathcal{V}_{\mu}^{(0)}-m\mathcal{F}^{(0)}= 0\,,\label{equ113review2}\\
& m\mathcal{P}^{(0)}=0\,,\label{equ114review2}\\
& k_{\mu}\mathcal{F}^{(0)}-m\mathcal{V}_{\mu}^{(0)}=0\,,\label{equ115review2}\\
& k^{\beta}\hstar{\mathcal{S}}_{\mu\beta}^{(0)}+m\mathcal{A}_{\mu}^{(0)}=0\,, \label{equ116review2}\\
& \epsilon_{\mu\nu\alpha\beta}k^{\alpha}\mathcal{A}^{\beta(0)}+m\mathcal{S}_{\mu\nu}^{(0)}=0\,,  
\label{equ117review2}
\end{align}
while Eqs.~\eqref{equ93review2}-\eqref{equ97review2} yield~\cite{Vasak:1987um}
\begin{align}
& k^{\mu}\mathcal{A}_{\mu}^{(0)}=0\,,\label{equ118review2}\\
& k^{\nu}\mathcal{S}_{\nu\mu}^{(0)}=0\,,\label{equ119review2}\\
& k_{\mu}\mathcal{P}^{(0)}=0\,,\label{equ120review2}\\
& k_{\mu}^{}\mathcal{V}_{\nu}^{(0)}-k_{\nu}^{}\mathcal{V}_{\mu}^{(0)}=0\,.\label{equ121review2}
\end{align}
From Eqs.~\eqref{equ113review2}--\eqref{equ121review2}, we observe that if we consider $\mathcal{F}^{(0)}$ and $\mathcal{A}_{\mu}^{(0)}$ as independent components, then other components of the Wigner function, i.e., $\mathcal{P}^{(0)}$, $\mathcal{V}_{\mu}^{(0)}$, and $\mathcal{S}_{\mu\nu}^{(0)}$ can be obtained in terms of $\mathcal{F}^{(0)}$ and $\mathcal{A}_{\mu}^{(0)}$. Note that $k^{\mu}\mathcal{A}_{\mu}^{(0)}=0$, which implies that only three components of $\mathcal{A}_{\mu}^{(0)}$ are independent, which are the three degrees of freedom canonical spin tensor (Eq.~\eqref{equ103review2}). We emphasize that at the leading order, i.e., $\hbar\rightarrow 0$, all terms involving $\hbar$ drop away, and Eqs.~\eqref{equ88review2}-\eqref{equ97review2} give rise to only algebraic equations, but no dynamics. In $\hbar\rightarrow 0$ limit, Eq.~\eqref{equ114review2} implies that 
$\mathcal{P}^{(0)}\!=\!0$, i.e., the leading-order pseudo-scalar component vanishes. On the other hand, using Eq.~\eqref{equ113review2} and \eqref{equ115review2}  one finds~\cite{Vasak:1987um,Florkowski:2018ahw,Gao:2020pfu} the on-shell condition for $\mathcal{F}^{(0)}_{}$~\cite{Gao:2020pfu}, i.e.,  
\begin{align}
    (k^2-m^2)\mathcal{F}^{(0)}=0.
    \label{equ46.1}
\end{align}
It is important to note that $k^{\mu}$ appearing in the definition of the Wigner function (Eq.~\eqref{eq:WigFunc}) is not the four-momentum of the particle. Only in the limit  $\hbar\rightarrow 0$ ,  can one identify $k^{\mu}$ as the momentum of the particle satisfying the on-shell condition, $k^2=m^2$. Interestingly, the energy-momentum tensor $T^{\mu\nu}$, as defined in Eq.~\eqref{equ102review2}, is by definition not symmetric under  change of $\mu\leftrightarrow\nu$. However, at the leading order of the semi-classical expansion, Eq.~\eqref{equ115review2} allows us to write the energy-momentum tensor as, 
\begin{align}
T^{\mu\nu}_{(0)}=\int\frac{d^4k}{(2\pi)^4} k^{\nu}\mathcal{V}^{\mu}_{(0)}= \frac{1}{m}\int\frac{d^4k}{(2\pi)^4} k^{\mu}k^{\nu}\mathcal{F}_{(0)},
\end{align}
which is manifestly symmetric with the change of $\mu\leftrightarrow\nu$. It is important to note that $\mathcal{F}_{(0)}$ is associated with $\delta(k^2-m^2)$ to manifest the on-shell condition. Similar on-shell condition for $\mathcal{V}^{\mu}_{(0)}$, $\mathcal{A}^{\mu}_{(0)}$, and $\mathcal{S}^{\mu\nu}_{(0)}$ can also be obtained using Eqs.~\eqref{equ113review2}-\eqref{equ121review2}~\cite{Das:2022azr}, 
\begin{align}
    & (k^2-m^2)\mathcal{V}^{\mu}_{(0)}=0, \label{equ124review2}\\
    & (k^2-m^2)\mathcal{A}^{\mu}_{(0)}=0, \label{equ125review2}\\
    & (k^2-m^2)\mathcal{S}^{\mu\nu}_{(0)}=0.\label{equ126review2}
\end{align}
Therefore, non-trivial solution of leading-order Wigner functions coefficients satisfy the on-shell condition, \textit{i.e.}, $k^{\mu}k_{\mu}=k^2=m^2$, where $k^{\mu}$ denotes the kinetic momentum. Moreover, $\mathcal{S}^{\mu\nu}_{(0)}$ and $\mathcal{A}^{\mu}_{(0)}$ are not independent, and they can be expressed in terms of each other. Eqs.~\eqref{equ116review2}-\eqref{equ117review2} allow us to write, 
\begin{align}
& \mathcal{S}_{\mu\nu}^{(0)} = -\frac{1}{m}\epsilon_{\mu\nu\alpha\beta}k^{\alpha}\mathcal{A}^{\beta}_{(0)},\label{equ127review2}\\
& \hstar{\mathcal{S}}^{\mu\nu}_{(0)}
= \frac{1}{m} \left(k^{\mu}\mathcal{A}^{\nu}_{(0)}-k^{\nu}\mathcal{A}^{\mu}_{(0)}\right), \label{equ128review2}\\
&  \mathcal{A}^{\rho}_{(0)}=-\frac{k_{\lambda }}{m}{\hstar{\mathcal{S}}}^{\rho\lambda}_{(0)}=-\frac{1}{2 m} \epsilon ^{\rho \lambda \alpha \beta }k_{\lambda}\mathcal{S}_{\alpha\beta}^{(0)}.\label{equ129review2}
\end{align}

The dynamical equations for leading order components of the Wigner function, $\mathcal{F}_{(0)}$, and $\mathcal{A}^{\mu}_{(0)}$ can be obtained at the $\hbar$ order.  In the first order in $\hbar$, Eqs.~\eqref{equ88review2}-\eqref{equ92review2} become,
\begin{align}
& k^{\mu}\mathcal{V}_{\mu}^{(1)}-m\mathcal{F}^{(1)}=\mathcal{D}_{\mathcal{F}}^{(0)},\label{equ130review2}\\
& \frac{1}{2}\partial^{\mu}\mathcal{A}_{\mu}^{(0)}+m\mathcal{P}^{(1)}=-\mathcal{D}_{\mathcal{P}}^{(0)},\label{equ131review2}\\
& k_{\mu}\mathcal{F}^{(1)}-\frac{1}{2}\partial^{\nu}\mathcal{S}_{\nu\mu}^{(0)}-m\mathcal{V}_{\mu}^{(1)}= \mathcal{D}_{\mathcal{V},\mu}^{(0)},\label{equ132review2}\\
& -\frac{1}{2}\partial_{\mu}\mathcal{P}^{(0)}+k^{\beta}\hstar{\mathcal{S}}_{\mu\beta}^{(1)}+m\mathcal{A}_{\mu}^{(1)}=-\mathcal{D}_{\mathcal{A},\mu}^{(0)},\label{equ133review2}\\
& \frac{1}{2}\left(\partial_{\mu}\mathcal{V}_{\nu}^{(0)}-\partial_{\nu}\mathcal{V}_{\mu}^{(0)}\right)-\epsilon_{\mu\nu\alpha\beta}k^{\alpha}\mathcal{A}^{\beta(1)}-m\mathcal{S}_{\mu\nu}^{(1)}=\mathcal{D}_{\mathcal{S},{\mu\nu}}^{(0)}\label{equ134review2},
\end{align}
while Eqs.~\eqref{equ93review2}-\eqref{equ97review2} yield~\cite{Vasak:1987um}
\begin{align}
& \partial^{\mu}\mathcal{V}_{\mu}^{(0)}=2 \mathcal{C}_{\mathcal{F}}^{(0)},\label{equ135review2}\\
& k^{\mu}\mathcal{A}_{\mu}^{(1)}=\mathcal{C}_{\mathcal{P}}^{(0)},\label{equ136review2}\\
& \frac{1}{2}\partial_{\mu}\mathcal{F}^{(0)}+k^{\nu}\mathcal{S}_{\nu\mu}^{(1)}=\mathcal{C}_{\mathcal{V},\mu}^{(0)},\label{equ137review2}\\
& k_{\mu}\mathcal{P}^{(1)}+\frac{1}{2}\partial^{\beta}
\hstar{\mathcal{S}}_{\mu\beta}^{(0)}=-\mathcal{C}_{\mathcal{A},\mu}^{(0)}, \label{equ138review2}\\
& \left(k_{\mu}\mathcal{V}_{\nu}^{(1)}-k_{\nu}\mathcal{V}_{\mu}^{(1)}\right)
+\frac{1}{2}\epsilon_{\mu\nu\alpha\beta}\partial^{\alpha}\mathcal{A}^{\beta(0)}=- \mathcal{C}_{\mathcal{S},\mu\nu}^{(0)}.\label{equ139review2}
\end{align}
Using Eq.~\eqref{equ127review2}, \eqref{equ130review2}, and \eqref{equ132review2} 
one obtains the constraint equation satisfied by $\mathcal{F}^{(1)}$,
\begin{align}
(k^2-m^2)\mathcal{F}^{(1)}=k^{\mu}\mathcal{D}_{\mathcal{V},\mu}^{(0)}+m\mathcal{D}_{\mathcal{F}}^{(0)}.
\label{equ65}
\end{align}
Moreover, from Eqs.~\eqref{equ131review2}, \eqref{equ132review2} and \eqref{equ134review2} one obtains the first-order corrections to the pseudo-scalar, vector and tensor components of the Wigner function, 
\begin{align}
&  \mathcal{P}^{(1)}=-\frac{1}{2m}\left[\partial^{\mu}\mathcal{A}_{\mu}^{(0)}+2\mathcal{D}_{\mathcal{P}}^{(0)}\right]\label{equ141review2},\\
&  \mathcal{V}_{\mu}^{(1)}=\frac{1}{m}\left[k_{\mu}\mathcal{F}^{(1)}-\frac{1}{2}\partial^{\nu}\mathcal{S}_{\nu\mu}^{(0)}-\mathcal{D}^{(0)}_{\mathcal{V},\mu}\right]\label{equ142review2},\\
&  \mathcal{S}_{\mu\nu}^{(1)}=\frac{1}{2m}\left[\partial_{\mu}\mathcal{V}_{\nu}^{(0)}-\partial_{\mu}\mathcal{V}_{\nu}^{(0)}-2\epsilon_{\mu\nu\alpha\beta}k^{\alpha}\mathcal{A}^{\beta}_{(1)}-2\mathcal{D}_{\mathcal{S},{\mu\nu}}^{(0)}\right]\label{equ143review2},\\
& \hstar{\mathcal{S}}_{\mu\beta}^{(1)}=\frac{1}{m}\Big[\frac{1}{4}\epsilon_{\mu\beta \sigma \rho} \left(\partial^{\sigma}\mathcal{V}^{\rho}_{(0)}-\partial^{\rho}\mathcal{V}^{\sigma}_{(0)}\right) + \left(k_{\mu}\mathcal{A}_{\beta}^{(1)}-k_{\beta}\mathcal{A}_{\mu}^{(1)}\right)-\frac{1}{2}\epsilon_{\mu\beta \sigma \rho }\mathcal{D}_{\mathcal{S}(0)}^{\sigma\rho}\Big].\label{equ144review2}
\end{align}
Constrained equations for $\mathcal{P}^{}_{(1)}$, $\mathcal{V}^{\mu}_{(1)}$, $\mathcal{A}^{\mu}_{(1)}$, and $\mathcal{S}^{\mu\nu}_{(1)}$ can also be obtained using Eqs.~\eqref{equ130review2}-\eqref{equ139review2}, and Eqs.~\eqref{equ141review2}-\eqref{equ144review2}~\cite{Das:2022azr}, 
\begin{align}
& (k^2-m^2)\mathcal{A}_{\mu}^{(1)}=k_{\mu}\mathcal{C}_{\mathcal{P}}^{(0)}-\frac{1}{2}\epsilon_{\mu\beta \sigma\rho}k^{\beta}\mathcal{D}_{\mathcal{S}(0)}^{\sigma\rho}+m\mathcal{D}_{\mathcal{A},\mu}^{(0)},\label{equ145review2}\\
& \big(k^2-m^2\big)\mathcal{V}^{\rho}_{(1)}=m\mathcal{D}^{\rho}_{\mathcal{V}(0)}+k^{\rho}\mathcal{D}_{\mathcal{F}(0)}^{}-k_{\lambda}\mathcal{C}^{\lambda\rho}_{\mathcal{S}(0)},\label{equ146review2}\\
& (k^2-m^2)\mathcal{P}^{(1)}  =  -k_{\rho}\mathcal{C}^{\rho}_{\mathcal{A}(0)}+m\mathcal{D}_{\mathcal{P}}^{(0)},\label{equ147review2}\\
& (k^2-m^2)\mathcal{S}^{\rho\lambda}_{(1)} = \left(k_{}^{\rho}\mathcal{C}_{\mathcal{V}(0)}^{\lambda}-k_{}^{\lambda}\mathcal{C}_{\mathcal{V}(0)}^{\rho}\right)+m\mathcal{D}^{\rho\lambda}_{\mathcal{S}(0)}-\epsilon^{\rho\lambda\sigma\alpha}k_{\sigma}\mathcal{D}^{(0)}_{\mathcal{A},\alpha}.\label{equ148review2}
\end{align}
These equations, indicate that, $\mathcal{F}_{(1)}$, $\mathcal{P}_{(1)}$, $\mathcal{V}^{\mu}_{(1)}$, $\mathcal{A}^{\mu}_{(1)}$, $\mathcal{S}^{\mu\nu}_{(1)}$ remain on-shell only in the collisionless limit~\cite{Weickgenannt:2020aaf,Weickgenannt:2021cuo}. 
Moreover, Eqs.~\eqref{equ135review2} along with Eq.~\eqref{equ115review2}, give us the kinetic equation satisfied by the leading-order scalar coefficient $\mathcal{F}_{(0)}$.
\begin{align}
k^{\mu}\partial_{\mu}\mathcal{F}^{(0)}=2m\,\mathcal{C}_{\mathcal{F}}^{(0)},
    \label{equ149review2}
\end{align}
Furthermore, using Eqs.~\eqref{equ128review2}, \eqref{equ131review2}, and \eqref{equ138review2}, one finds the kinetic equation satisfied by the leading-order axial-vector coefficient $\mathcal{A}^{\mu}_{(0)}$, 
\begin{eqnarray}
k^{\beta}\partial_{\beta}\mathcal{A}_{\mu}^{(0)}&=& 2m\,\mathcal{C}_{\mathcal{A},\mu}^{(0)}-2k_{\mu}\mathcal{D}_{\mathcal{P}}^{(0)}.
\label{equ150review2} 
\end{eqnarray}
From the above discussion, it is evident that to obtain the kinetic equation of Wigner function components at leading order, we need to look into the evolution equation at $\hbar$ order. More precisely, to obtain the kinetic equation for $\mathcal{F}^{(0)}$, and $\mathcal{A}^{\mu}_{(0)}$, we need to write down Eqs.~\eqref{equ88review2}-\eqref{equ97review2} at the order of $\hbar$. Similarly, to obtain the kinetic equation for $\mathcal{F}^{(1)}$, and $\mathcal{A}^{\mu}_{(1)}$, we need to write down Eqs.~\eqref{equ88review2}-\eqref{equ97review2} at the order of $\hbar^2$. At the order $\hbar^2$, Eqs.~\eqref{equ88review2}-\eqref{equ97review2} boil down to, 
\begin{align}
& k^{\mu}\mathcal{V}_{\mu}^{(2)}-m\mathcal{F}^{(2)}=\mathcal{D}_{\mathcal{F}}^{(1)},\label{equ151review2}\\
& \frac{1}{2}\partial^{\mu}\mathcal{A}_{\mu}^{(1)}+m\mathcal{P}^{(2)}=-\mathcal{D}_{\mathcal{P}}^{(1)},\label{equ152review2}\\
& k_{\mu}\mathcal{F}^{(2)}-\frac{1}{2}\partial^{\nu}\mathcal{S}_{\nu\mu}^{(1)}-m\mathcal{V}_{\mu}^{(2)}= \mathcal{D}_{\mathcal{V},\mu}^{(1)},\label{equ153review2}\\
& -\frac{1}{2}\partial_{\mu}\mathcal{P}^{(1)}+k^{\beta}\hstar{\mathcal{S}}_{\mu\beta}^{(2)}+m\mathcal{A}_{\mu}^{(2)}=-\mathcal{D}_{\mathcal{A},\mu}^{(1)},\label{equ154review2}\\
& \frac{1}{2}\left(\partial_{\mu}\mathcal{V}_{\nu}^{(1)}-\partial_{\nu}\mathcal{V}_{\mu}^{(1)}\right)-\epsilon_{\mu\nu\alpha\beta}k^{\alpha}\mathcal{A}^{\beta(2)}-m\mathcal{S}_{\mu\nu}^{(2)}=\mathcal{D}_{\mathcal{S},{\mu\nu}}^{(1)}\label{equ155review2},\\
& \partial^{\mu}\mathcal{V}_{\mu}^{(1)}=2 \mathcal{C}_{\mathcal{F}}^{(1)},\label{equ156review2}\\
& k^{\mu}\mathcal{A}_{\mu}^{(2)}=\mathcal{C}_{\mathcal{P}}^{(1)},\label{equ157review2}\\
& \frac{1}{2}\partial_{\mu}\mathcal{F}^{(1)}+k^{\nu}\mathcal{S}_{\nu\mu}^{(2)}=\mathcal{C}_{\mathcal{V},\mu}^{(1)}\,,\label{equ158review2}\\
& k_{\mu}\mathcal{P}^{(2)}+\frac{1}{2}\partial^{\beta}
\hstar{\mathcal{S}}_{\mu\beta}^{(1)}=-\mathcal{C}_{\mathcal{A},\mu}^{(1)}\,, \label{equ159review2}\\
& \left(k_{\mu}\mathcal{V}_{\nu}^{(2)}-k_{\nu}\mathcal{V}_{\mu}^{(2)}\right)
+\frac{1}{2}\epsilon_{\mu\nu\alpha\beta}\partial^{\alpha}\mathcal{A}^{\beta(1)}=- \mathcal{C}_{\mathcal{S},\mu\nu}^{(1)}\,.\label{equ160review2}
 \end{align}
Now using Eqs.~\eqref{equ142review2} in Eq.~\eqref{equ156review2} we obtain the kinetic equation for $\mathcal{F}_{(1)}$,
\begin{align}
k^{\mu}\partial_{\mu}\mathcal{F}^{(1)}=2m\,\mathcal{C}_{\mathcal{F}}^{(1)}+\partial^{\mu}\mathcal{D}_{\mathcal{V},\mu}^{(0)}.\label{equ161review2}
\end{align}
Moreover, one can find a closed form expression for $\mathcal{P}^{(2)}$ from Eq.~\eqref{equ152review2}. Using the expression of $\mathcal{P}^{(2)}$, and expression of $\hstar{\mathcal{S}}_{\mu\beta}^{(1)}$, back into Eq.~\eqref{equ159review2} we find 
the kinetic equation for $\mathcal{A}_{\mu}^{(1)}$,
\begin{align}
    k^{\beta}\partial_{\beta}\mathcal{A}_{\mu}^{(1)}=2m\,\mathcal{C}_{\mathcal{A},\mu}^{(1)}-2k_{\mu}\mathcal{D}_{\mathcal{P}}^{(1)}-\frac{1}{2}\epsilon_{\mu\beta \gamma\delta}\partial^{\beta}\mathcal{D}_{\mathcal{S}(0)}^{\gamma\delta}.
~~~\label{equ162review2}
\end{align}
Combining Eqs.~\eqref{equ149review2} and \eqref{equ161review2}, we get the kinetic equation for the scalar coefficient up to $\hbar$ correction,
\begin{align}
    & k^{\mu}\partial_{\mu}\left(\mathcal{F}^{(0)}+\hbar\,\mathcal{F}^{(1)}\right)=2m \,\left[\mathcal{C}_{\mathcal{F}}^{(0)}+\hbar\left(\mathcal{C}_{\mathcal{F}}^{(1)}+\frac{1}{2m}\partial^{\mu}\mathcal{D}_{\mathcal{V},\mu}^{(0)}\right)\right]\nonumber\\
   \implies &  k^{\mu}\partial_{\mu}\widetilde{\mathcal{F}} = 2m\,\widetilde{\mathcal{C}}_{\mathcal{F}}
    \label{equ163review2}
\end{align}
Similarly, combining Eqs.~\eqref{equ150review2} and \eqref{equ162review2}, we get the kinetic equation for the axial vector coefficient up to $\hbar$ correction, 
\begin{align}
    & k^{\beta}\partial_{\beta}\left(\mathcal{A}_{\mu}^{(0)}+\hbar\,\mathcal{A}_{\mu}^{(1)}\right)=2m~\left[\mathcal{C}_{\mathcal{A},\mu}^{(0)}+\hbar \,  \mathcal{C}_{\mathcal{A},\mu}^{(1)}-\frac{k_{\mu}}{m}\left(\mathcal{D}_{\mathcal{P}}^{(0)}+\hbar \mathcal{D}_{\mathcal{P}}^{(1)}\right)-\frac{\hbar}{4m}\epsilon_{\mu\beta \gamma\delta}\partial^{\beta}\mathcal{D}_{\mathcal{S}(0)}^{\gamma\delta}\right]\nonumber\\
    \implies & k^{\beta}\partial_{\beta}\widetilde{\mathcal{A}}_{\mu} = 2m~ \widetilde{\mathcal{C}}_{\mathcal{A},\mu}   \label{equ164review2}
\end{align}

The kinetic equations for $\widetilde{\mathcal{F}}$, and $\widetilde{\mathcal{A}}_{\mu}$ have been obtained in Ref.~\cite{Das:2022azr} in a general form that does not assume any special conditions. In Refs.~\cite{Weickgenannt:2020aaf,Weickgenannt:2021cuo} authors consider the effect of spin polarization at first order in $\hbar$ in the semi-classical expansion of the Wigner function. The underlying assumption is that the spin polarization effects are small and they arise through microscopic scattering processes, which appear at least at $\hbar$ order. In the Wigner function approach, the axial-vector ($\mathcal{A}^{\mu}$) component determines the spin polarization effects ~\cite{Weickgenannt:2019dks,Weickgenannt:2020aaf,Weickgenannt:2021cuo}. Since there is no spin polarization effect at the leading order, this implies $\mathcal{A}^{(0)}_{\mu}=0$. Consequently, the leading order tensor component of the Wigner function $\mathcal{S}^{(0)}_{\mu \nu}=0$. Moreover Eq.~\eqref{equ114review2} implies $\mathcal{P}^{(0)}=0$. Therefore, all the Wigner function components that carry pseudo-scalar quantum numbers vanish. Furthermore, for the consistency of the framework, collision terms involving the pseudo-scalar, axial-vector, and tensor components must vanish, \textit{i.e.}, $\mathcal{C}_{\mathcal{P}}^{(0)}=0$; $\mathcal{C}_{\mathcal{A}}^{\mu(0)}=0$;
$\mathcal{C}_{\mathcal{S}}^{\mu\nu(0)}=0$;
$\mathcal{D}_{\mathcal{P}}^{(0)}=0$;
$\mathcal{D}_{\mathcal{A}}^{\mu(0)}=0$;
$\mathcal{D}_{\mathcal{S}}^{\mu\nu(0)}=0$~\cite{Weickgenannt:2020aaf,Weickgenannt:2021cuo}. At $\hbar\rightarrow 0$, we know $k_{\mu}\mathcal{A}^{\mu}_{(0)}=0$, moreover if we consider that $\mathcal{C}_{\mathcal{P}}^{(0)}=0$, then Eq.~\eqref{equ136review2} implies that $k_{\mu}\mathcal{A}^{\mu}_{(1)}=0$. Finally, Eq.~\eqref{equ145review2} implies that due to the vanishing collision term, $\mathcal{A}^{\mu}_{(1)}$ also remains on-shell, due to the vanishing collision terms.

Note that under the Lorentz transformation $\widetilde{\mathcal{F}}$ and $\widetilde{\mathcal{A}}^{\mu}$ transform like scalar and pseudo-vector, respectively. If we introduce another pseudo-vector $\mathfrak{s}^{\alpha}$, the spin four-vector, then $\mathfrak{s}^{\alpha}\widetilde{\mathcal{A}}_{\alpha}$ transforms like a scalar.  
This allows us to introduce a generalized distribution function $\mathfrak{f}(x,k,\mathfrak{s})$ where spin is a part of the extended phase-space~\cite{Zamanian:2010zz,Ekman:2017kxi,Florkowski:2018fap,Ekman:2019vrv,Weickgenannt:2020aaf}.
\begin{align}
  \mathfrak{f}(x,k,\mathfrak{s})=\frac{1}{2}\left(\widetilde{\mathcal{F}}(x,k)-\mathfrak{s}^{\mu}\widetilde{\mathcal{A}}_{\mu}(x,k)\right) ,\label{equ165review2}
\end{align}
Eq.~\eqref{equ165review2} can be inverted to obtain $\widetilde{\mathcal{F}}(x,k)$, and 
$\widetilde{\mathcal{A}}_{\mu}(x,k)$ in terms of $\mathfrak{f}(x,k,\mathfrak{s})$. To do so one introduces spin measure~\cite{Weickgenannt:2020aaf,Weickgenannt:2021cuo}
\begin{align}
    \int \mathrm{dS}(k) \equiv \frac{1}{\pi}\sqrt{\frac{k^2}{3}} \int d^4\mathfrak{s} \, \delta(\mathfrak{s}\cdot \mathfrak{s}+3) \, \delta(k\cdot \mathfrak{s}).
    \label{equ166review2}
\end{align}
The spin measure defined in Eq.~\eqref{equ166review2} is similar to the spin measure defined in Eq.~\eqref{equ66review} up to some normalization factors. Moreover, the spin measure satisfies the following identities~\cite{Weickgenannt:2020aaf,Weickgenannt:2021cuo}
\begin{align}
\int \mathrm{dS}(k) &= 2\,,\label{equ167review2}\\
\int \mathrm{dS}(k)    \, \mathfrak{s}^{\mu} &= 0\,,\label{equ168review2}\\
\int \mathrm{dS}(k) \, \mathfrak{s}^{\mu}\mathfrak{s}^{\nu} &=
-2\bigg(g^{\mu\nu}-\frac{k^{\mu}k^{\nu}}{k^2}\bigg)\,.\label{equ169review2}
\end{align}
Using Eqs.~\eqref{equ167review2}-\eqref{equ169review2} we can invert Eq.~\eqref{equ165review2} to write down $\widetilde{\mathcal{F}}(x,k)$ and $\widetilde{\mathcal{A}}(x,k)$ in terms of the generalized distribution function, 
\begin{align}
    \int \mathrm{dS}(k) \, \mathfrak{f}(x,k,\mathfrak{s}) &= \widetilde{\mathcal{F}}(x,k)\,,
    \label{equ170review2}\\
    \int \mathrm{dS}(k) \, \mathfrak{s}^{\mu}~\mathfrak{f}(x,k,\mathfrak{s}) &= \widetilde{\mathcal{A}}^{\mu}(x,k)\,,
    \label{equ171review2}
\end{align}
It is important to note that, if we use Eq.~\eqref{equ165review2} on the left hand side of Eq.~\eqref{equ171review2} then we obtain, 
\begin{align}
    \int \mathrm{dS}(k) \, \mathfrak{s}^{\mu} \, \mathfrak{f}(x,k,\mathfrak{s}) = \widetilde{\mathcal{A}}^{\mu}(x,k)-\frac{k^{\mu}}{k^2}\left(k_{\alpha}\widetilde{\mathcal{A}}^{\alpha}\right),
    \label{amuinversion}
\end{align}
where $k_{\alpha}\widetilde{\mathcal{A}}^{\alpha}=k_{\alpha}\mathcal{A}^{\alpha}_{(0)}+\hbar k_{\alpha}\mathcal{A}^{\alpha}_{(1)} = \hbar \, \mathcal{C}_{\mathcal{P}}^{(0)}$. Here we have used Eq.~\eqref{equ118review2}, and \eqref{equ136review2}. Therefore in general $k_{\alpha}\widetilde{\mathcal{A}}^{\alpha}\neq 0$. But the assumption that spin polarization effect only appears at $\hbar$ order, which implies $\mathcal{C}_{\mathcal{P}}^{(0)}=0$, also preserves the orthogonality of $\widetilde{\mathcal{A}}^{\alpha}$ with respect to $k^{\alpha}$, i.e., $k_{\alpha}\widetilde{\mathcal{A}}^{\alpha}=0$. Therefore one obtains $\widetilde{\mathcal{A}}^{\mu}(x,k)$, in terms of the generalized distribution function as given in Eq.~\eqref{equ171review2}. The inversion of Eq.~\eqref{equ165review2}, i.e, the expression of $\widetilde{\mathcal{F}}(x,k)$ and $\widetilde{\mathcal{A}}^{\mu}(x,k)$ in terms of $\mathfrak{f}(x,k,\mathfrak{s})$ are crucial because, in the quantum kinetic theory approach, macroscopic currents, i.e., energy-momentum tensor ($T^{\mu\nu}$) and the spin tensor ($S^{\lambda\mu\nu}$) can be expressed in terms of various independent components of the Wigner function. Therefore using Eqs.~\eqref{equ170review2}-\eqref{equ171review2}
energy-momentum tensor ($T^{\mu\nu}$) and the spin tensor ($S^{\lambda\mu\nu}$)
can be written in terms of the spin-dependent distribution function $\mathfrak{f}(x,k,\mathfrak{s})$ ~\cite{Speranza:2020ilk,Florkowski:2017ruc,Florkowski:2017dyn,Florkowski:2018ahw}. Now using Eqs.~\eqref{equ163review2} and \eqref{equ164review2}, one obtains the Boltzmann equation for the generalized distribution function $\mathfrak{f}(x,k,\mathfrak{s})$~\cite{Weickgenannt:2020aaf,Weickgenannt:2021cuo}.
\begin{eqnarray}
k^{\mu}\partial _{\mu }\mathfrak{f}(x,k,\mathfrak{s})= m~\left(\widetilde{\mathcal{C}}_{\mathcal{F}}-\mathfrak{s}_{\alpha} \widetilde{\mathcal{C}}^{\alpha}_{\mathcal{A}}\right)=m\, \mathfrak{C}(\mathfrak{f}). \label{equ173review2}
\end{eqnarray}
$\mathfrak{C}(\mathfrak{f})$ is the generalized spin dependent collision term. Within the quasi-particle approximation, one can separate the on-shell singularity from the distribution function using the representation,  $\mathfrak{f}(x,k,\mathfrak{s})=m\delta(k^2-M^2)f(x,k,\mathfrak{s})$~\cite{Weickgenannt:2020aaf,Weickgenannt:2021cuo}.
The distribution $f(x,k,\mathfrak{s})$ does not contain an on-shell singularity. In this representation, the on-shell singularity is completely contained within the Dirac delta function $\delta(k^2-M^2)$. $M$ is the quasi-particle mass, which contains quantum corrections, and is different from the bare mass $m$. Using the quasi-particle representation of $\mathfrak{f}(x,k,\mathfrak{s})$ in Eq.~\eqref{equ173review2}, and canceling off-shell correction, one obtains the on-shell Boltzmann equation involving the distribution function $f(x,k,\mathfrak{s})$~\cite{Weickgenannt:2020aaf,Weickgenannt:2021cuo}, 
\begin{eqnarray}
k^{\mu}\partial _{\mu }f(x,k,\mathfrak{s})=\mathfrak{C}_{\rm {on-shell}}[f]. \label{equ174review2}
\end{eqnarray}
In the above equation, one can explicitly write $\delta(k^2-m^2)$ on both sides to indicate the on-shell conditions. $\mathfrak{C}_{\rm {on-shell}}[f]$ is the on-shell collision term. The collision term $\mathfrak{C}_{\rm {on-shell}}[f]$ can be separated into local collision term $\mathfrak{C}_{{\rm on-shell},l}[f]$, and non-local collision term $\mathfrak{C}_{{\rm on-shell},nl}[f]$ where particles
scatter with a finite impact parameter. Explicit expressions of different collision kernels, i.e., $\mathfrak{C}_{{\rm on-shell},l}[f]$, and $\mathfrak{C}_{{\rm on-shell},nl}[f]$ have been discussed in details in Refs.~\cite{Weickgenannt:2020aaf, Weickgenannt:2021cuo}. The non-local collision term plays a crucial role in generating spin polarization~\cite{Weickgenannt:2020aaf, Weickgenannt:2021cuo}, as well as in producing purely vorticity-induced contributions to dissipation at second order~\cite{Jaiswal:2012qm}. For completeness, we write the explicit expression of the non-local collision term, 
\begin{align}
& \mathfrak{C}_{{\rm on-shell},nl}[f] =\int \mathrm{{d\Gamma_1}} \mathrm{d\Gamma_2} \mathrm{{d\Gamma}}' \, \widetilde{W}_{k\mathfrak{s}}
\big[f\left(x+\Delta _1,k_1,{\mathfrak{s}}_1\right)~f\left(x+\Delta _2,k_2,{\mathfrak{s}}_2\right)-f\left(x+\Delta ,k,{\mathfrak{s}}\right)~f\left(x+\Delta',k',{\mathfrak{s}}'\right)\big]\nonumber\\
&~~~~~~~~~~~~~+\int \mathrm{{d\Gamma_2}} \mathrm{{dS_1}}(k) \, \mathrm{{dS'}}(k_2) \, {W_\mathfrak{s}}\big[f\left( x+\Delta _1,k,{\mathfrak{s}}_1\right) ~f\left(x+\Delta _2,k_2,{\mathfrak{s}}_2\right)-f\left(x+\Delta ,k,{\mathfrak{s}}\right)~f\left(x+\Delta',k_2,{\mathfrak{s}}'\right)\big],
\label{equ175review2}
\end{align}
here, $\widetilde{W}_{k\mathfrak{s}}$ represents the transition matrix elements for a collision where spin and momentum exchange take place. On the other hand, $W_\mathfrak{s}$ represents the collision term where only spin exchange takes place. The explicit expression of $\widetilde{W}_{k\mathfrak{s}}$, and $W_\mathfrak{s}$ can be found in Ref.~\cite{Weickgenannt:2020aaf,Weickgenannt:2021cuo}.  The phase space measure is~\cite{Weickgenannt:2020aaf,Weickgenannt:2021cuo}, 
\begin{align}
\int \mathrm{{d\Gamma}} = \int d^4k \, \delta  \left(k^2-m^2\right)\int \mathrm{dS}(k)\,. \label{equ176review2}
\end{align}
In the non-local collision term Eq.~\eqref{equ175review2}, the distribution function $f$ of different particles is evaluated at different spacetime points. The spacetime shifts ($\Delta^\alpha$) characterizes the non-localness of the collision term. The spacetime shifts are of order $\hbar$ and are functions of momentum and spin~\cite{DeGroot:1980dk,Weickgenannt:2020aaf,Weickgenannt:2021cuo}. The collision term as given in Eq.~\eqref{equ175review2} can be used to check the form of the equilibrium distribution function. We start with the following form of the local equilibrium distribution function, 
\begin{align}
f_{\rm eq}\left(x,k,{\mathfrak{s}}\right)= \frac{1}{(2\pi\hbar)^3}\exp\left[-\beta^{\mu}(x)k_{\mu}+\frac{\hbar}{4}~\bar{\Omega}_{\mu\nu}(x)\Sigma^{\mu\nu}_{\mathfrak{s}}\right].
\label{equ177review2}
\end{align}
The factor of $(2\pi\hbar)^3$ may not be written with the distribution function, but can be absorbed in the integration measure of the momentum (Eq.~\eqref{equ176review2}). In Eq.~\eqref{equ177review2} $\bar{\Omega}_{\mu\nu}(x)$ is the spin chemical potential. 
$\Sigma^{\mu\nu}_{\mathfrak{s}}\equiv -\epsilon^{\mu\nu\alpha\beta}k_{\alpha}\mathfrak{s}_{\beta}/m$ is the dipole tensor. The local equilibrium distribution function given in Eq.~\eqref{equ64review}, and in Eq.~\eqref{equ177review2} are equivalent. Using Eq.~\eqref{equ177review2}, we can write the distribution function including the non-local effect, 
\begin{align}
f_{\rm eq}\left(x+\Delta,k,{\mathfrak{s}}\right)= \frac{1}{(2\pi\hbar)^3}\exp\left[-\beta^{\mu}(x+\Delta)k_{\mu}+\frac{\hbar}{4}~\bar{\Omega}_{\mu\nu}(x+\Delta)\Sigma^{\mu\nu}_{\mathfrak{s}}\right].
\label{}
\end{align}
We can expand the exponential in $f_{\rm eq}\left(x+\Delta,k,{\mathfrak{s}}\right)$ in the powers of $\hbar$, and keeping only linear terms in $\hbar$ we find, 
\begin{align}
 f_{\rm eq}\left(x+\Delta,k,{\mathfrak{s}}\right)\simeq & \frac{1}{(2\pi\hbar)^3}\exp\left[-\beta^{\mu}(x)k_{\mu}\right]\exp\left[-k^{\mu}\Delta^{\nu}\partial_{\nu}\beta_{\mu}(x)\right]\exp\left[\frac{\hbar}{4}\bar{\Omega}_{\mu\nu}(x)\Sigma^{\mu\nu}_{\mathfrak{s}}\right]\nonumber\\
\simeq & \frac{1}{(2\pi\hbar)^3}\exp\left[-\beta^{\mu}(x)k_{\mu}\right]\left(1-k^{\mu}\Delta^{\nu}\partial_{\nu}\beta_{\mu}(x)+\frac{\hbar}{4}\bar{\Omega}_{\mu\nu}(x)\Sigma^{\mu\nu}_{\mathfrak{s}}\right)
\label{equ179review2}
\end{align}
Using Eq.~\eqref{equ179review2} back into Eq.~\eqref{equ175review2} the non-local collision term can be expressed as, 
\begin{align}
& \mathfrak{C}_{{\rm on-shell},nl}[f] =-\int \mathrm{{d\Gamma_1}} \mathrm{d\Gamma_2} \mathrm{{d\Gamma}}' \, \frac{\widetilde{W}_{k\mathfrak{s}}}{(2\pi\hbar)^6} \exp\left(-\beta^{\mu}(k_1+k_2)_{\mu}\right)
\big[\partial_{\mu}\beta_{\nu}\left(\Delta^{\mu}_1k^{\nu}_1+\Delta^{\mu}_2k^{\nu}_2-\Delta^{\mu}k^{\nu}-\Delta^{\prime\mu}k^{\prime\nu}\right)\nonumber\\
& ~~~~~~~~~~~~~~~~~~~~~~~~~~~~~~~~~~~~~~~~~~~~~~~~~~~~~~~~~~~~~-\frac{\hbar}{4}\bar{\Omega}_{\mu\nu}\left(\Sigma^{\mu\nu}_{\mathfrak{s}_1}+\Sigma^{\mu\nu}_{\mathfrak{s}_2}-\Sigma^{\mu\nu}_{\mathfrak{s}}-\Sigma^{\mu\nu}_{\mathfrak{s}^{\prime}}\right)\big]\nonumber\\
&~~~~~~~~~~~~~-\int \mathrm{{d\Gamma_2}} \mathrm{{dS_1}}(k) \, \mathrm{{dS'}}(k_2) \, \frac{{W_\mathfrak{s}}}{(2\pi\hbar)^6}\exp\left(-\beta^{\mu}(k+k_2)_{\mu}\right)
\big[\partial_{\mu}\beta_{\nu}\left(\Delta^{\mu}_1k^{\nu}+\Delta^{\mu}_2k^{\nu}_2-\Delta^{\mu}k^{\nu}-\Delta^{\prime\mu}k^{\nu}_2\right)\nonumber\\
& ~~~~~~~~~~~~~~~~~~~~~~~~~~~~~~~~~~~~~~~~~~~~~~~~~~~~~~~~~~~~~-\frac{\hbar}{4}\bar{\Omega}_{\mu\nu}\left(\Sigma^{\mu\nu}_{\mathfrak{s}_1}+\Sigma^{\mu\nu}_{\mathfrak{s}_2}-\Sigma^{\mu\nu}_{\mathfrak{s}}-\Sigma^{\mu\nu}_{\mathfrak{s}^{\prime}}\right)\big]
\label{equ180review2}
\end{align}
To obtain the above expression, we have used the conservation of linear momentum, i.e., $k_1^{\mu}+k_2^{\mu}=k^{\mu}+k^{\prime\mu}$ when momentum exchange is involved in the scattering process. Let us look into the integrand of the term involving $\widetilde{W}_{k\mathfrak{s}}$, which can be simplified to, 
\begin{align}
& -\int \mathrm{{d\Gamma_1}} \mathrm{d\Gamma_2} \mathrm{{d\Gamma}}' \, \frac{\widetilde{W}_{k\mathfrak{s}}}{(2\pi\hbar)^6} \exp\left(-\beta^{\mu}(k_1+k_2)_{\mu}\right)
\big[\partial_{\{\mu}\beta_{\nu\}}\left(\Delta^{\{\mu}_1k^{\nu\}}_1+\Delta^{\{\mu}_2k^{\nu\}}_2-\Delta^{\{\mu}k^{\nu\}}-\Delta^{\prime\{\mu}k^{\prime\nu\}}\right)\nonumber\\
& ~~~~~~~~~~~~~~~~~~~~~~~~~~~~+\frac{1}{2}\partial_{[\mu}\beta_{\nu]}\left(J^{\mu\nu}_1+J^{\mu\nu}_2-J^{\mu\nu}-J^{\prime\mu\nu}\right)-\frac{\hbar}{4}\left(\partial_{[\mu}\beta_{\nu]}+\bar{\Omega}_{\mu\nu}\right)\left(\Sigma^{\mu\nu}_{\mathfrak{s}_1}+\Sigma^{\mu\nu}_{\mathfrak{s}_2}-\Sigma^{\mu\nu}_{\mathfrak{s}}-\Sigma^{\mu\nu}_{\mathfrak{s}^{\prime}}\right)\big].
\label{}
\end{align}
Here the total angular momentum tensor is $J^{\mu\nu}=2\Delta^{[\mu}k^{\nu]}+\frac{\hbar}{2}\Sigma^{\mu\nu}_{\mathfrak{s}^{}}$. In microscopic collisions, the conservation of the total angular momentum tensor implies,  $J^{\mu\nu}_1+J^{\mu\nu}_2=J^{\mu\nu}+J^{\prime\mu\nu}$. Therefore, the collision term vanishes when, 
\begin{align}
& \partial_{\mu}\beta_{\nu}+\partial_{\nu}\beta_{\mu}=0, \label{equ182rewview2} \\
& \bar{\Omega}_{\mu\nu}=-\frac{1}{2}\left(\partial_{\mu}\beta_{\nu}-\partial_{\nu}\beta_{\mu}\right)=\varpi_{\mu\nu}=\text{Constant.} \label{equ183rewview2}
\end{align}
These are the conditions for the global equilibrium, which can also be obtained by demanding that the full non local collision term vanish. 

Since all the macroscopic currents can be defined in terms of the distribution function $f(x,k,\mathfrak{s})$, using the kinetic equation Eq.~\eqref{equ174review2}, one can develop the spin hydrodynamic description within the framework of spin-kinetic theory~\cite{Weickgenannt:2022qvh,Wagner:2024fry,Sapna:2025yss}. In the kinetic theory approach one expands the out of equilibrium distribution function ($f_{}\left(x,k,{\mathfrak{s}}\right)$) in terms of irreducible moments of the deviation with respect to the local equilibrium distribution function ($f_{\rm eq}\left(x,k,{\mathfrak{s}}\right)$)~\cite{Denicol:2021clh}. In the spinless fluid, the collision term vanishes in local equilibrium. However, in spin hydrodynamics, the local equilibrium can be defined differently. In Refs.~\cite{Weickgenannt:2022qvh,Wagner:2024fry,Sapna:2025yss}, the local equilibrium is defined when the local collision term vanishes, but the non-local collision term is non-vanishing, i.e., $\mathfrak{C}_{\rm {on-shell}}[f_{\rm eq}\left(x,k,{\mathfrak{s}}\right)]\neq 0$, but  $\mathfrak{C}_{\rm {on-shell},l}[f_{\rm eq}\left(x,k,{\mathfrak{s}}\right)]= 0$. Only in the global equilibrium the full collision term  $\mathfrak{C}_{\rm {on-shell}}[f_{\rm eq}\left(x,k,{\mathfrak{s}}\right)]$ vanishes. In the global equilibrium $\beta^{\mu}$ must satisfy the Killing equation (Eq.~\eqref{equ182rewview2}), and the spin chemical potential $\bar{\Omega}^{\mu\nu}$ must be related to the thermal vorticity (Eq.~\eqref{equ183rewview2}). The numerical implementation of the spin hydrodynamic equation derived from the quantum kinetic theory has also been done in Ref.~\cite{Sapna:2025yss}. In Ref.~\cite{Sapna:2025yss}, the spin evolution is performed on a (3+1) dimensional hydrodynamic background. Numerical results show that relevant spin degrees of freedom can be generated due to the 
spin-orbital coupling. In the numerical framework, the initial orbital angular momentum originates from
the initial condition set for the hydrodynamic background. The microscopic interactions also play an important role, and the information of these microscopic interactions is encoded in the transport coefficients. In Ref.~\cite{Sapna:2025yss}, the authors report that for a certain form of the microscopic interaction, numerical simulations give results which are in close agreement with the local and global polarization measurements of the Lambda hyperons.   

Although given a description of microscopic interaction, one can explicitly compute the collision term $\mathfrak{C}_{\rm {on-shell}}[f]$. But obtaining the hydrodynamic equations, solving kinetic theory equations with the exact
collision integrals, can be non trivial. In such a case, one can avoid the explicit computation of the collision term and use the relaxation time approximation (RTA) to obtain the spin hydrodynamic framework. Note that we have already introduced the relaxation time approximation for spin kinetic theory in Section~\ref{subsec3b}. However, the framework discussed in Section~\ref{subsec3b} inherently considers only local collisions. This is also evident from the fact that the energy-momentum tensor is symmetric and the spin tensor is separately conserved. However, in the presence of non-local collisions, spin may not be separately conserved. In this case, the relaxation-time approximation needs to be improved to account for the effect of non-local interactions in the collision term~\cite{Weickgenannt:2024ibf}. If we start with the standard form of the kinetic equation within the relaxation time approximation, then one writes,  
\begin{align}
k^{\mu}\partial_{\mu}f\left(x,k,{\mathfrak{s}}\right) = -(k^{\mu}u_{\mu})\frac{f\left(x,k,{\mathfrak{s}}\right)-f_{\rm eq}\left(x,k,{\mathfrak{s}}\right)}{\tau_R}.
\label{equ184review2}
\end{align}
Here $f_{\rm eq}\left(x,k,{\mathfrak{s}}\right)$ is the local equilibrium distribution function (Eq.~\eqref{equ177review2}), and $\tau_R$ is the relaxation time. Note that the collision term appearing in Eq.~\eqref{equ184review2} can not be used to approximate the local and non-local collision terms together. In local equilibrium, the RHS of Eq.~\eqref{equ184review2} must vanish, and the LHS will be non-vanishing. Only in the global equilibrium will both LHS and RHS vanish together. We have already seen that for the local equilibrium distribution function $f_{\rm eq}\left(x,k,{\mathfrak{s}}\right)$, the non-local collision term does not vanish, i.e., $\mathfrak{C}_{{\rm on-shell},nl}[f_{\rm eq}]\neq 0$ in local equilibrium. In fact $\mathfrak{C}_{{\rm on-shell},nl}[f_{\rm eq}]=0$, only in the global equilibrium, where $\beta^{\mu}$, and $\bar{\Omega}^{\mu\nu}$ are determined by Eq.~\eqref{equ182rewview2}, and Eq.~\eqref{equ183rewview2}. In local equilibrium, there is no requirement that Eq.~\eqref{equ182rewview2}, and \eqref{equ183rewview2} will be satisfied. Hence, we can conclude that the collision term in the relaxation time approximation does not incorporate the effect of the non-local collision term. In Ref.~\cite{Weickgenannt:2024ibf}, the authors introduce a generalized relaxation time kinetic equation, which also models the non-local interactions, 
\begin{align}
k^{\mu}\partial_{\mu}f\left(x,k,{\mathfrak{s}}\right) = -k^{\mu}u_{\mu}\frac{f\left(x,k,{\mathfrak{s}}\right)-f_{\rm eq}\left(x,k,{\mathfrak{s}}\right)}{\tau_R}+\widetilde{\xi}~\frac{k^{\alpha}u_{\alpha}}{\tau_R}\Delta^{\mu}k^{\nu}\left(\partial_{\mu}\beta_{\nu}+\bar{\Omega}_{\mu\nu}\right)f^{(0)}
\label{equ185review2}
\end{align}
The above equation is called the non-local relaxation time approximation (NLRTA). On the RHS of Eq.~\eqref{equ185review2}, the first term corresponds to the local collisions, and the second term incorporates the non-local interaction which is directly proportional to the space-time shift $\Delta^{\mu}$. $f^{(0)}$ is the distribution function without the spin potential, i.e., $f^{(0)}=\frac{1}{(2\pi\hbar)^3}\exp(-\beta^{\mu}k_{\mu})$. Once again, one can absorb the $(1/(2\pi\hbar)^3)$ factor in the measure of the momentum integration. Note that in the local equilibrium, the first term on the RHS of Eq.~\eqref{equ185review2} vanishes, but the second term does not vanish. However, in the global equilibrium, using Eqs.~\eqref{equ182rewview2}-\eqref{equ183rewview2}, one can easily show that the second term on the RHS of Eq.~\eqref{equ185review2} vanishes. This is the generic feature of the spin kinetic equation, i.e., the local collision term vanishes in local and global equilibrium, but the non-local collision term does not vanish in local equilibrium, but only vanishes in global equilibrium. In Eq.~\eqref{equ185review2}, the ratio $\widetilde{\xi}/\tau_R$ determines how fast the global equilibrium is achieved compared to the local equilibrium. In the second term on the RHS of Eq.~\eqref{equ185review2} one keeps $f^{(0)}$, this is because $\Delta^{\mu}$ is already of the order of $\hbar$. If we keep terms at the linear order of $\hbar$ then only $f^{(0)}$ contributes in the second term. The NLRTA collision term, 
\begin{align}
\mathfrak{C}_{NLRTA}[f]=-k^{\mu}u_{\mu}\frac{f\left(x,k,{\mathfrak{s}}\right)-f_{\rm eq}\left(x,k,{\mathfrak{s}}\right)}{\tau_R}+\widetilde{\xi}~\frac{k^{\alpha}u_{\alpha}}{\tau_R}\Delta^{\mu}k^{\nu}\left(\partial_{\mu}\beta_{\nu}+\bar{\Omega}_{\mu\nu}\right)f^{(0)}
\end{align}
must also be consistent with the conservation equation, i.e., the conservation of the energy-momentum tensor, and the conservation of the total angular momentum tensor. This implies the following constraints on the collision term, 
\begin{align}
& \int d\Gamma~k^{\mu}\mathfrak{C}_{NLRTA}[f] = 0,\label{equ187review2}\\ 
& \int d\Gamma~\left(2\Delta^{[\mu}k^{\nu]}+\frac{\hbar}{2}\Sigma^{\mu\nu}_{\mathfrak{s}}\right)\mathfrak{C}_{NLRTA}[f]=0.\label{equ188review2}
\end{align}
Here, $d\Gamma$ includes the momentum integration measure and spin integration measure. Eqs.~\eqref{equ187review2}, and \eqref{equ188review2} determines the Lagrange multipliers $\beta^{\mu}$, and $\bar{\Omega}^{\mu\nu}$, that appear in the local equilibrium distribution function. Given the kinetic equation, one considers the method of gradient expansion of the out-of-equilibrium distribution function around the local equilibrium to derive the hydrodynamic equation~\cite{Weickgenannt:2024ibf}.

\section{Kubo relations for transport coefficients}
\label{sec4}

New transport coefficients, also known as the spin transport coefficients, appear in the dissipative spin hydrodynamic framework~\cite{Hattori:2019lfp,Daher:2022wzf}. The dynamical evolution of spin fluid will crucially depend on these transport coefficients. In standard hydrodynamics, several complementary methods have been used to calculate transport coefficients, e.g., the kinetic theory frameworks~\cite{FReif:1965,DeGroot:1980dk}, Green-Kubo framework~\cite{Green:1954,Kubo:1957mj}, etc. In the kinetic theory approach, one often considers the 
relaxation time approximation (RTA)~\cite{Danielewicz:1984ww,Sasaki:2008fg,Sasaki:2008um}, where one assumes that the microscopic collisions always restore the system to local equilibrium with a relaxation time $\tau$. Although the kinetic theory description within the RTA is much simpler due to the simplified collision kernel, this approach often does not yield correct conservation equations for the macroscopic currents~\cite{PhysRev.94.511,Rocha:2021zcw}. To overcome such problems novel relaxation time approach has been proposed~\cite{PhysRev.94.511,Rocha:2021zcw}. 
Apart from the RTA, the Chapman-Enskog approach has also been considered for the estimation of transport coefficients. This method is also consistent with the macroscopic conservation equations. This apart, Green-Kubo formalism provides an alternative 
quantum field theory (QFT) based approach for the estimation of transport coefficients, e.g. bulk viscosity, shear viscosity, etc.~\cite{Demir:2008tr,Muronga:2003tb,Fuini:2010xz,Wesp:2011yy,Reining:2011xn,Hosoya:1983id,Becattini:2019dxo,Karsch:1986cq,Landsman:1986uw,Ilin:1989nj,Ilin:1992xg,Jeon:1992kk,Jeon:1994if,Wang:1995qg,Jeon:1995zm,Wang:1999gv,Arnold:2000dr,Policastro:2001yc,Kraemmer:2003gd,Nakamura:2004sy,Sakai:2007cm,Fernandez-Fraile:2008sjf,Demir:2008tr,Hidaka:2009ma,Pal:2010es,Huang:2011dc,Lang:2012tt,Becattini:2012pp,Becattini:2013fla,Becattini:2014yxa,Ghosh:2014ija,Becattini:2018duy,Ghiglieri:2020dpq,Rose:2020lfc,Hu:2021lnx,Satapathy:2021cjp,Harutyunyan:2021rmb,Satapathy:2021wex,Hu:2022azy,Rocha:2023ilf,Dong:2023cng,Dey:2024hhc}. 

In the Green-Kubo formalism, one studies a statistical theory of irreversible processes, where one considers a statistical ensemble representing a system in a non-equilibrium state. Such a state is represented by the 
non-equilibrium statistical operators (NESO). We consider a system in the hydrodynamic regime, which is 
characterized by spin hydrodynamic variables, temperature ($T$), chemical potential ($\mu$), and spin chemical potential ($\omega^{\alpha\beta}$). In the previous section, we discussed different spin hydrodynamic frameworks depending on the gradient ordering of the spin chemical potential. The gradient ordering of the spin chemical potential also distinguishes the Green-Kubo framework. We first discuss the case where the spin chemical potential is a leading-order term in the gradient expansion, i.e, $\omega^{\alpha\beta}\sim\mathcal{O}(1)$. We start with the non-equilibrium statistical operator (NESO)~\cite{alma991021547569703276,Huang:2011dc,Hu:2021lnx,Hosoya:1983id}
\begin{align}
\widehat{\rho}(t)=\frac{1}{\mathcal{Z}}\exp\left[-\int d^3x~\widehat{Z}\left(\vec{x},t\right)\right],
\label{equ28ver1}
\end{align}
The factor $\mathcal{Z}$ is the normalization,
\begin{align}
\mathcal{Z}=\Tr \exp\left[-\int d^3x~\widehat{Z}\left(\vec{x},t\right)\right].
\label{equ29ver1}
\end{align}
Eqs.~\eqref{equ28ver1}, and ~\eqref{equ29ver1} implies that $\Tr\widehat{\rho}(t)=1$. The operator $\widehat{Z}(\vec{x},t)$ is defined as,
\begin{align}
\widehat{Z}(\vec{x},t) &=\epsilon \int _{-\infty}^t~dt^{\prime}~e^{\epsilon(t^{\prime}-t)}\bigg[\beta(\vec{x},t^{\prime}) u_{\nu}(\vec{x},t^{\prime})\widehat{T}^{0\nu}(\vec{x},t^{\prime})-\beta(\vec{x},t^{\prime})\mu (\vec{x},t^{\prime})\widehat{J}^0(\vec{x},t^{\prime})-\beta(\vec{x},t^{\prime})\omega_{\rho\sigma}(\vec{x},t^{\prime})\widehat{S}^{0\rho\sigma}(\vec{x},t^{\prime})\bigg]\nonumber\\
& =\epsilon \int _{-\infty}^t~dt^{\prime}~e^{\epsilon(t^{\prime}-t)}\bigg[\beta_{\nu}(\vec{x},t^{\prime})\widehat{T}^{0\nu}(\vec{x},t^{\prime})-\alpha (\vec{x},t^{\prime})\widehat{J}^0(\vec{x},t^{\prime})-\widetilde{\Omega}_{\rho\sigma}(\vec{x},t^{\prime})\widehat{S}^{0\rho\sigma}(\vec{x},t^{\prime})\bigg],
\label{equ30ver1}
\end{align}
with $\epsilon\rightarrow +0$. Due to the presence of the time-dependent factor $e^{\epsilon(t^{\prime}-t)}$ 
one can show that $d \widehat{Z}(\vec{x},t)/dt=0$ in the limit $\epsilon\rightarrow +0$. In general, $\widehat{\rho}$ can be time-dependent, however in the limit $\epsilon\rightarrow +0$ the operator $\widehat{Z}$ and the density operator $\widehat{\rho}$ are stationary~\cite{Li:2025pef,alma991021547569703276,Hosoya:1983id}. In this limit, the density operator also preserves the covariant structure. The $e^{\epsilon(t^{\prime}-t)}$ becomes relevant to write down the Kubo relation in terms of the derivative of the retarded Green function~\cite{Hosoya:1983id,Huang:2011dc}. 
$\widehat{T}^{\mu\nu}$ is the energy-momentum tensor, which is symmetric because we consider the spin chemical potential $\omega^{\alpha\beta}\sim\mathcal{O}(1)$.
$\widehat{J}^{\mu}$, and $\widehat{S}^{\mu\alpha\beta}$ are conserved four current, and spin tensor. The microscopic conservation equations are,
\begin{align}
\partial_{\mu}\widehat{T}^{\mu\nu}(\vec{x},t^{})=0; ~~\partial_{\mu}\widehat{J}^{\mu}(\vec{x},t^{})=0; ~~\partial_{\lambda}\widehat{S}^{\lambda\mu\nu}(\vec{x},t^{})=0.\label{equ34ver1}
\end{align}
Since $\widehat{T}^{\mu\nu}(\vec{x},t^{})$ is symmetric the spin tensor $\widehat{S}^{\lambda\mu\nu}$ is separately conserved. Using the spin hydrodynamic equations (Eq.~\eqref{equ34ver1}), it can be shown that~\cite{Dey:2024cwo,alma991021547569703276,Huang:2011dc}, 
\begin{align}
\widehat{\rho}(t)=\frac{1}{\mathcal{Z}}\exp\left(-\widehat{\mathcal{A}}+\widehat{\mathcal{B}}\right), \text{with},~~~\mathcal{Z}=\Tr \exp\left(-\widehat{\mathcal{A}}+\widehat{\mathcal{B}}\right). 
\label{equ38ver1}
\end{align}
here $\widehat{\mathcal{A}}(t)$, and $\widehat{\mathcal{B}}(t)$ are, 
\begin{align}
\widehat{\mathcal{A}}(t) =\int d^3x\bigg[\beta_{\nu}(\vec{x},t^{})\widehat{T}^{0\nu}(\vec{x},t^{})-\alpha (\vec{x},t^{})\widehat{J}^0(\vec{x},t^{})-\widetilde{\Omega}_{\rho\sigma}(\vec{x},t^{})\widehat{S}^{0\rho\sigma}(\vec{x},t^{})\bigg],
\label{equ36ver1}
\end{align}
\begin{align}
\widehat{\mathcal{B}}(t) & = \int d^3x~\int _{-\infty}^t~dt^{\prime}~e^{\epsilon(t^{\prime}-t)}\bigg[\partial_{\mu}\beta_{\nu}(\vec{x},t^{\prime})\widehat{T}^{\mu\nu}_{}(\vec{x},t^{\prime})-\partial_{\mu}\widetilde{\Omega}_{\rho\sigma}(\vec{x},t^{\prime})\widehat{S}^{\mu\rho\sigma}_{}(\vec{x},t^{\prime})-\partial_{\mu}\alpha(\vec{x},t^{\prime})\widehat{J}^{\mu}(\vec{x},t^{\prime})\bigg]\nonumber\\
& = \int d^3x~\int _{-\infty}^t~dt^{\prime}~e^{\epsilon(t^{\prime}-t)}~\widehat{\mathcal{C}}(\vec{x},t^{\prime}).
\label{equ37ver1}
\end{align}
The operator $\widehat{\mathcal{C}}(\vec{x},t)$, and $\widehat{\mathcal{B}}(t)$ contains \textit{thermodynamic forces}, i.e., gradient of  local thermodynamic quantities, $\partial_{\mu}\beta_{\nu}(\vec{x},t^{})$, $\partial_{\mu}\alpha(\vec{x},t^{})$, and $\partial_{\mu}\widetilde{\Omega}_{\rho\sigma}(\vec{x},t^{})$. These thermodynamic forces vanish in equilibrium. Therefore, the operator $\widehat{\mathcal{B}}$ is identified as the non-equilibrium part of the statistical operator.  which vanishes in equilibrium. $\widehat{\mathcal{A}}$ is the  equilibrium part which can be used to define the local equilibrium statistical operator~\cite{alma991021547569703276,Huang:2011dc} 
\begin{align}
\widehat{\rho}_{l}(t)=\frac{1}{\mathcal{Z}_l}\exp\left(-\widehat{\mathcal{A}}\right), \text{with},~~~\mathcal{Z}_l=\Tr \exp\left(-\widehat{\mathcal{A}}\right).
\end{align}
In the linear response theory, we can expand the non-equilibrium statistical operator ($\widehat{\rho}(t)$) around the local equilibrium and keep only the linear terms in the gradient of  local thermodynamic quantities. In the linear response theory, it can be shown that~\cite{Dey:2024cwo}
\begin{align}
\widehat{\rho}(t)
& \simeq \bigg[1+\int_0^1 d\tau \bigg\{e^{-\tau\widehat{\mathcal{A}}}\widehat{\mathcal{B}}e^{\tau\widehat{\mathcal{A}}}-\langle \widehat{\mathcal{B}}_{\tau}\rangle_l\bigg\}\bigg]\widehat{\rho}_l,
\label{equ91review}
\end{align}
here $\langle \widehat{\mathcal{B}}_{\tau}\rangle_l\equiv \langle e^{-\tau\widehat{\mathcal{A}}}\widehat{\mathcal{B}} e^{\tau\widehat{\mathcal{A}}}\rangle_l$, and $\langle \widehat{\mathcal{O}}\rangle_l=\Tr\left(\widehat{\rho}_l\widehat{\mathcal{O}}\right)$ is the 
statistical average of the operator $\widehat{\mathcal{O}}$ with respect to the local equilibrium statistical 
operator. Using the expression of the 
out of equilibrium density operator (Eq.~\eqref{equ91review}), we can calculate the statistical average of $\widehat{T}^{\mu\nu}(\vec{x},t)$, $ \widehat{J}^{\mu}(\vec{x},t)$, and  $\widehat{S}^{\mu\alpha\beta}(\vec{x},t)$,
\begin{align}
& \langle \widehat{T}^{\mu\nu}(\vec{x},t)\rangle = \langle \widehat{T}^{\mu\nu}(\vec{x},t)\rangle_l+\delta \langle \widehat{T}^{\mu\nu}(\vec{x},t)\rangle,\\
& \langle \widehat{J}^{\mu}(\vec{x},t)\rangle  = \langle \widehat{J}^{\mu}(\vec{x},t)\rangle_l+\delta \langle \widehat{J}^{\mu}(\vec{x},t)\rangle, \\
& \langle \widehat{S}^{\mu\alpha\beta}(\vec{x},t)\rangle = \langle \widehat{S}^{\mu\alpha\beta}(\vec{x},t)\rangle_l+\delta \langle \widehat{S}^{\mu\alpha\beta}(\vec{x},t)\rangle.
\end{align}
The local equilibrium quantities can be identified as, 
\begin{align}
& \langle \widehat{T}^{\mu\nu}(\vec{x},t)\rangle_l = \Tr\left(\widehat{\rho}_l(t)\widehat{T}^{\mu\nu}(\vec{x},t)\right), \\
& \langle \widehat{J}^{\mu}(\vec{x},t)\rangle_l = \Tr\left(\widehat{\rho}_l(t)\widehat{J}^{\mu}(\vec{x},t)\right),\\
&  \langle \widehat{S}^{\mu\alpha\beta}(\vec{x},t)\rangle_l = \Tr\left(\widehat{\rho}_l(t)\widehat{S}^{\mu\alpha\beta}(\vec{x},t)\right),
\end{align}
and the out-of-equilibrium contribution to the energy-momentum tensor, conserved current, and the spin tensor are, 
\begin{align}
& \delta \langle \widehat{T}^{\mu\nu} (\vec{x},t)\rangle = \Tr\left(\int_0^1 d\tau \bigg\{e^{-\tau\widehat{\mathcal{A}}}\widehat{\mathcal{B}}e^{\tau\widehat{\mathcal{A}}}-\langle \widehat{\mathcal{B}}_{\tau}\rangle_l\bigg\}\widehat{\rho}_l~\widehat{T}^{\mu\nu}(\vec{x},t)\right),\\
& \delta\langle \widehat{J}^{\mu}(\vec{x},t)\rangle = \Tr\left(\int_0^1 d\tau \bigg\{e^{-\tau\widehat{\mathcal{A}}}\widehat{\mathcal{B}}e^{\tau\widehat{\mathcal{A}}}-\langle \widehat{\mathcal{B}}_{\tau}\rangle_l\bigg\}\widehat{\rho}_l~\widehat{J}^{\mu}(\vec{x},t)\right),\\
& \delta\langle \widehat{S}^{\mu\alpha\beta}(\vec{x},t)\rangle = \Tr\left(\int_0^1 d\tau \bigg\{e^{-\tau\widehat{\mathcal{A}}}\widehat{\mathcal{B}}e^{\tau\widehat{\mathcal{A}}}-\langle \widehat{\mathcal{B}}_{\tau}\rangle_l\bigg\}\widehat{\rho}_l~\widehat{S}^{\mu\alpha\beta}(\vec{x},t)\right).
\end{align}
Using the expression of $\widehat{\mathcal{B}}(t)$ as given in Eq.~\eqref{equ37ver1}, the out-of-equilibrium contributions can be written as, 
\begin{align}
& \delta\langle \widehat{T}^{\mu\nu}(\vec{x},t)\rangle = \int d^3x^{\prime}\int_{-\infty}^t dt^{\prime}~e^{\epsilon(t^{\prime}-t)}\bigg(\widehat{T}^{\mu\nu}(\vec{x},t),\widehat{\mathcal{C}}^{}(\vec{x}^{\prime},t^{\prime})\bigg)_l,\label{equ101review}\\
& \delta\langle \widehat{J}^{\mu}(\vec{x},t)\rangle = \int d^3x^{\prime}\int_{-\infty}^t dt^{\prime}~e^{\epsilon(t^{\prime}-t)}\bigg(\widehat{J}^{\mu}(\vec{x},t),\widehat{\mathcal{C}}^{}(\vec{x}^{\prime},t^{\prime})\bigg)_l,\label{equ102review}\\
& \delta\langle \widehat{S}^{\mu\alpha\beta}(\vec{x},t)\rangle  = \int d^3x^{\prime}\int_{-\infty}^t dt^{\prime}~e^{\epsilon(t^{\prime}-t)}\bigg(\widehat{S}^{\mu\alpha\beta}(\vec{x},t),\widehat{\mathcal{C}}^{}(\vec{x}^{\prime},t^{\prime})\bigg)_l.\label{equ103review}
\end{align}
In the above equation, we have used the following shorthand notation,
\begin{align}
\bigg(\widehat{\mathcal{X}}(\vec{x},t),\widehat{\mathcal{Y}}^{}(\vec{x}^{\prime},t^{\prime})\bigg)_l = \int_0^1 d\tau \bigg\langle \widehat{\mathcal{X}}^{}(\vec{x},t)\bigg\{e^{-\tau\widehat{\mathcal{A}}}\widehat{\mathcal{Y}}(\vec{x}^{\prime},t^{\prime})e^{\tau\widehat{\mathcal{A}}}-\langle\widehat{\mathcal{Y}}(\vec{x}^{\prime},t^{\prime})_{\tau}\rangle_l\bigg\}\bigg\rangle_l.
\end{align}

The out-of-equilibrium contributions $\delta\langle \widehat{T}^{\mu\nu}(\vec{x},t)\rangle $, $\delta\langle \widehat{J}^{\mu}(\vec{x},t)\rangle$, and $\delta\langle \widehat{S}^{\mu\alpha\beta}(\vec{x},t)\rangle$ can be identified with dissipative corrections of the macroscopic currents, i.e., $T^{\mu\nu}_{(1)}$, $J^{\mu}_{(1)}$, and $S^{\mu\alpha\beta}_{(1)}$. Such an identification allows us to write down the Kubo relations for various transport coefficients arising in the dissipative spin-hydrodynamic framework. In the Green-Kubo framework, once we find $\widehat{\mathcal{C}}^{}(\vec{x}^{},t^{})$ in terms of $\widehat{T}^{\mu\nu}(\vec{x}^{},t^{})$, $\widehat{J}^{\mu}(\vec{x}^{},t^{})$, $\widehat{S}^{\mu\alpha\beta}(\vec{x}^{},t^{})$, and the thermodynamic forces $\partial_{\mu}\beta_{\nu}(\vec{x},t^{})$, $\partial_{\mu}\alpha(\vec{x},t^{})$, $\partial_{\mu}\widetilde{\Omega}_{\rho\sigma}(\vec{x},t^{})$, the Kubo relations can be obtained for different transport coefficients.
We emphasize that, in the Kubo framework, the irreducible tensor decomposition of microscopic currents is the same as the decomposition of its macroscopic counterpart. For $\omega^{\alpha\beta}\sim\mathcal{O}(1)$, the irreducible tensor decomposition of the microscopic currents are, 

\begin{align} 
& \widehat{T}^{\alpha\beta}= \widehat{\varepsilon} u^{\mu}u^{\nu}-\widehat{P}\Delta^{\mu\nu}
+ \widehat{h}^{\alpha}u^{\beta}+\widehat{h}^{\beta}u^{\alpha}+\widehat{\pi}^{\mu\nu}+\widehat{\Pi}\Delta^{\mu\nu},\label{equ105review}\\
& \widehat{J}^{\mu}=\widehat{n}u^{\mu}+ \widehat{J}_{(1)}^{\mu},\label{equ106review}\\
& \widehat{S}^{\mu\alpha\beta}=u^{\mu}\widehat{S}^{\alpha\beta}+2u^{[\alpha}\Delta^{\mu\beta]}\widehat{\Phi}+2u^{[\alpha}\widehat{\tau}^{\mu\beta]}_{(s)}+2u^{[\alpha}\widehat{\tau}^{\mu\beta]}_{(a)}+\widehat{\Theta}^{\mu\alpha\beta}. \label{equ107review}
\end{align}
Using Eqs.~\eqref{equ105review}-\eqref{equ107review} in the expression of $\widehat{\mathcal{C}}(\vec{x},t)$ as defined in Eq.~\eqref{equ37ver1} we find, 
\begin{align}
\widehat{\mathcal{C}}(\vec{x}^{},t^{})
& = \widehat{\varepsilon}D\beta-\widehat{P}\beta\theta-\beta\widehat{h}^{\mu}\left(\beta\nabla_{\mu}T-Du_{\mu}\right)+\beta\widehat{\pi}^{\mu\nu}\sigma_{\mu\nu}+\beta\widehat{\Pi}\theta\nonumber\\
& -\widehat{n}D\alpha-\widehat{J}^{\mu}_{(1)}\nabla_{\mu}\alpha-\widehat{S}^{\rho\sigma}D\widetilde{\Omega}_{\rho\sigma}-\widehat{S}^{\mu\rho\sigma}_{(1)}\nabla_{\mu}\widetilde{\Omega}_{\rho\sigma}.
\label{appenE1ver1}
\end{align}
The above expression of $\widehat{\mathcal{C}}(\vec{x}^{},t^{})$ can be further simplified to, 
\begin{align}
\widehat{\mathcal{C}} & =-\widehat{P}^{\star}\beta\theta
-\widehat{\mathcal{J}}^{\mu}\nabla_{\mu}\alpha+\widehat{h}^{\mu}\frac{S^{\alpha\beta}}{\varepsilon+P}\nabla_{\mu}(\beta\omega_{\alpha\beta})+\beta \widehat{\pi}^{\mu\nu}\sigma_{\mu\nu}\nonumber\\
&~~~~ -2\widehat{\Phi} u^{\alpha}\nabla^{\beta}(\beta\omega_{\alpha\beta})-2\widehat{\tau}^{\mu\beta}_{(s)}u^{\alpha}\Delta^{\gamma\rho}_{\mu\beta}\nabla_{\gamma}(\beta\omega_{\alpha\rho})-2\widehat{\tau}^{\mu\beta}_{(a)}u^{\alpha}\Delta^{[\gamma\rho]}_{[\mu\beta]}\nabla_{\gamma}(\beta\omega_{\alpha\rho}) -\widehat{\Theta}_{\mu\alpha\beta}\Delta^{\alpha\delta}\Delta^{\beta\rho}\Delta^{\mu\gamma}\nabla_{\gamma}(\beta\omega_{\delta\rho})
\label{equ53ver1}
\end{align}
where,
\begin{align}
& \widehat{P}^{\star}=\left(\widehat{P}-\widehat{\Pi}-\widehat{\varepsilon}\gamma+\widehat{n}\gamma^{\prime}+\widehat{S}^{\alpha\beta}\gamma_{\alpha\beta}\right),\label{bulk}\\
& \widehat{\mathcal{J}}^{\mu} = \widehat{J}^{\mu}_{(1)}-\frac{n}{\varepsilon+P}\widehat{h}^{\mu}.
\end{align}
Explicit expressions of $\gamma$, $\gamma^{\prime}$, and $\gamma^{\alpha\beta}$ are given below~\cite{Dey:2024cwo}, 

\begin{align}
& \gamma=\bigg[\left(\frac{\partial\varepsilon}{\partial\beta}\right)^{-1}_{n,S_{\alpha\beta}}\left(\frac{\partial P}{\partial\beta}\right)_{\alpha,\widetilde{\Omega}_{\alpha\beta}}-\left(\frac{\partial n}{\partial\beta}\right)^{-1}_{\varepsilon,S_{\alpha\beta}}\left(\frac{\partial P}{\partial\alpha}\right)_{\beta,\widetilde{\Omega}_{\alpha\beta}}-\left(\frac{\partial S_{\alpha\beta}}{\partial\beta}\right)^{-1}_{n,\varepsilon}\left(\frac{\partial P}{\partial\widetilde{\Omega}^{\alpha\beta}}\right)_{\alpha,\beta}\bigg]
\label{equE9new}
\end{align}
\begin{align}
& \gamma^{\prime}= \bigg[\left(\frac{\partial\varepsilon}{\partial\alpha}\right)^{-1}_{n,S_{\alpha\beta}}\left(\frac{\partial P}{\partial\beta}\right)_{\alpha,\widetilde{\Omega}_{\alpha\beta}}-\left(\frac{\partial n}{\partial\alpha}\right)^{-1}_{\varepsilon,S_{\alpha\beta}}\left(\frac{\partial P}{\partial\alpha}\right)_{\beta,\widetilde{\Omega}_{\alpha\beta}}-\left(\frac{\partial S_{\alpha\beta}}{\partial\alpha}\right)^{-1}_{n,\varepsilon}\left(\frac{\partial P}{\partial\widetilde{\Omega}^{\alpha\beta}}\right)_{\alpha,\beta}\bigg]
\label{equE10new}
\end{align}
\begin{align}
& \gamma^{\alpha\beta}=\bigg[\left(\frac{\partial\varepsilon}{\partial\widetilde{\Omega}^{\alpha\beta}}\right)^{-1}_{n,S_{\alpha\beta}}\left(\frac{\partial P}{\partial\beta}\right)_{\alpha,\widetilde{\Omega}_{\alpha\beta}}-\left(\frac{\partial n}{\partial\widetilde{\Omega}^{\alpha\beta}}\right)^{-1}_{\varepsilon,S_{\alpha\beta}}\left(\frac{\partial P}{\partial\alpha}\right)_{\beta,\widetilde{\Omega}_{\alpha\beta}}-\left(\frac{\partial S_{\alpha^{\prime}\beta^{\prime}}}{\partial\widetilde{\Omega}^{\alpha\beta}}\right)^{-1}_{n,\varepsilon}\left(\frac{\partial P}{\partial\widetilde{\Omega}^{\alpha^{\prime}\beta^{\prime}}}\right)_{\alpha,\beta}\bigg].
\label{equE11new}
\end{align}
In order to obtain Eq.~\eqref{equ53ver1} from Eq.~\eqref{appenE1ver1} one uses the spin hydrodynamic equations, 
\begin{align}
& D\varepsilon+(\varepsilon+P)\theta = -u_{\nu}\partial_{\mu}T^{\mu\nu}_{(1)}\\
& (\varepsilon+P)Du^{\alpha}=\nabla^{\alpha}P-\Delta^{\alpha}_{~\nu}\partial_{\mu}T^{\mu\nu}_{(1)}\\
& Dn+n\theta=-\partial_{\mu}J^{\mu}_{(1)}\\
& D S^{\alpha\beta}+S^{\alpha\beta}\theta=-\partial_{\mu}S^{\mu\alpha\beta}_{(1)},
\end{align}
and generalised, thermodynamic relation, 
\begin{align}
dP =-\frac{1}{\beta}(\varepsilon+P)d\beta+\frac{n}{\beta}d\alpha+\frac{S^{\alpha\beta}}{\beta}d\widetilde{\Omega}_{\alpha\beta}.
\label{equE6ver1}
\end{align}
Note that $\widehat{\mathcal{C}}(\vec{x},t)$ given in Eq.~\eqref{equ53ver1} contains the all dissipative currents. Using the expression of $\widehat{\mathcal{C}}(\vec{x},t)$ in Eqs.~\eqref{equ101review}-\eqref{equ103review} and using the appropriate projector, in the linear response theory one finds~\cite{Dey:2024cwo},
\begin{align}
    & \Pi=-\langle \widehat{P}^{\star}(\vec{x},t)\rangle
     = \beta\theta\int d^3x^{\prime}\int_{-\infty}^t dt^{\prime}~e^{\epsilon(t^{\prime}-t)}\bigg(\widehat{P}^{\star}(\vec{x},t),\widehat{P}^{\star}(\vec{x}^{\prime},t^{\prime})\bigg)_l\label{Pid}\\
   &  \pi^{\mu\nu}(\vec{x},t)=\langle\widehat{\pi}^{\mu\nu}(\vec{x},t)\rangle
     = \beta\sigma_{\rho\delta}\int d^3x^{\prime}\int_{-\infty}^t dt^{\prime}~e^{\epsilon(t^{\prime}-t)}\bigg(\widehat{\pi}^{\mu\nu}(\vec{x},t),\widehat{\pi}^{\rho\delta}(\vec{x}^{\prime},t^{\prime})\bigg)_l.\label{sigd}
\end{align}
\begin{align}
    \mathcal{J}^{\mu}(\vec{x},t)=\langle\widehat{\mathcal{J}}^{\mu}(\vec{x},t)\rangle 
    & = -\nabla_{\rho}\alpha\int d^3x^{\prime}\int_{-\infty}^t dt^{\prime}~e^{\epsilon(t^{\prime}-t)}\bigg(\widehat{\mathcal{J}}^{\mu}(\vec{x},t),\widehat{\mathcal{J}}^{\rho}(\vec{x}^{\prime},t^{\prime})\bigg)_l\nonumber\\
    &\,\,+\nabla_{\rho}(\beta\omega_{\alpha\beta})\int d^3x^{\prime}\int_{-\infty}^t dt^{\prime}~e^{\epsilon(t^{\prime}-t)}\bigg(\widehat{\mathcal{J}}^{\mu}(\vec{x},t),\widehat{h}^{\rho}(\vec{x}^{\prime},t^{\prime})\bigg)_l\Big(\frac{S^{\alpha\beta}}{\varepsilon+P}\Big).
    \label{Jid}
\end{align}
\begin{align}
    h^{\mu}(\vec{x},t)=\langle\widehat{h}^{\mu}(\vec{x},t)\rangle 
    & = -\nabla_{\rho}\alpha\int d^3x^{\prime}\int_{-\infty}^t dt^{\prime}~e^{\epsilon(t^{\prime}-t)}\bigg(\widehat{h}^{\mu}(\vec{x},t),\widehat{\mathcal{J}}^{\rho}(\vec{x}^{\prime},t^{\prime})\bigg)_l\nonumber\\
    &\,\,+\nabla_{\rho}(\beta\omega_{\alpha\beta})\int d^3x^{\prime}\int_{-\infty}^t dt^{\prime}~e^{\epsilon(t^{\prime}-t)}\bigg(\widehat{h}^{\mu}(\vec{x},t),\widehat{h}^{\rho}(\vec{x}^{\prime},t^{\prime})\bigg)_l\Big(\frac{S^{\alpha\beta}}{\varepsilon+P}\Big).
    \label{hid}
\end{align}
 \begin{align}
        &\Phi=\langle\widehat{\Phi}\rangle= -2 \nabla^{\delta}(\beta\omega_{\gamma\delta})u^{\gamma}\int d^3x^{\prime}\int_{-\infty}^t dt^{\prime}~e^{\epsilon(t^{\prime}-t)}\bigg(\widehat{\Phi}(\vec{x},t),\widehat{\Phi}(\vec{x}^{\prime},t^{\prime})\bigg)_l,\label{operatorcorre1}\\
        &\tau^{\mu\alpha}_{(s)}=\left\langle\widehat{\tau}_{(s)}^{\mu\alpha}\right\rangle=-2 \nabla_{\rho}(\beta\omega_{\gamma\delta})u^{\gamma}\int d^3x^{\prime}\int_{-\infty}^t dt^{\prime}~e^{\epsilon(t^{\prime}-t)}\bigg(\widehat{\tau}_{(s)}^{\mu\alpha}(\vec{x},t),\widehat{\tau}_{(s)}^{\rho\delta}(\vec{x}^{\prime},t^{\prime})\bigg)_l,\label{operatorcorre2}\\
        &\tau^{\mu\alpha}_{(a)}=\left\langle\widehat{\tau}_{(a)}^{\mu\alpha}\right\rangle=-2 \nabla_{\rho}(\beta\omega_{\gamma\delta})u^{\gamma}\int d^3x^{\prime}\int_{-\infty}^t dt^{\prime}~e^{\epsilon(t^{\prime}-t)}\bigg(\widehat{\tau}_{(a)}^{\mu\alpha}(\vec{x},t),\widehat{\tau}_{(a)}^{\rho\delta}(\vec{x}^{\prime},t^{\prime})\bigg)_l,\label{operatorcorre3}\\
       & \Theta^{\mu\alpha\beta}= \langle\widehat{\Theta}^{\mu\alpha\beta}\rangle=-\nabla_{\rho}(\beta\omega_{\gamma\delta})\int d^3x^{\prime}\int_{-\infty}^t dt^{\prime}~e^{\epsilon(t^{\prime}-t)}\bigg(\widehat{\Theta}^{\mu\alpha\beta}(\vec{x},t),\widehat{\Theta}^{\rho\gamma\delta}(\vec{x}^{\prime},t^{\prime})\bigg)_l\,.
       \label{operatorcorre4}
    \end{align}
To obtain Eqs.~\eqref{Pid}-\eqref{operatorcorre4} one uses Curie's principle, i.e., the correlation function between operators of different ranks and spatial parity vanishes~\cite{Landau_Physical_kinetics}. Now comparing Eqs.~\eqref{equ56review}-\eqref{equ63review} with Eqs.~\eqref{Pid}-\eqref{operatorcorre4} and using appropriate projectors, e.g., $\Delta^{\mu\nu}$, $\Delta^{\mu\nu}_{\alpha\beta}$, $\Delta^{[\mu\nu]}_{[\alpha\beta]}$, etc.,  one find the Kubo
correlation functions for the various transport coefficients~\cite{Harutyunyan:2017lrm,Huang:2011dc,Dey:2024cwo}.
\begin{align}
    & \eta=\frac{\beta}{10}\int d^3x^{\prime}\int_{-\infty}^t dt^{\prime}~e^{\epsilon(t^{\prime}-t)}\Big(\widehat{\pi}^{\mu\nu}(\vec{x},t),\widehat{\pi}_{\mu\nu}(\vec{x}^{\prime},t^{\prime})\Big)_l,\label{equ128review}\\
    & \zeta=\beta\int d^3x^{\prime}\int_{-\infty}^t dt^{\prime}~e^{\epsilon(t^{\prime}-t)}\Big(\widehat{P}^{\star}(\vec{x},t),\widehat{P}^{\star}(\vec{x}^{\prime},t^{\prime})\Big)_l\,,\label{equ129review}\\
    & \widetilde{\kappa}_{11}=-\frac{1}{3}\int d^3x^{\prime}\int_{-\infty}^t dt^{\prime}~e^{\epsilon(t^{\prime}-t)}\bigg(\widehat{\mathcal{J}}^{\mu}(\vec{x},t),\widehat{\mathcal{J}}_{\mu}(\vec{x}^{\prime},t^{\prime})\bigg)_l\,,\label{equ130review}\\
    & \widetilde{\kappa}_{12}=\kappa_{12}=\frac{1}{3}\int d^3x^{\prime}\int_{-\infty}^t dt^{\prime}~e^{\epsilon(t^{\prime}-t)}\bigg(\widehat{\mathcal{J}}^{\mu}(\vec{x},t),\widehat{h}_{\mu}(\vec{x}^{\prime},t^{\prime})\bigg)_l\,,\label{equ131review}\\
    & \kappa_{11}=-\frac{1}{3}\int d^3x^{\prime}\int_{-\infty}^t dt^{\prime}~e^{\epsilon(t^{\prime}-t)}\bigg(\widehat{h}^{\mu}(\vec{x},t),\widehat{h}_{\mu}(\vec{x}^{\prime},t^{\prime})\bigg)_l\,,\label{equ132review}\\
    & \chi_{1}=\int d^3x^{\prime}\int_{-\infty}^t dt^{\prime}~e^{\epsilon(t^{\prime}-t)}\bigg(\widehat{\Phi}(\vec{x},t),\widehat{\Phi}(\vec{x}^{\prime},t^{\prime})\bigg)_l\,,\label{equ133review}\\
    & \chi_{2}=\frac{1}{5}\int d^3x^{\prime}\int_{-\infty}^t dt^{\prime}~e^{\epsilon(t^{\prime}-t)}\bigg(\widehat{\tau}_{(s)}^{\lambda\nu}(\vec{x},t),\widehat{\tau}_{(s)\,\lambda\nu}(\vec{x}^{\prime},t^{\prime})\bigg)_l\,,\label{equ134review}\\
    & \chi_{3}=\frac{1}{3}\int d^3x^{\prime}\int_{-\infty}^t dt^{\prime}~e^{\epsilon(t^{\prime}-t)}\bigg(\widehat{\tau}_{(a)}^{\lambda\nu}(\vec{x},t),\widehat{\tau}_{(a)\,\lambda\nu}(\vec{x}^{\prime},t^{\prime})\bigg)_l\,,\label{equ135review}\\
    & \chi_{4}=-\frac{1}{9}\int d^3x^{\prime}\int_{-\infty}^t dt^{\prime}~e^{\epsilon(t^{\prime}-t)}\bigg(\widehat{\Theta}^{\lambda\eta\zeta}(\vec{x},t),\widehat{\Theta}_{\lambda\eta\zeta}(\vec{x},t)\bigg)_l\,\label{equ136review}
    \end{align}
Eqs.~\eqref{equ128review}-\eqref{equ129review} represent standard expressions for the shear and bulk viscosities, Eqs.~\eqref{equ130review}-\eqref{equ132review} represent transport coefficients for the cross-conductivities, and Eqs.~\eqref{equ133review}-\eqref{equ136review} represent spin transport coefficients. 

Some comment regarding the Kubo relations of different transport coefficients, and the hydrodynamic ordering of spin chemical potential is in order here. The above description of the Kubo formula is for the case when the spin chemical potential is considered to be a leading order term, i.e., $\mathcal{O}(1)$. Kubo relations get modified if we consider the spin chemical potential $\omega^{\alpha\beta}\sim\mathcal{O}(\partial)$~\cite{Hu:2021lnx}. In this case, $\widehat{T}^{\mu\nu}$ contains symmetric as well as anti-symmetric parts, which affects the conservation equations. The conservation equations when $\omega^{\alpha\beta}\sim\mathcal{O}(\partial)$ can be written as, 
\begin{align}
\partial_\mu \widehat{T}^{\mu\nu}(\vec{x},t) = 0, 
\qquad
\partial_\mu \widehat{J}^{\mu}(\vec{x},t) = 0, 
\qquad
\partial_\mu \widehat{S}^{\mu\rho\sigma}(\vec{x},t)
= \widehat{T}^{\sigma\rho}(\vec{x},t) - \widehat{T}^{\rho\sigma}(\vec{x},t).
\end{align}
In this case, the structure of the density operator remains the same,
\begin{align}
\widehat{\rho}(t)=\frac{1}{\mathcal{Z}}\exp\left(-\widehat{\mathcal{A}}+\widehat{\mathcal{B}}\right), \text{with},~~~\mathcal{Z}=\Tr \exp\left(-\widehat{\mathcal{A}}+\widehat{\mathcal{B}}\right). 
\label{}
\end{align}
but the operators $\widehat{\mathcal{A}}(t)$, and $\widehat{\mathcal{B}}(t)$ gets modified. Now $\widehat{\mathcal{A}}(t)$, and $\widehat{\mathcal{B}}(t)$ can be expressed as, 
\begin{align}
\widehat{\mathcal{A}}(t) =\int d^3x\bigg[\beta_{\nu}(\vec{x},t^{})\widehat{T}^{0\nu}(\vec{x},t^{})-\alpha (\vec{x},t^{})\widehat{J}^0(\vec{x},t^{})-\widetilde{\Omega}_{\rho\sigma}(\vec{x},t^{})\widehat{S}^{0\rho\sigma}(\vec{x},t^{})\bigg],
\label{equ139review}
\end{align}

\begin{align}
\widehat{\mathcal{B}}(t) & = \int d^3 x \int_{-\infty}^{t} dt'\, e^{\epsilon (t' - t)}
\left[\widehat{T}_{\mu\nu}(\vec{x},t')\, \partial^{\mu} \beta^{\nu}(\vec{x},t')
- \widehat{J}^{\mu}(\vec{x},t')\, \partial_{\mu} \alpha(\vec{x},t')
+ 2\widetilde{\Omega}_{\mu\nu}(\vec{x},t')\, \widehat{T}^{[\mu\nu]}(\vec{x},t')\right]\nonumber\\
&  = \int d^3x~\int _{-\infty}^t~dt^{\prime}~e^{\epsilon(t^{\prime}-t)}~\widehat{\mathcal{C}}(\vec{x},t^{\prime}).
\label{equ140review}
\end{align}
Let us compare Eqs.~\eqref{equ36ver1}, and Eq.~\eqref{equ139review}. In Eqs.~\eqref{equ36ver1} in the integrand $\widehat{T}^{0\nu}$ is symmetric, but in Eq.~\eqref{equ139review}, $\widehat{T}^{0\nu}$ is asymmetric. Moreover, in Eqs.~\eqref{equ36ver1} $\widetilde{\Omega}^{\mu\nu}$ is $\mathcal{O}(1)$ term, but in Eqs.~\eqref{equ139review} $\widetilde{\Omega}^{\mu\nu}$ is $\mathcal{O}(\partial)$ term.  Similarly, in Eq.~\eqref{equ37ver1} the energy-momentum tensor is symmetric, hence $\widehat{T}^{[\mu\nu]}(\vec{x},t')$ does not appear in this equation. On the other hand, in Eq.~\eqref{equ140review}, the energy-momentum tensor is asymmetric.  Moreover Eq.~\eqref{equ37ver1} contains a term of the form $(\partial_{\mu}\widetilde{\Omega}_{\rho\sigma})\widehat{S}^{\mu\rho\sigma}$, such a term does not appear in Eq.~\eqref{equ140review}. This is because in Eq.~\eqref{equ140review} the spin chemical potential is a $\mathcal{O}(\partial)$, hence  $\partial_{\mu}\widetilde{\Omega}_{\mu\nu}\sim \mathcal{O}(\partial^2)$ which is a higher order term in the gradients. Since the term ($\partial_{\mu}\widetilde{\Omega}_{\rho\sigma})\widehat{S}^{\mu\rho\sigma}$ does not appear in Eq.~\eqref{equ140review} dissipative parts of the spin tensor do not contribute to the density operator when we consider that spin chemical potential $\sim\mathcal{O}(\partial)$, within the framework of first order theory. Therefore, in this framework, one obtains Kubo relations only for transport coefficients appearing in the dissipative parts of energy-momentum tensor, and conserved current~\cite{Hu:2021lnx}. Using the method discussed above, for a baryon rich medium \footnote{In Eqs.~\eqref{equ44review}-\eqref{equ48review} transport coefficients are given for a baryon free system, i.e., $J^{\mu}=0$. Those transport coefficients can also be obtained in the presence of the conserved current $J^{\mu}$~\cite{Hu:2021lnx}.}, the Kubo relations for different transport coefficients that appear in Eqs.~\eqref{equ44review}-\eqref{equ48review} can be summarized as~\cite{Hu:2021lnx}, 
\begin{align}
\eta &= \frac{\beta}{10} \int d^3x' \int_{-\infty}^{t} dt'\,
e^{\epsilon (t' - t)}
\bigg( \widehat{\pi}^{\mu\nu}(\vec{x},t), \widehat{\pi}_{\mu\nu}(\vec{x}',t') \bigg)_l, \label{equ141review} \\
\kappa &= -\frac{\beta}{3} \int d^3x' \int_{-\infty}^{t} dt'\,
e^{\epsilon (t' - t)}
\bigg( \widehat{h}^{\prime\mu}(\vec{x},t), \widehat{h}^{\prime}_{\mu}(\vec{x}',t') \bigg)_l, \label{equ142review}\\
\zeta &= \beta \int d^3x' \int_{-\infty}^{t} dt'\,
e^{\epsilon (t' - t)}
\bigg( \widehat{P}^{*}(\vec{x},t), \widehat{P}^{*}(\vec{x}',t') \bigg)_l, \label{equ143review}\\
\widetilde{\gamma} &= -\frac{\beta}{6} \int d^3x' \int_{-\infty}^{t} dt'\,
e^{\epsilon (t' - t)}
\bigg( \widehat{\phi}^{\mu\nu}(\vec{x},t), \widehat{\phi}_{\mu\nu}(\vec{x}',t') \bigg)_l, 
 \label{equ144review}\\
\lambda &= -\frac{\beta}{3} \int d^3x' \int_{-\infty}^{t} dt'\,
e^{\epsilon (t' - t)}
\bigg( \widehat{q}^{\mu}(\vec{x},t), \widehat{q}_{\mu}(\vec{x}',t') \bigg)_l. \label{equ145review}
\end{align}
Here, 
\begin{align}
& \widehat{h}^{\prime\mu} = \widehat{h}^{\mu}-\frac{\varepsilon+P}{n}\widehat{J}^{\mu}_{(1)};~~\widehat{P}^{*} = \widehat{P}' + \frac{\widehat{n}}{\beta \theta} D\alpha;~~\widehat{P}' = \widehat{P} - \widehat{\Pi}- \frac{1}{J\beta}\left[-(\varepsilon+P)\,\frac{\partial n}{\partial \alpha}+ n\,\frac{\partial \varepsilon}{\partial \alpha}\right]\widehat{\varepsilon},\\
& D\alpha = \frac{\theta}{J^{\prime}}\left(
-(\varepsilon+P)\frac{\partial n}{\partial \beta}
+ n\frac{\partial \varepsilon}{\partial \beta}
\right),~~J^{\prime} = \frac{\partial \varepsilon}{\partial \alpha}\frac{\partial n}{\partial \beta}
- \frac{\partial n}{\partial \alpha}\frac{\partial \varepsilon}{\partial \beta},\\
& D\beta = \frac{\theta}{J^{}}\left(
-(\varepsilon+P)\frac{\partial n}{\partial \alpha}
+ n\frac{\partial \varepsilon}{\partial \alpha}
\right),~~J^{} = \frac{\partial \varepsilon}{\partial \beta}\frac{\partial n}{\partial \alpha}
- \frac{\partial n}{\partial \beta}\frac{\partial \varepsilon}{\partial \alpha}.
\end{align}
In above equations $\widehat{\varepsilon}$, $\widehat{P}$, $\widehat{n}$, $\widehat{\Pi}$, $\widehat{\pi}^{\mu\nu}$, $\widehat{h}^{\mu}$, $\widehat{\phi}^{\mu\nu}$, $\widehat{q}^{\mu}$, are microscopic counterparts of $\varepsilon$, $P$, $n$, $\Pi$, $\pi^{\mu\nu}$, $h^{\mu}$, $\phi^{\mu\nu}$, $q^{\mu}$.  
The transport coefficients associated with the anti-symmetric component for the energy-momentum tensor, i.e,  the boost heat conductivity ($\lambda$) and
rotational viscosity ($\widetilde{\gamma}$), are the novel features of the spin hydrodynamic framework where the spin chemical potential is a $\mathcal{O}(\partial)$ term.


\section{Rotational Brownian motion in relativistic plasma}
\label{sec5}

\subsection{An introduction to Brownian motion}

An important probe of a fluid is the diffusion of a dilute gas of impurities suspended in it. In fluids made of quasiparticles, impurity diffusion and shear viscosity (which encodes momentum diffusion) are closely connected: both scale the same way with the coupling strength and share the same temperature dependence up to kinematic prefactors. However, in a most-perfect fluid \footnote{Here, the term ‘perfect’ limit of a fluid is used to denote a fluid that dissipates the minimal possible amount of energy and remains well described by hydrodynamic equations over the largest possible domain. Quantitatively, such a limit may be characterized by the ratio of shear viscosity to entropy density, $\eta/s$, approaching its lower bound. In strongly coupled quantum field theories with holographic duals, this bound is predicted to be $\eta/s \ge \frac{1}{4\pi}$ \cite{Policastro:2001yc,Kovtun:2004de}. A fluid that saturates or approaches this bound is often referred to as nearly perfect, reflecting its extremely low dissipation.}, this linkage can fail: the diffusion constant can vanish while the shear viscosity remains finite.

In the following, we review the Brownian motion of a heavy particle in a medium of light particles. The scalar (dot) product of three dimensional vectors is defined as $\vec{A}\cdot\vec{B} \equiv A_i B_i$. Here, the totally antisymmetric tensor in three dimensions is denoted by $\epsilon_{ijk}$ (Levi-Civita symbol). Repeated indices are implicitly summed over according to the Einstein summation convention. Natural units are employed, with $c=\hbar=k_B=1$.

We assume the number of impurity particles is conserved, so the number density obeys
\begin{align}
  \partial_t n + \vec{\nabla} \cdot \vec{j} = 0 \, .
\end{align}
For smoothly varying number density, the leading–gradient constitutive relation is $\vec{j} = - D \, \vec{\nabla} n$, where $D$ is the diffusion constant~\footnote{The diffusion constant $D$ should not be confused with the comoving derivative $D=u^{\mu}\partial_{\mu}$ that we introduced earlier.}. Substituting into the continuity equation gives
the diffusion equation
\begin{align}
  \partial_t n = D \, \vec{\nabla}^{2} n \, .\label{diffn}
\end{align}

A more microscopic perspective on diffusion comes from analyzing the Brownian
motion of a single suspended particle, whose dynamics is governed by a
stochastic (Langevin) equation
\begin{align}
    \frac{d\vec{p}}{dt}=-\eta_{D}\vec{p}+\vec{\xi}(t)\,,\quad\quad\langle\xi_{i}(t)\xi_{j}(t^{\prime})\rangle=\kappa\delta_{ij}\delta(t-t^{\prime})\,.\label{LanEq}
\end{align}
Here, $\vec{p}$ is the momentum \sout{particle} of the suspended impurity. $\eta_{D}$ and $\xi(t)$ are the drag coefficient and stochastic force, respectively. The (momentum-space) diffusion coefficient $\kappa$ is related to the mean-square
momentum change per unit time. In three spatial dimensions $3\kappa=\langle\Delta\vec{p}^{~2}\rangle/\Delta t$, where $\langle\Delta\vec{p}^{~2}(t)\rangle=\langle(\vec{p}(t)-\vec{p}(0))^{2}\rangle$. Integrating the Eq.~\eqref{LanEq} yields the momentum correlation function as 
\begin{align}
    \langle {p}_{i}(t){p}_{j}(t^{\prime})\rangle=\langle{p}_{i}(t)\rangle\langle{p}_{j}(t^{\prime})\rangle+\delta_{ij}\,\frac{\kappa}{2\eta_D}\Big(e^{-\eta_{D}(t-t^{\prime})}-e^{-\eta_{D}(t+t^{\prime})}\Big)\label{CorrP}
\end{align}
For very large times, $t, t^{\prime}\rightarrow\infty$, the last term and the average momentum $\langle\vec{p}(t)\rangle\rightarrow0$ vanishes as well, that the late time behavior of  correlation between momentum  takes the exponentially decaying form as
\begin{align}
    \langle{p}_{i}(t){p}_{j}(t^{\prime})\rangle\rightarrow\delta_{ij}\,\frac{\kappa}{2\eta_D}e^{-\eta_{D}(t-t^{\prime})}
\end{align}

From the result above,
\begin{align}
  \langle \vec{p}^{~2}(t)\rangle
  = \langle \vec{p}^{~2}(0)\rangle e^{-2\eta_D t}
    + 3\,\frac{\kappa}{2\eta_D}\big(1-e^{-2\eta_D t}\big)
  \;\;\Longrightarrow\;\;
  \langle \vec{p}^{~2}(t)\rangle \xrightarrow[t\to\infty]{} \frac{3\kappa}{2\eta_D}\,.
\end{align}
On the other hand, for late times $t\gg \eta_D^{-1}$ the particle equilibrates, and
equipartition gives
\begin{align}
  \langle \vec{p}^{~2}\rangle_{\rm eq} = 3 M T \, .
\end{align}
Consistency then implies the fluctuation--dissipation relation
\begin{align}
  \kappa = 2\,\eta_D\,M T \, .
\end{align}
The relation between the diffusion coefficient $D$ and the drag coefficient $\eta_D$
follows by deriving the mean–squared displacement $\langle \Delta \vec{x}^{2}(t)\rangle
\equiv \big\langle\big(\vec{x}(t)-\vec{x}(0)\big)^{2}\big\rangle$ for a suspended particle in a thermalized medium from Eq.~\eqref{diffn}, and a standard
calculation shows that at late times the solution of the diffusion equation gives 
\begin{align}
  \langle \Delta \vec{x}^{2}(t)\rangle = 6 D\,|t| \label{MeSD}\,.
\end{align}
On the other hand, the Einstein relation between diffusion and drag coefficients,
\begin{align}
  D = \frac{T}{m\,\eta_{D}} = \frac{2 T^{2}}{\kappa}\,, \label{Enr}
\end{align}
can be obtained directly from the Langevin equation
Eq.~\eqref{LanEq} by computing the position correlator
$\langle \vec{x}(t)\,\vec{x}(t')\rangle$, in analogy with the momentum case (Eq.~\eqref{CorrP}).
The diffusion constant $D$ measures how strongly random collisions in the medium
kick a particle around, while the mobility quantifies how readily the particle
drifts under an applied force (i.e.\ how small the drag is). Both coefficients arise
from the same microscopic bombardment by medium particles. In particular, they are related,
and the diffusion constant is inversely proportional to the drag coefficient,
$D \propto 1/\eta_D$. Eq.~\eqref{Enr} is the Einstein relation, a canonical example of the fluctuation--dissipation theorem (FDT): the particle’s random-walk fluctuations
are directly linked to the dissipative drag (momentum loss) it experiences as it
moves through the fluid.

The Einstein relation offers a practical route to determine Boltzmann’s constant ($k_B$):
by tracking a particle’s Brownian motion. For large spherical particles (radius $a$)
suspended in a Newtonian fluid of shear viscosity $\eta$, the Navier-Stokes equation gives the friction
coefficient $\zeta=6\pi\eta a$, and the corresponding drag rate per unit mass is
$\eta_D=\zeta/m$. The Einstein relation then implies
\begin{align}
  D \;=\; \frac{k_{\mathrm B}T}{\zeta}
  \;=\; \frac{k_{\mathrm B}T}{6\pi\eta a}\,.\label{FDT}
\end{align}
This leads to a relation between the diﬀusion constant and the shear
viscosity, $D=T/(6\pi\eta a)$.
Consequently, the mean–squared displacement in three dimensions is
\begin{align}
  \langle \Delta \vec{x}^2(t) \rangle
  \;=\; 6 D\, t
  \;=\; \frac{k_{\mathrm B}T}{\pi \eta a}\, t,
\end{align}
in agreement with Eq.~\eqref{MeSD}. This prediction was verified in 1909 by
Jean Baptiste Perrin, work that led to his 1926 Nobel Prize. 

Because the medium exerts a stochastic force on the particle, a probabilistic
description is natural. We introduce a time–dependent probability density
$P(\vec{x},y,t)$ (where $y$ denotes internal degrees of freedom, such as the momentum $\vec{p}$ or the orientation of the particle’s dipole moment in an external field) for finding
the impurity in a given configuration at time $t$. The evolution of this probability
is governed by the Fokker–Planck equation, which is equivalent to the underlying
Langevin dynamics. In the next section, we briefly outline this correspondence.

\subsection{A general formalism}
For a single degree of freedom, the Langevin equation is a stochastic differential equation of the form,
\begin{equation}
  \frac{dy}{dt} \;=\; A(y,t) \;+\; C(y,t)\,\xi(t)
  \label{eq:5.1}
\end{equation}
where $\xi(t)$ is a given stochastic force. The choice of $\xi(t)$ that makes $y(t)$ a Markov process is \emph{Gaussian white noise}
(the Langevin process), whose statistics are
\begin{subequations}\label{eq:5.2}
\begin{align}
  \langle \xi(t) \rangle &= 0, \label{eq:5.2a}\\
  \langle \xi(t_1)\,\xi(t_2) \rangle &= 2\mathcal{D}\,\delta(t_1 - t_2). \label{eq:5.2b}
\end{align}
\end{subequations}
Since Eq.~\eqref{eq:5.1} is a first-order differential equation, for each sample function (realization) of $\xi(t)$, it determines $y(t)$ uniquely for a given initial profile $y(t_{0})$. Moreover, since $\xi(t)$ is $\delta$-correlated, its values at different times are statistically independent. Consequently, the values of $\xi(t)$ for $t<t_{0}$ do not affect the conditional probabilities at later times $t>t_{0}$. Hence, the solution of the Langevin Eq.~\eqref{eq:5.1} is Markovian.

The presence of $\xi(t)$ acts as a random forcing with prescribed statistics, 
Eq.~\eqref{eq:5.1} belongs to the class of stochastic differential equations. 
In this framework, one aims for a probabilistic characterization of $y(t)$—its 
probability density, moments, and correlators—rather than a single deterministic
trajectory. In the decomposition of Eq.~\eqref{eq:5.1}, $A(y,t)$ supplies the 
deterministic drift/transport, while $C(y,t)\,\xi(t)$ provides the diffusive, 
multiplicative noise contribution. 

Since $\xi(t)$ is Gaussian with zero mean, all odd moments vanish, and every
$2n$–point moment factorizes into products of two–point functions (Wick/Isserlis theorem).
Using Eq.~\eqref{eq:5.2} one finds
\begin{align}
  \big\langle \xi(t_1)\,\xi(t_2)\cdots \xi(t_{2n}) \big\rangle
  \;=\;
  \sum_{\text{pairings } \mathcal{P}}
  \prod_{(i,j)\in \mathcal{P}} \langle \xi(t_i)\,\xi(t_j)\rangle
  \;=\;
  (2\mathcal{D})^{n}
  \sum_{\text{pairings } \mathcal{P}}
  \prod_{(i,j)\in \mathcal{P}} \delta(t_i-t_j),
\end{align}
so the overall scaling of the $2n$–point correlator is $(2\mathcal{D})^{n}$.

Since the solution of the Langevin equation is a Markov process, it obeys a
master equation. The master equation can be expressed as 
\begin{equation}
\frac{\partial P(y,t)}{\partial t}
= \int \mathrm{d}y'\,
\big[\, W(y\!\mid\!y')\,P(y',t)
      - W(y'\!\mid\!y)\,P(y,t) \,\big]\,.
\label{eq:4.8}
\end{equation}
This equation governs the time evolution of the probability $P(y,t)$ of finding the
system in state $y$ at time $t$. It has the standard gain–loss (balance) structure
of a master equation: the term $\int \! \mathrm dy'\, W(y\!\mid\!y')\,P(y',t)$ accounts for
probability flowing \emph{into} state $y$ from all other states $y'$, whereas
$\int \! \mathrm dy'\, W(y'\!\mid\!y)\,P(y,t)$ represents probability \emph{leaving}
state $y$ toward other configurations $y'$. Here $W(a\!\mid\!b)\Delta t $ denotes the
transition rate (probability per unit time) for a jump $b \to a$ for a small time interval $\Delta t$, where $W(a\!\mid\!b)\geq0$. Since the Langevin dynamics is Markovian, the probability density satisfies a
master equation. Using the Kramers–Moyal expansion, the integro–differential
Eq.~\eqref{eq:4.8} can be rewritten as an infinite series of differential operators.
This representation is not inherently simpler, but under suitable conditions, the
series may be truncated. In what follows, we keep terms up to second order,
arriving at the Fokker–Planck form that captures diffusive effects. In order to do such an expansion, one needs to first express the transition probability $W$ as a function of the size $\Delta y$ of the jump from one configuration $y'$ to another one $y$:
\begin{align}
    W(y\!\mid\!y')=W(y'\!;\Delta y)\,, \quad \quad \Delta y=y-y'\,.
\end{align}
The master equation \eqref{eq:4.8} then reads,
\begin{align}
    \frac{\partial P(y,t)}{\partial t}
= \int \mathrm{d}\Delta y\,
\big[\, W(y-\Delta y;\Delta y)\,P(y-\Delta y,t)
      - W(y;-\Delta y)\,P(y,t) \,\big]\,.
\end{align}
This form of the master equation is attainable only when the change of variables
$y' \to \Delta y$ can be absorbed into the integration limits. This is not possible
for finite integration limits. In this review, we work with unbounded domains,
so this concern does not affect our analysis.

We assume that changes in $y$ occur through \emph{small jumps} so that $W(y';\Delta y)$ is sharply peaked in $\Delta y$ and varies only slowly with $y'$. Further, assuming that $P(y,t)$ is also slowly varying in $y$, one arrives at the Kramers–Moyal expansion of the master equation
\begin{equation}
\frac{\partial P(y,t)}{\partial t}
= \sum_{m=1}^{\infty} \frac{(-1)^m}{m!}\,
\frac{\partial^{m}}{\partial y^{m}}
\!\left[\, a^{(m)}(y,t)\,P(y,t) \,\right]
\label{eq:4.14}
\end{equation}
with the jump moments
\begin{align}
  a^{(m)}(y,t) \;=\; \int \! d(\Delta y)\, (\Delta y)^{m}\, W(y;\Delta y).
\end{align}
In many situations, the higher moments are negligible or vanish for $m>2$.
Truncating the series at second order yields the Fokker–Planck equation
\begin{equation}
\frac{\partial P(y,t)}{\partial t}
= -\,\frac{\partial}{\partial y}\!\left[\, a^{(1)}(y,t)\,P(y,t) \,\right]
+ \frac{1}{2}\,\frac{\partial^{2}}{\partial y^{2}}
\!\left[\, a^{(2)}(y,t)\,P(y,t) \,\right]
\label{eq:4.15}
\end{equation}
where the first term represents the \emph{drift/transport} contribution and the second is the \emph{diffusion} term; correspondingly, $a^{(1)}(y,t)$ and $a^{(2)}(y,t)$ are the drift and diffusion coefficients.

Since the Langevin dynamics is Markovian, the probability density obeys a master
equation that can be written in the Kramers–Moyal form, Eq.~\eqref{eq:4.14}.
To determine the coefficients $a^{(m)}(y,t)$ appearing in that expansion, we first
rewrite the differential Eq.~\eqref{eq:5.1} as an integral equation,
\begin{equation}
  y(t+\Delta t) - y
  = \int_{t}^{t+\Delta t} \! dt_1\, A\!\big(y(t_1),t_1\big)
  + \int_{t}^{t+\Delta t} \! dt_1\, C\!\big(y(t_1),t_1\big)\,\xi(t_1)\,,
  \label{eq:5.7}
\end{equation}
where $y$ denotes the initial value $y(t)$.
Expanding around $y$ gives
\begin{subequations}\label{}
\begin{align}
  A\!\big[y(t_1),t_1\big]
  &= A(y,t_1) + A'(y,t_1)\,\big[y(t_1)-y\big] + \cdots , \\
  C\!\big[y(t_1),t_1\big]
  &= C(y,t_1) + C'(y,t_1)\,\big[y(t_1)-y\big] + \cdots ,
\end{align}
\end{subequations}
with primes indicating partial derivatives with respect to $y$, evaluated at the
initial point:
\begin{subequations}\label{eq:5.9}
\begin{align}
  A'(y,t) &\equiv \left.\frac{\partial A(y,t)}{\partial y}\right|_{y}\,, \\
  C'(y,t) &\equiv \left.\frac{\partial C(y,t)}{\partial y}\right|_{y}\,.
\end{align}
\end{subequations}

One gets

\begin{equation}
\begin{aligned}
  y(t+\Delta t) - y
  &= \int_{t}^{t+\Delta t} \! dt_1\, A(y,t_1)
   + \int_{t}^{t+\Delta t} \! dt_1\, A'(y,t_1)\,[\,y(t_1)-y\,] \\
  &\quad
   + \int_{t}^{t+\Delta t} \! dt_1\, C(y,t_1)\,\xi(t_1)
   + \int_{t}^{t+\Delta t} \! dt_1\, C'(y,t_1)\,[\,y(t_1)-y\,]\,\xi(t_1)
   + \cdots
\end{aligned}
\label{eq:5.8}
\end{equation}

If we now use iteration method to express $y(t_{1})-y$ in the integrand of the Eq.~\eqref{eq:5.8}, which gives 

\begin{equation}
\begin{aligned}
  y(t+\Delta t) - y
  &= \int_{t}^{t+\Delta t} \! dt_1\, A(y,t_1)
  + \int_{t}^{t+\Delta t} \! dt_1\, A'(y,t_1)
     \int_{t}^{t_1} \! dt_2\, A(y,t_2) \\
 &\quad
  + \int_{t}^{t+\Delta t} \! dt_1\, A'(y,t_1)
     \int_{t}^{t_1} \! dt_2\, C(y,t_2)\,\xi(t_2) \\
  &\quad
   + \int_{t}^{t+\Delta t} \! dt_1\, C(y,t_1)\,\xi(t_1) \\
  &\quad
  + \int_{t}^{t+\Delta t} \! dt_1\, C'(y,t_1)\,\xi(t_1)
     \int_{t}^{t_1} \! dt_2\, A(y,t_2) \\
 &\quad
   + \int_{t}^{t+\Delta t} \! dt_1\, C'(y,t_1)\,\xi(t_1)
     \int_{t}^{t_1} \! dt_2\, C(y,t_2)\,\xi(t_2)
   + \cdots
\end{aligned}
\label{}
\end{equation}

Fix the initial value $y=y(t)$ and average the iterated form using the
noise statistics in Eq.~\eqref{eq:5.2}. The conditional mean needed for
$a^{(1)}(y,t)$ is
\begin{equation}
\begin{aligned}
\big\langle y(t+\Delta t)-y \big\rangle
&= \int_{t}^{t+\Delta t}\!dt_{1}\,A(y,t_{1})
  + \int_{t}^{t+\Delta t}\!dt_{1}\,A'(y,t_{1})
    \int_{t}^{t_{1}}\!dt_{2}\,A(y,t_{2}) \\
&\quad
  + 2\mathcal D \int_{t}^{t+\Delta t}\!dt_{1}\,C'(y,t_{1})
    \int_{t}^{t_{1}}\!dt_{2}\,C(y,t_{2})\,\delta(t_{2}-t_{1})
  + \cdots .
\end{aligned}
\end{equation}
Using
$a^{(1)}=\lim_{\Delta t\to 0}\frac{1}{\Delta t}\,
\langle y(t+\Delta t)-y(t)\rangle\big|_{y(t)=y}$
and retaining only terms of order $\Delta t$, one obtains
\begin{align}
  a^{(1)}(y,t)
  = A(y,t) + \mathcal D\,C(y,t)\,\partial_y C(y,t)\,.\label{a1}
\end{align}

By the same reasoning, the second Kramers--Moyal coefficient,
$a^{(2)}=\lim_{\Delta t\to 0}\frac{1}{\Delta t}\,
\langle [y(t+\Delta t)-y(t)]^{2}\rangle\big|_{y(t)=y}$, is
\begin{equation}
\begin{aligned}
  a^{(2)}(y,t)
  &= \lim_{\Delta t\to 0}\frac{1}{\Delta t}
     \int_{t}^{t+\Delta t}\!dt_{1}\,C(y,t_{1})
     \int_{t}^{t+\Delta t}\!dt_{2}\,C(y,t_{2})\,
       2\mathcal D\,\delta(t_{1}-t_{2}) \\
  &= 2\mathcal D\,C^{2}(y,t)\,, \label{a2}
\end{aligned}
\end{equation}
while all higher coefficients vanish, $a^{(m)}(y,t)=0$ for $m\ge 3$.

From Eqs.~\eqref{a1}–\eqref{a2} we conclude that, for the Markov process generated by the Langevin equation \eqref{eq:5.1} with Gaussian $\delta$–correlated noise, all Kramers–Moyal coefficients beyond second order vanish. Hence, the probability density satisfies the Fokker–Planck equation, Eq.~\eqref{eq:4.15}, which in terms of the jump moments reads
\begin{equation}
\frac{\partial P(y,t)}{\partial t}
= -\,\frac{\partial}{\partial y}\!\left\{\!\Big[A(y,t)
      + \mathcal{D}\,C(y,t)\,\frac{\partial C(y,t)}{\partial y}\Big]\,P(y,t)\!\right\}
  + D\,\frac{\partial^{2}}{\partial y^{2}}\!\left[ C^{2}(y,t)\,P(y,t) \right]
\label{eq:5.12}
\end{equation}
Besides the deterministic drift $A(y,t)$, the effective drift contains the
\emph{noise–induced drift} term $D\,C(y,t)\,\partial_y C(y,t)$. This form is especially useful: it lets one read off the Fokker–Planck operator
directly from the coefficients in the stochastic equation of motion. In simple
situations, the result can even be written down almost by inspection. 

These results can be extended to a multi-component stochastic process $y = (y_1, y_2, \ldots, y_N)$, for which Eq.~\eqref{eq:5.1} generalizes to
\begin{equation}
\frac{dy_i}{dt}
= A_i(y, t) + \sum_{k} C_{ik}(y, t)\, \xi_k(t),
\label{eq:5.13}
\end{equation}
where the $\xi_k(t)$ represent independent white noise terms. Their statistical properties are given by:
\begin{subequations}\label{eq:5.14}
\begin{align}
  \langle \xi_k(t) \rangle &= 0, \label{eq:5.14a} \\
  \langle \xi_k(t_1)\, \xi_\ell(t_2) \rangle &= 2\mathcal{D}\, \delta_{k\ell}\, \delta(t_1 - t_2). \label{eq:5.14b}
\end{align}
\end{subequations}
Higher-order moments can be constructed under the assumption that the noise follows a multivariate Gaussian distribution.

The successive coefficients in Eqs.~\eqref{a1} and \eqref{a2} for a single-variable process can be generalized to the multi-variable case using the Kramers–Moyal expansion. For the stochastic process defined by Eq.~\eqref{eq:5.13}, the corresponding coefficients take the form:
\begin{equation}
\label{eq:5.15}
\begin{aligned}
  a_i^{(1)}(y, t) &= A_i(y, t)
  + \mathcal{D} \sum_{j,k} C_{jk}(y, t)\,
      \frac{\partial C_{ik}(y, t)}{\partial y_j}, \\[4pt]
  a_{ij}^{(2)}(y, t) &= 2\mathcal{D} \sum_{k}
      C_{ik}(y, t)\,C_{jk}(y, t), \\[4pt]
  a^{(m)}_{j_1,\ldots,j_m}(y, t) &= 0,
  \qquad m \geq 3.
\end{aligned}
\end{equation}

\noindent
For the Markov process described by the Langevin equation in Eq.~\eqref{eq:5.13}, the Kramers–Moyal expansion terminates at second order, yielding the following Fokker–Planck equation for the probability distribution $P(y, t)$:
\begin{align}
\frac{\partial P(y, t)}{\partial t}
= &-\sum_i \frac{\partial}{\partial y_i}
  \left\{ \left[ A_i(y, t)
  + \mathcal{D} \sum_{j,k} C_{jk}(y, t)\,
        \frac{\partial C_{ik}(y, t)}{\partial y_j} \right] P(y,t) \right\}\nonumber\\
        &
  + \mathcal{D} \sum_{i,j} \frac{\partial^2}{\partial y_i \partial y_j}
  \left\{ \left[ \sum_k C_{ik}(y, t)\,C_{jk}(y, t) \right] P(y,t) \right\} 
\label{eq:5.16}
\end{align}

\noindent
This Fokker–Planck equation is fully determined by the drift and diffusion coefficients that appear in the Langevin formulation.

It is often useful to express the Fokker–Planck equation in the form of a continuity equation:
\begin{align}
    \frac{\partial P(y,t)}{\partial t} = -\sum_{i} \frac{\partial J_i}{\partial y_i}, \label{GenFP1}
\end{align}
where $J_i$ is the probability current, given by
\begin{align}
    J_i = \left[ A_i(y, t)
    + \mathcal{D} \sum_{j,k} C_{jk}(y, t)\,
        \frac{\partial C_{ik}(y, t)}{\partial y_j} \right] P(y, t)
    + \mathcal{D} \sum_j \frac{\partial}{\partial y_j}
    \left[ \sum_k C_{ik}(y, t)\, C_{jk}(y, t)\, P(y, t) \right]. \label{GenFP2}
\end{align}

The second term represents the diffusion contribution, arising from thermal fluctuations; its effect is to smooth out the probability distribution, driving it toward uniformity. The first term corresponds to the drift component, which is present even in the absence of thermal noise and reflects the deterministic motion of particles under the influence of background forces. The continuity equation ensures that the total probability is conserved over time.
Now, if we recall the special case of momentum diffusion ($y=\vec{p}$) of the impurity particle in the thermal medium as given in Eq.~\eqref{LanEq}, one can identify $A_{i}$ and $C_{ij}$ as
\begin{align}
    A_{i}=-\eta_{D}p_i\,,\quad\quad C_{ij}=\delta_{ij}\,,\quad\text{and}\,\,\quad \mathcal{D}=\kappa\,,
\end{align}
which helps us to write the Fokker-Planck equation in the given form
\begin{align}
    \frac{\partial P(\vec{p},t)}{\partial t}=\frac{\partial}{\partial\vec{p}}\cdot\Big[\eta_{D}\,\vec{p}~P(\vec{p},t)\Big]+\kappa\nabla^{2}_{\vec{p}}~P(\vec{p},t)
\end{align}
The solution of this equation produces the same two point correlation function of the momentum variables as it has been shown in Eq.~\eqref{CorrP}, and the highers.

\subsection{Rotational Brownian motion}

While Brownian motion, as discussed in the previous subsection, typically involves translational motion, there exist studies that consider rotational degrees of freedom. The importance of such rotational motion was first highlighted by Peter Debye in 1913 in the context of the response of polar molecules to an external electric field.

The rotational Brownian motion of polar molecules was first systematically studied by Debye in his theory of dielectric relaxation~\cite{Debye1929}. In this model, each molecule possesses a permanent electric dipole moment and is free to rotate in space. One may visualize each dipole as rigidly attached to a spherical particle that undergoes rotational motion under the influence of both deterministic and stochastic torques. In the absence of an external electric field, the orientations of the dipoles are completely random due to thermal fluctuations, and there is no preferred direction of alignment.
To describe the statistical properties of such a system, one introduces the distribution function $P(\vartheta,t)$, which represents the probability density of finding a dipole oriented at an angle $\vartheta$ with respect to the direction of an external electric field. The number of dipoles whose orientations lie within the solid angle element $d\Omega = \sin\vartheta\, d\vartheta\, d\phi$ is given by $P(\vartheta,t)\, d\Omega$. The time evolution of this distribution function can be obtained from the Langevin equation describing rotational Brownian motion. In analogy with translational Brownian motion discussed earlier, the rotational dynamics is governed by the equation
\begin{align}
I\ddot{\vartheta}
= -\zeta\dot{\vartheta}
- \frac{\partial V(\vartheta,t)}{\partial \vartheta}
+ \lambda(t)\,,
\label{RotLan}
\end{align}
where $I$ is the moment of inertia, $\zeta$ is the rotational friction coefficient, and $V(\vartheta,t)$ is the interaction potential. The stochastic torque $\lambda(t)$ represents random interactions with the surrounding medium and is characterized by the statistical properties
\begin{align}
\langle \lambda(t) \rangle &= 0, \\
\langle \lambda(t_1)\lambda(t_2) \rangle
&= 2\mathcal{D}_{R}\,\delta(t_1 - t_2),
\label{noisRot}
\end{align}
where $\mathcal{D}_{R}$ is the rotational diffusion coefficient. This coefficient determines the rate at which the dipole orientation becomes randomized due to thermal fluctuations. The deterministic torque, given by $-\partial_{\vartheta}V(\vartheta,t)$, tends to align the dipole along the external field direction.
In Debye’s original treatment, the moment of inertia was neglected, corresponding to the overdamped limit of rotational motion.\footnote{The neglect of inertial effects restricts the applicability of Debye’s theory at high frequencies. Extensions incorporating inertia have been discussed in later works, see Ref.~\cite{coffey_1980}.} Under this approximation, Eq.~\eqref{RotLan} reduces to a first-order stochastic differential equation. The corresponding Fokker–Planck equation for the distribution function can then be obtained using the general formalism introduced earlier, yielding
\begin{equation}
\frac{\partial P(\vartheta,t)}{\partial t}
= \frac{1}{\sin \vartheta}
\frac{\partial}{\partial \vartheta}
\left[
\sin \vartheta
\left(
\zeta^{-1}
\frac{\partial V(\vartheta,t)}{\partial \vartheta} P(\vartheta,t)
+ \mathcal{D}_{R}\frac{\partial P(\vartheta,t)}{\partial \vartheta}
\right)
\right].
\label{eq:DebyeFP}
\end{equation}
This equation represents the rotational analogue of the Fokker–Planck equation for translational Brownian motion. At long times, the drag and diffusion coefficients are related through the fluctuation–dissipation theorem, ensuring relaxation toward thermal equilibrium.

We now consider a system subjected to an external electric field $\vec{E}$, with interaction potential
\begin{equation}
V(\vartheta) = -\vec{q}\cdot\vec{E} = -qE\cos\vartheta,
\end{equation}
where $\vec{q}$ is the electric dipole moment. Suppose the system is initially equilibrated in the presence of a weak constant field $E_0$ for $t<0$, and the field is switched off at $t=0$. The initial equilibrium distribution is given by the Boltzmann distribution
\begin{equation}
P_0(\vartheta)
\propto
\exp\!\left(\frac{qE_0\cos\vartheta}{k_B T}\right).
\end{equation}
For weak fields, one may introduce the small parameter
\begin{equation}
\alpha = \frac{qE_0}{k_B T} \ll 1,
\end{equation}
and expand the distribution as
\begin{align}
P_0(\vartheta)
\simeq
N\left(1 + \alpha\cos\vartheta\right).
\label{InP}
\end{align}
For times $t>0$, when the field is removed, the distribution evolves toward an isotropic equilibrium state. Assuming a solution of the form
\begin{align}
P(\vartheta,t)
=
N\left(1 + \alpha g(t)\cos\vartheta\right),
\label{LateP}
\end{align}
and substituting into Eq.~\eqref{eq:DebyeFP}, one finds
\begin{align}
g(t)
=
e^{-t/\tau_D},
\end{align}
where
\begin{equation}
\tau_D
=
\frac{\zeta}{2k_B T}
\end{equation}
is the Debye relaxation time.
The time-dependent distribution function therefore becomes
\begin{equation}
P(\vartheta,t)
=
\frac{1}{4\pi}
\left[
1
+
\frac{qE_0}{k_B T}
e^{-t/\tau_D}
\cos\vartheta
\right].
\label{eq:DebyeSolution}
\end{equation}
The average dipole moment along the field direction is given by
\begin{equation}
\langle q\cos\vartheta \rangle
=
\frac{q^2E_0}{3k_B T}
e^{-t/\tau_D}.
\label{eq:5.37}
\end{equation}
This result describes the relaxation of the polarization from its initial equilibrium value toward zero after the external field is removed. More specifically, this result interpolates between the Curie-law response of a paramagnet 
$\frac{q^2 E_{0}}{3k_B T}$, at $t=0$, and the vanishing polarization at
$t\!\to\!\infty$ once the field is removed.  
Hence, the solution~\eqref{eq:5.37} describes the full relaxation from the
initial non-equilibrium state to the final equilibrium configuration. The characteristic timescale governing this relaxation is the Debye relaxation time $\tau_D$, which plays a central role in dielectric response theory. 

Here, the purpose of discussing these two specific cases of physical systems is to
set the stage for the following section, where we will demonstrate how both
translational (momentum) and rotational Brownian motion become relevant in the
context of heavy-ion collisions, particularly in relation to heavy-quark
transport and phenomenology.


\subsection*{Motivation in the Context of Heavy-Ion Collisions:}

Relativistic viscous hydrodynamics has become a standard framework for modeling the evolution of the quark–gluon plasma (QGP) in heavy-ion collisions. The assumption that the QGP behaves as an expanding, nearly perfect fluid has been remarkably successful in reproducing light-flavor observables—such as particle spectra and flow harmonics~\cite{Gale:2013da, Heinz:2013th, Dubla:2018czx}—indicating that the mean free path of quarks and gluons is much smaller than the spatial extent of the produced fireball. Elliptic flow, in particular, provides a crucial observable for studying the collective behavior of the medium produced in heavy-ion collisions. It arises as a hydrodynamic response to the anisotropy in the initial geometry and is therefore highly sensitive to the early and strongly interacting stages of the system’s evolution. Interestingly, recent measurements~\cite{ALICE:2020iug, ALICE:2020pvw} have shown that open heavy-flavor hadrons and charmonium states—such as $D$ mesons and $J/\psi$—exhibit a pronounced positive elliptic flow, suggesting that heavy quarks may undergo substantial interactions with the medium, possibly approaching local thermalization within the QGP~\cite{Andronic:2021erx, Andronic:2024oxz, Bhaduri:2018iwr, Bhaduri:2020lur, Kumar:2023acr}.

Heavy quarks serve as valuable probes of the quark–gluon plasma (QGP) created
in heavy-ion collisions. Owing to their large masses, they are produced
predominantly in initial hard scatterings and subsequently experience the entire
evolution of the expanding medium. The characteristic timescale for heavy quarks
to achieve local kinetic equilibration is expected to be roughly a factor of $\sim m_{Q}/T$ longer than that of light quarks \cite{Moore:2004tg}, where $m_{Q}$ denotes the heavy-quark mass and $T$ the temperature of the medium.

Over the past years, numerous transport approaches—based on the Boltzmann
equation and its simplified forms (see recent reviews \cite{Prino:2016cni})—have been developed to
describe the in-medium dynamics of heavy quarks. In these frameworks, heavy
quarks are modeled as Brownian probes that interact with the surrounding
partons of the QGP through elastic scattering and, in some cases, radiative scatterings.
Since each collision transfers only a small amount of momentum, many such
interactions are needed for a substantial change in the heavy-quark momentum.
Consequently, charm and bottom quarks can achieve local kinetic equilibration
only over relatively long timescales. Modern theoretical efforts therefore focus on
computing the relevant transport coefficients, which encapsulate the medium’s
properties and quantify the coupling between heavy and light partons in the QGP. From a phenomenological standpoint, an important advantage is that several key
quantities describing heavy-flavor interactions in the medium can be calculated
using thermal lattice QCD. Examples include the heavy-flavor diffusion
coefficient~\cite{Banerjee:2011ra,Kaczmarek:2014jga}, heavy-quark
susceptibilities~\cite{Petreczky:2008px}, and mesonic correlation
functions~\cite{Bazavov:2014cta}. Although these observables are not directly
measurable in experiments, they provide essential theoretical benchmarks and
constraints for transport models, thereby establishing a crucial link between
first-principles QCD calculations and experimental data~\cite{Riek:2010py}.

In studies of QCD matter created in heavy-ion collisions, the large mass of heavy
quarks, $m_{Q}$, combined with flavor conservation in strong interactions, lead to
several important consequences. These features provide the theoretical foundation
for formulating the Brownian motion framework of heavy quarks propagating through a thermalized medium:
\begin{enumerate}
    \item The production of heavy-quark pairs occurs almost exclusively during the initial
hard nucleon–nucleon scatterings of the collision, as their production threshold
far exceeds the typical temperatures achieved in the medium.
\item Because the heavy-quark mass is much larger than the pseudo-critical temperature,
charm and bottom quarks preserve their identity across the hadronization transition
in ultra-relativistic heavy-ion collisions. This makes them ideal probes of the
hadronization process—revealing whether they combine with light quarks from the
medium or hadronize through independent fragmentation, even at low momenta.

\item The momentum exchanged between heavy-flavor particles and the thermal medium is,
in general, much smaller than their typical thermal momentum, $p_{\text{th}} \simeq \sqrt{2m_Q T}$. Consequently, heavy quarks undergo a Brownian-like motion, experiencing numerous small momentum kicks as they interact with the surrounding medium.

\item The thermal relaxation time of heavy-flavor particles,
$\tau_{Q} \simeq \tau_{\text{th}}\, m_{Q}/T$, is significantly longer than the
thermalization time $\tau_{\text{th}}$ of the bulk medium constituents. If $\tau_{Q}$ is
comparable to or exceeds the lifetime of the fireball, the resulting heavy-flavor
spectra in heavy-ion collisions preserve a memory of their interaction history within the evolving QGP.
\end{enumerate}

As discussed earlier, the propagation of heavy-flavor particles—whether quarks or
hadrons—through the QCD medium at temperatures typical of ultra-relativistic
heavy-ion collisions resemble a Brownian process, in which a massive probe moves
through a thermal bath composed of much lighter constituents. The ensuing diffusion of the heavy probe is determined by its interaction strength
with the surrounding medium and can be characterized, in a schematic sense, by its
mean squared displacement over time as given in Eq.~\eqref{MeSD}. A small diffusion coefficient $D$ indicates strong coupling between the heavy
particle and the surrounding medium—frequent collisions restrict the probe’s
spatial diffusion. The corresponding thermal relaxation time is related by
$\tau_{Q} = D\,m_{Q}/T$, illustrating how the large mass-to-temperature ratio
$m_{Q}/T$ delays the equilibration of heavy quarks. The coefficient $D$ thus
encapsulates an intrinsic transport property of the QGP. When normalized by the
thermal wavelength $\lambda_{\text{th}} = 1/(2\pi T)$, it yields a dimensionless
measure that has been conjectured to scale with the medium’s shear viscosity to
entropy density ratio, $\eta/s$ \cite{Rapp:2009my,Moore:2004tg},
\begin{align}
    D2\pi T\propto\frac{\eta}{s}4\pi\,.
\end{align}
The proportionality constant in this relation is not universal; for instance, it is $\sim 1$ in strongly coupled conformal field theories \cite{Herzog:2006gh,Gubser:2006bz,Casalderrey-Solana:2006fio}, while in kinetic
theory for a weakly coupled ultra-relativistic gas it takes a smaller value $2/5$ \cite{Danielewicz:1984ww}. In the preceding discussion, we focused on momentum diffusion and its associated
timescales. However, this review also explores whether an analogous framework of
rotational Brownian motion may possess relevant phenomenological implications in
the context of heavy-ion collisions.

Before going further into the discussion of rotational Brownian motion, it is
worth highlighting another remarkable feature of heavy-ion collisions. In
non-central collisions, the fast-moving nuclei generate extremely strong
electromagnetic fields. Each nucleus, carrying $Z=79$ (Au) or $Z=82$ (Pb)
protons, involves charge distributions confined within a very small spatial
scale of about $1\,\mathrm{fm}$. The spectator protons—those passing outside the
almond-shaped overlap region—make the dominant contribution to the magnetic
field. At ultrarelativistic energies, where the nuclei approach the speed of
light and collide with a finite impact parameter, the resulting magnetic field
can reach magnitudes of $eB \sim 10^{18}\ \text{G}$,
as estimated in
Refs.~\cite{Kharzeev:2007jp,Skokov:2009qp,Voronyuk:2011jd,Bloczynski:2012en}.
Such values correspond to some of the strongest magnetic fields ever produced
in the known Universe.
\begin{figure}[t]  
    \centering
    \includegraphics[width=0.9\textwidth]{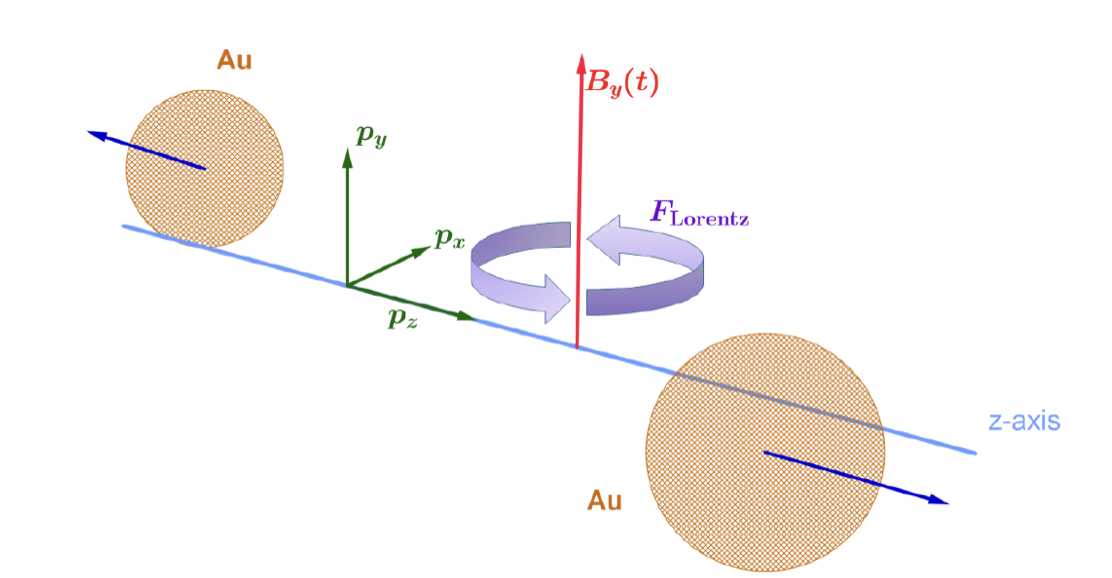}  
    \caption{A schematic depiction of a non-central heavy-ion collision is shown, where the magnetic field $\vec{B}$ is oriented perpendicular to the reaction plane, a consequence of left–right symmetry inherent in the collision geometry. Figure adapted from Ref.~\cite{Greif:2017irh}.}
    \label{fig:MagneticF}
\end{figure}

A natural starting point for estimating the magnetic field is to evaluate its strength
generated by the fast-moving spectator nucleons at the moment when the two nuclei
pass through each other. In general, the electromagnetic fields produced by
$N_{\text{ch}}$ charged particles with charges $Q_i = q_i e$ and velocities
$\vec{v}_i$ can be obtained from the relativistic Liénard–Wiechert potentials \cite{Bzdak:2011yy,Deng:2012pc},
which represent the exact solution of Maxwell’s equations for moving point charges
\begin{align}
e\vec{E}(t,\vec{r})
&= \alpha_{\mathrm{EM}} \sum_{i=1}^{N_{\mathrm{ch}}}
q_i \frac{ \vec{R}_i }
{ R_i^3 \gamma_i^2 (1 - (\vec{R}_i \times \vec{v}_i)^2 / R_i^2 )^{3/2} },
\\[4pt]
e\vec{B}(t,\vec{r})
&= \alpha_{\mathrm{EM}} \sum_{i=1}^{N_{\mathrm{ch}}}
q_i \frac{ \vec{v}_i \times \vec{R}_i }
{ R_i^3 \gamma_i^2 (1 - (\vec{R}_i \times \vec{v}_i)^2 / R_i^2 )^{3/2} },
\label{eq:1.35}
\end{align}
Here,
\[
\vec{R}_i = \vec{r} - \vec{r}_i(t_R)
= \vec{r} - \vec{r}_i\big(t - R_n(t)\big)
\]
represents the vector distance between the observation point $\vec{r}$ and the
instantaneous position of the $i$-th charged particle $\vec{r}_i(t)$, evaluated
at the retarded time $t_R = t - R_n(t)$, where
$R_n(t) = |\vec{r} - \vec{r}_n(t)|$.
The quantity $\alpha_{\mathrm{EM}}$ stands for the electromagnetic fine-structure constant. A schematic view of the collision geometry is illustrated in Fig.~\ref{fig:MagneticF}. 
In this setup, protons move approximately along straight-line trajectories 
parallel to the beam direction (the $z$-axis) with velocities 
$\vec{v} = (0,0,\pm v_z)$. 
The speed of the protons is determined by the collision energy, expressed as
\[
v_z^2 = 1 - \left( \frac{2m_p}{\sqrt{s_{NN}}} \right)^2,
\qquad
\gamma_p = \frac{\sqrt{s_{NN}}}{2m_p},
\]
\begin{figure}[t]  
    \centering
    \includegraphics[width=0.5\textwidth]{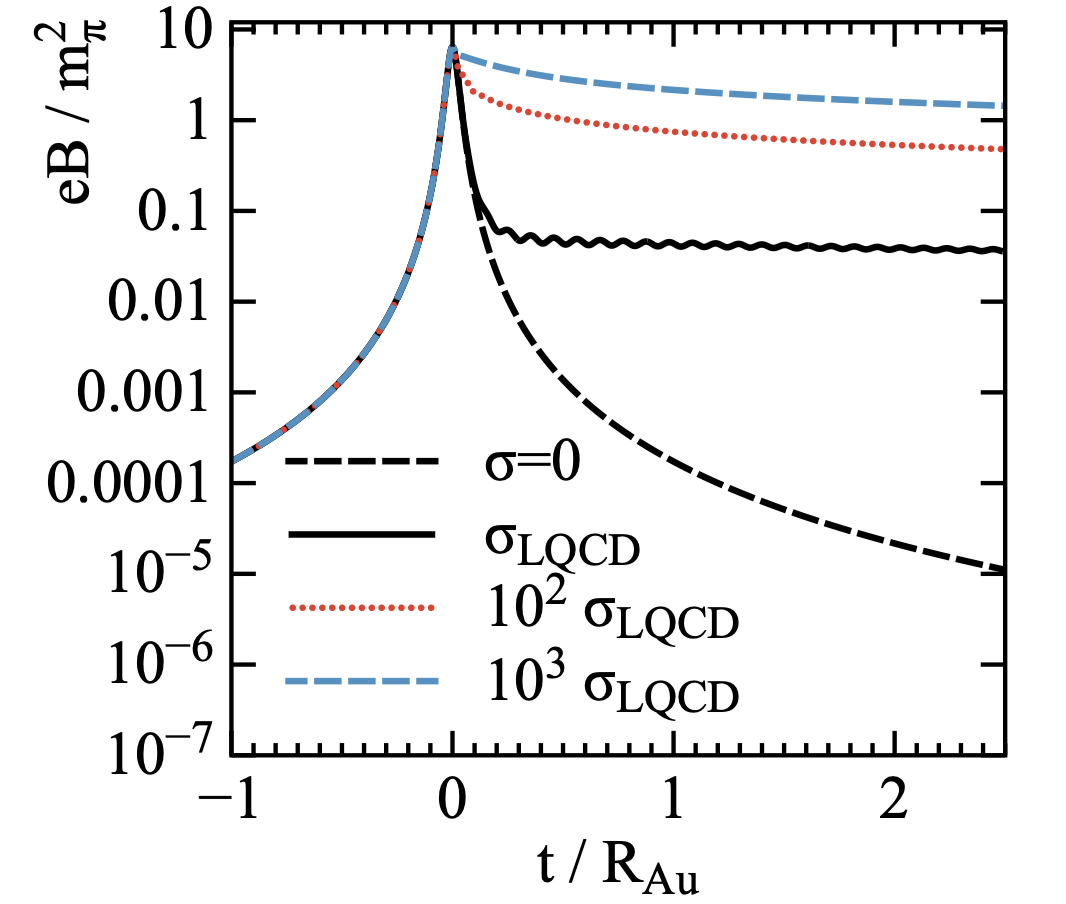}  
    \caption{Time evolution of the magnetic field and its modification due to the finite electrical conductivity of the QGP at top RHIC energy (graph adapted from Ref.~\cite{McLerran:2013hla}.) The black solid line corresponds to $\sigma_{\text{LQCD}}$ which exhibits an oscillating structure due to the damped harmonic evolution of the magnetic field -- one can refer to \cite{PhysRevC.107.034901} for details. The other two cases correspond to $10^{3}\sigma_{\text{LQCD}}$ and $10^{2}\sigma_{\text{LQCD}}$ which represent the overdamped limit of the evolution equation of the magnetic field. Here one assumes that the QGP medium with finite electrical conductivity formed just after the collision at time $t=0$.}
    \label{fig:MagneticFde}
\end{figure}
where $\sqrt{s_{NN}}$ represents the nucleon–nucleon center-of-mass energy, 
and $m_p$ denotes the proton mass. At the space–time point $\vec{r}=0$ and $t=0$, the event-averaged electromagnetic field
has a single non-vanishing component, namely $e\langle B_y \rangle \neq 0$. For impact
parameters $b < 2R_A$, where $R_A$ denotes the nuclear radius, the magnitude of the
event-averaged magnetic field $e\langle B_y \rangle$ grows roughly linearly with $b$ and
reaches its maximum value near $b \simeq 2R_A$. The following formula approximately expresses the event-averaged magnetic field, $e\langle B_{y}\rangle$, as a function of the impact parameter $b$, collision energy $\sqrt{s}$, charge number $Z$ and atomic number $A$
dependence \cite{Huang:2015oca}:
\begin{align}
    e\langle B_{y}\rangle\propto\frac{\sqrt{s}}{2m_{p}}\frac{Z}{A^{2/3}}\frac{b}{2R_{A}}m_{\pi}^{2}\,\quad \text{for}\,\,\,b<2R_{A}\,.
\end{align}
where $m_{\pi}$ is the pion mass.

Although the generated magnetic field is extremely strong, its lifetime is much shorter
than any typical thermalization timescale—the initial transient field decays rapidly.
However, the presence of a conducting medium can slow down this decay due to the
finite electrical conductivity of the QGP. Fig~\ref{fig:MagneticFde} illustrates the time evolution
of the magnetic field and its modification arising from the medium’s conductivity at
top RHIC energies \cite{McLerran:2013hla}.

Despite the theoretical prediction of extremely strong electromagnetic fields in
non-central heavy-ion collisions, a clear experimental signal of such magnetic
fields has not yet been conclusively identified. Several proposed observables
are expected to be sensitive to the presence of these fields, most notably the
Chiral Magnetic Effect (CME)---a charge separation phenomenon along the magnetic
field direction arising from the interplay between the chiral anomaly and the
strong magnetic field~\cite{Kharzeev:2007jp,Fukushima:2008xe}. 
Although charge-dependent azimuthal correlations consistent with CME expectations
have been reported by the STAR~\cite{STAR:2009wot,STAR:2019bjg} and 
ALICE~\cite{ALICE:2020siw} collaborations, 
subsequent analyses, including the recent isobar run at RHIC, 
indicate that the observed signals can largely be attributed to background
effects related to elliptic flow, leaving the CME interpretation unconfirmed.
Similarly, the global polarization of $\Lambda$ and $\bar{\Lambda}$ hyperons
measured by STAR~\cite{STAR:2017ckg,STAR:2020xbm}, 
while providing evidence of large vorticity in the QGP, 
shows only a very small difference between particle and antiparticle 
polarizations; thus, no definitive contribution from magnetic-field-induced
effects has been established. In summary, while theoretical models robustly
predict intense and transient magnetic fields in heavy-ion collisions, 
their direct manifestation in experimental observables remains an open question. In this review, we investigate the rotational Brownian motion of heavy flavors by
analyzing the polarization of open heavy-flavor particles, which can serve as a probe of the initial magnetic field generated in heavy-ion collisions. It is generally accepted that the magnetic field decays rapidly, well before the
formation of the thermalized medium. However, as previously discussed, the large
mass threshold of heavy quarks confines their production to very early times
($\sim 0.1\,\mathrm{fm}/c$) through hard scattering processes. This makes heavy
quarks excellent probes of the early-time dynamics in heavy-ion collisions. The initially strong magnetic field can polarize heavy quarks during their brief
lifetime, leading to an initial alignment of their magnetic moments. Once the
quark–gluon plasma is formed and the magnetic field weakens, the heavy quarks
interact with the surrounding light partons, gradually losing their polarization
through multiple scatterings with the medium. This scenario differs from the rotational Brownian motion of polar molecules
discussed in the previous section. In heavy-ion collisions, heavy quarks experience
a strong transient magnetic field but do not reach thermal equilibrium during its
lifetime. It is therefore expected that the initial spin alignment induced by the
magnetic field may persist and manifest as a residual polarization of the produced
hadrons along the event-averaged magnetic field direction. In this review we discuss the observational consequences of such strong field environment in heavy ion collisions on the polarization of heavy quarks.

\subsection{Heavy quarks Brownian motion in QGP}

The Brownian propagation of heavy quarks in the quark--gluon plasma (QGP) produced in ultrarelativistic heavy-ion collisions has long served as one of the most informative probes of the medium’s transport characteristics~\cite{Moore:2004tg, Gubser:2006bz, Das:2010tj, Akamatsu:2008ge, Banerjee:2011ra, Ding:2012iy, vanHees:2007me, PhysRevD.37.2484, dong2019heavy, PhysRevC.73.034913, das2015toward, PhysRevLett.100.192301, ALICE:2021rxa, STAR:2017kkh}. A central aim of phenomenological investigations of heavy-quark dynamics is the extraction of the corresponding drag and diffusion coefficients, which encode the strength and nature of the interaction between a heavy quark and the surrounding deconfined medium. Most existing analyses are concerned with the translational aspect of Brownian motion, in which the interaction with the QGP modifies the heavy quark’s linear momentum and spatial trajectory through momentum drag and momentum-space diffusion. In light of the recent experimental evidence for hadron polarization in relativistic heavy-ion collisions~\cite{Liang:2004ph, Liang:2004xn, STAR:2017ckg, STAR:2018pps, STAR:2018fqv, Niida:2018hfw, STAR:2018gyt, ALICE:2019aid, STAR:2019erd, ALICE:2019onw, Singha:2020qns, Chen:2020pty, STAR:2020xbm, ALICE:2021pzu, STAR:2021beb, Mohanty:2021vbt, STAR:2022fan, STAR:2023eck}, it is natural to extend this framework to include the rotational Brownian motion of heavy quarks during their propagation through the QGP. Rotational Brownian motion describes the stochastic evolution of the orientation and angular velocity of a microscopic object arising from thermal kicks imparted by the surrounding medium. In contrast to ordinary Brownian motion, which concerns fluctuations of position and linear momentum, rotational Brownian motion is governed by random torques generated through repeated collisions with nearby particles. 

The earliest treatment of this phenomenon was given by Debye~\cite{Debye1929}, who extended Einstein's theory of Brownian motion in order to account for the unusual dielectric dispersion observed at radio frequencies. Debye considered an ensemble of molecules possessing permanent electric dipole moments and restricted to rotate about an axis normal to the dipole direction. Neglecting intermolecular electric interactions, he argued that the molecules are statistically equivalent, so that the behavior of the entire system may be inferred from that of a single representative molecule. The resulting problem is therefore equivalent to the two-dimensional rotational Brownian motion of an individual dipole, or, equivalently, to the dynamics of a rigid rotator driven by an external oscillating electric field. Following Debye's pioneering work, rotational Brownian motion has been investigated extensively in a wide variety of contexts, particularly in connection with molecular dynamics, dielectric relaxation, and spin systems~\cite{Perrin1934, Landau:1935qbc, Furry1957, Favro1960, Evanov1964, Hubbard1972, Valiev_1973, COFFEY1980, Coffey_2003,KuboSPINFIRST}. By comparison, the application of these ideas to the polarization dynamics of heavy quarks remains largely unexplored, with only a few recent studies addressing the subject~\cite{Liu:2024hii, Dey:2025ail, Li:2025ipk, Li:2025ugb, Jaiswal:2026ixt}.

In contrast to light hadrons, which are predominantly emitted near the freeze-out surface when the medium hadronizes, heavy quarks are created mainly in the earliest stage of the collision through hard partonic scatterings. Because they are produced before the quark--gluon plasma fully develops and subsequently propagate through its entire evolution, heavy quarks provide a particularly sensitive probe of the properties of the hot deconfined matter at early times. This feature is especially relevant in non-central relativistic heavy-ion collisions, where extremely intense but short-lived magnetic fields are generated. These fields reach their maximum strength shortly after the collision and diminish rapidly thereafter, making them primarily an early-time phenomenon. During this period, the magnetic field can align the spins of heavy quarks along its direction. Such polarization may survive hadronization and become observable through the spin structure of heavy-flavor hadrons. In particular, the polarization acquired by charm or bottom quarks may be reflected in open heavy-flavor states, including $D$ mesons and $\Lambda_c$ baryons. Measurements of the polarization of these hadrons therefore provide a potential window into the magnitude and time evolution of the strong magnetic fields produced in relativistic heavy-ion collisions. Consequently, heavy-quark polarization observables constitute a unique and valuable probe of the electromagnetic fields present during the earliest stage of the collision. In the following, we review the investigation of the rotational Brownian motion of heavy quarks in a QCD medium~\cite{Dey:2025ail, Jaiswal:2026ixt}, and discuss comparison with experimental results for the polarization of open heavy-flavor hadrons~\cite{ALICE:2025cdf}.

\subsection{Rotational Brownian motion for heavy quarks}

For a classical magnetic moment, the corresponding Langevin description is provided by the stochastic Landau--Lifshitz--Gilbert equation, which describes the time evolution of the spin of a particle in its rest frame in the presence of both deterministic torques and thermal fluctuations~\cite{Landau:1935qbc, Gilbert_2004, Nishino_2015, Meo_2023}. Expressed in terms of the particle spin vector, the stochastic Landau--Lifshitz--Gilbert equation takes the form
\begin{equation}\label{SpinLangevin}
\frac{d\vec{s}}{d\tau} = \vec{s} \times \left[ \vec{\widetilde{B}} + \vec{\xi}(\tau) \right] - \lambda\, \vec{s} \times \left( \vec{s} \times \vec{\widetilde{B}} \right),
\end{equation}
where, $\vec{s}$ denotes the classical spin vector of the particle, while $\tau$ is the proper time in the particle rest frame. The quantity $\vec{\widetilde{B}}$ is the effective magnetic field entering the spin dynamics and is determined from the spin Hamiltonian $\mathcal{H}(\vec{s})$ through
\begin{equation}
\vec{\widetilde{B}} \equiv \gamma \vec{B}
= -\frac{\partial \mathcal{H}}{\partial \vec{s}} \, ,
\end{equation}
where $\vec{B}$ is the external magnetic field measured in the rest frame of the particle. The constant $\gamma$ is the gyromagnetic ratio, defined through the relation between the magnetic moment and the spin,
\begin{equation}
\vec{\mu}=\gamma\,\vec{s}.
\end{equation}
The contribution proportional to $\vec{s}\times \vec{\widetilde{B}}$ generates the precession of the spin around the direction of the effective field. The second term, in the Landau--Lifshitz--Gilbert equation containing the double cross product, introduces dissipation: it drives the spin toward energetically preferred orientations while preserving the magnitude of $\vec{s}$. The vector $\vec{\xi}(\tau)$ represents a stochastic torque arising from interactions of the particle with the surrounding medium. Its components are treated as random variables whose statistical properties satisfy the usual correlation relations, discussed later in Eq.~\eqref{Correlation}. The parameter $\lambda$ determines the relative importance of damping and precessional motion, thereby controlling the competition between relaxation and coherent spin rotation. 

Note that the evolution of a relativistic classical spin is governed by the Thomas--Bargmann--Michel--Telegdi (Thomas--BMT) equation~\cite{Leader_2001}. In the particle rest frame, this equation becomes independent of the particle momentum. In that limit, and in the absence of both the stochastic torque and damping terms, the Thomas--BMT equation reduces exactly to Eq.~\eqref{SpinLangevin}. The noise terms introduced in Eq.~\eqref{SpinLangevin} are assumed to have the statistical properties of white noise~\cite{PhysRevB.91.134411,PhysRevB.83.054432}
\begin{equation}\label{Correlation}
    \langle \xi_k(\tau) \rangle = 0, \quad
    \langle \xi_k(\tau_1)\, \xi_{l}(\tau_2) \rangle = 2\, D_{s}\, \delta_{kl} \,\delta(\tau_1 - \tau_2),
\end{equation}
where $D_s$ is the spin diffusion coefficient. Further, Eq.~\eqref{SpinLangevin} may be viewed as a particular realization of the generalized multivariate Langevin equation introduced in Eq.~\eqref{eq:5.16}~\cite{risken1996fokker, Balakrishnan}. In the present context, that general framework is specialized to describe spin dynamics in the particle rest frame, with the evolution parameter taken to be the proper time $\tau$. Subsequently, Eq.~\eqref{SpinLangevin} can be written in the form~\cite{Dey:2025ail}
\begin{equation}\label{mult_langevin}
    \frac{d y_i}{d\tau} = A_i(y, \tau) + C_{ik}(y, \tau) \, \xi_k(\tau),
\end{equation}
where $y=s_i$ and,
\begin{align}
    A_i &=  \epsilon_{ijk} s_j \widetilde{B}_k + \lambda\, (s^2 \delta_{ik} - s_i s_k) \widetilde{B}_k, \label{eq:A_coefficient} \\
    C_{ik} &= \epsilon_{ijk} s_j. \label{eq:B_coefficient}
\end{align}
Using the standard Kramers–Moyal expansion in Eq.~\eqref{eq:4.14} along with  Eqs.~\eqref{mult_langevin} and \eqref{Correlation}, one arrives at the Fokker–Planck equation~\cite{PhysRevB.58.14937, PhysRevB.91.134411, PhysRevB.83.054432, risken1996fokker, Livi_Politi_2017, Balakrishnan}
\begin{align}
    \frac{\partial P}{\partial \tau} =
    - \frac{\partial}{\partial s_i} \bigg(
    \left[
        \epsilon_{ijk}\, s_j \widetilde{B}_k + \lambda
        (s^2 \delta_{ik} - s_i\, s_k) \widetilde{B}_k \!
        - 2D_{s} s_i \right] P \bigg)
     + D_{s} \, \frac{\partial^2}{\partial s_i\, \partial s_j} 
    \bigg(\left[ s^2 \delta_{ij} - s_i s_j \right] P\bigg), \label{delPdelt}
\end{align}
where $P \equiv P(\vec{s},\tau)$ denotes the probability density for the spin vector to be oriented along $\vec{s}$ at proper time $\tau$. An important feature of the above equation is that its form is fixed completely by the coefficients appearing in the corresponding Langevin equation. The derivation of these expressions makes use of the following identities
\begin{align}
\frac{\partial C_{ik}}{\partial s_j} =&\, \epsilon_{ijk}, \quad  C_{jk} \, \frac{\partial C_{ik}}{\partial s_j} = -2\,s_i,  
\label{eq:diffusion_derivative}\\
C_{ik} C_{jk} =&\, s^2 \delta_{ij} - s_i s_j. \label{CikCjk}
\end{align}
It should be emphasized that incorporating the stochastic field into the damping contribution of Eq.~\eqref{SpinLangevin} generates additional terms in Eq.~\eqref{delPdelt}. These corrections are proportional to $D_s \lambda^2$ and therefore arise only beyond leading order. Consequently, they may be consistently neglected in a first analysis.

\begin{figure}[t]  
    \centering
    \includegraphics[width=0.4\textwidth]{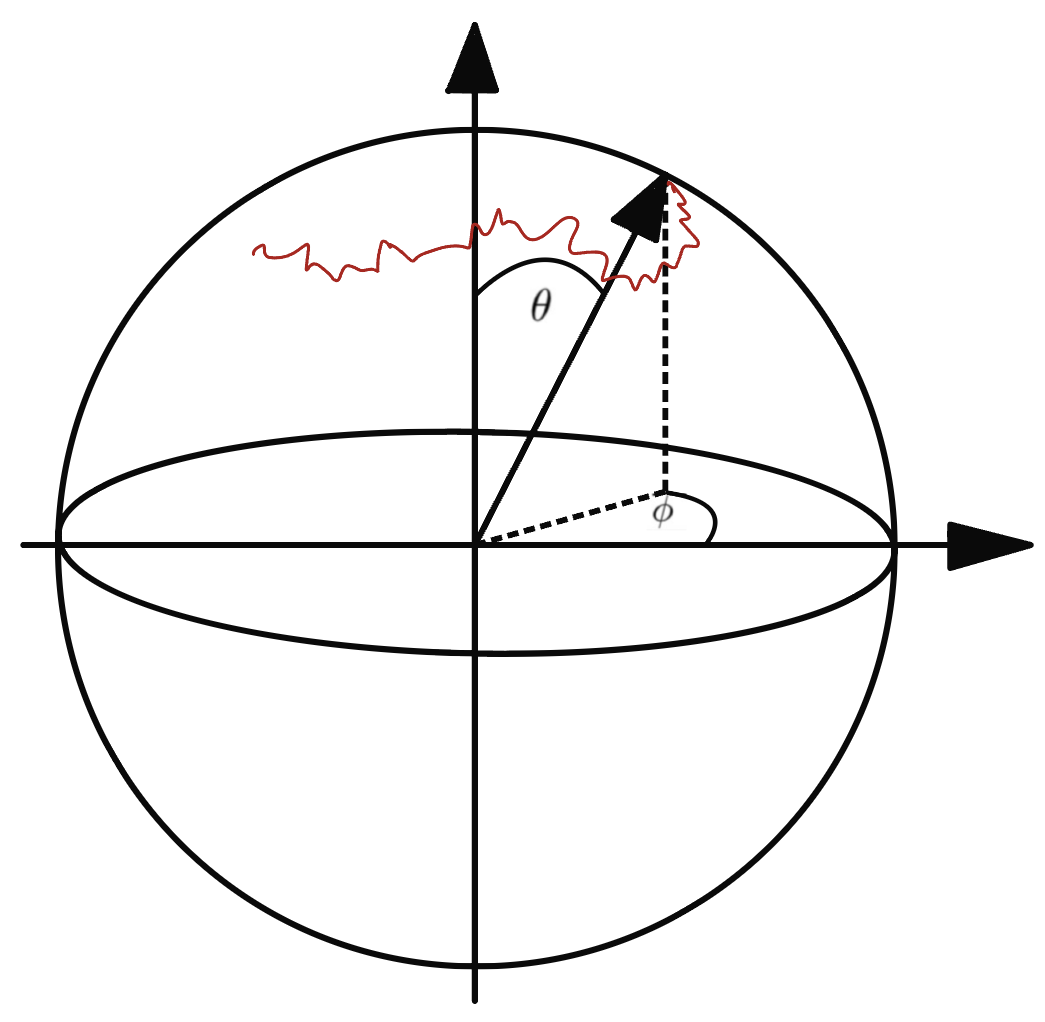}  
    \caption{Each point on the unit sphere corresponds to a distinct orientation of the heavy-quark spin vector. 
The evolution of the spin in time can be represented as a trajectory on this sphere, driven by deterministic torques and random fluctuations arising from the surrounding medium. 
The coordinates $(\theta,\phi)$ specify the instantaneous alignment of the spin axis with respect to a fixed reference frame.
The red stochastic curve depicts a representative realization of the spin’s Brownian motion on the spin-orientation manifold—an analogue, in spin space, of the random walk of a classical Brownian particle in position space—whose irregularity reflects the action of fluctuating torques.}
    \label{fig:SpinSph}
\end{figure}

When the effective field $\vec{\widetilde{B}}$ is treated as an externally prescribed field that does not depend on the spin vector $\vec{s}$, Eq.~\eqref{delPdelt} simplifies to a Smoluchowski-type equation~\cite{LEcturenotes1}. This equation plays the role of the rotational analogue of the Klein--Kramers equation~\cite{RevModPhys.15.1}. Our interest is in the probability that a polarized spin is instantaneously oriented along a direction specified by the polar and azimuthal angles $(\theta,\phi)$. To describe this, we restrict the spin vector to have fixed magnitude $s$, so that the allowed configurations lie on a sphere in spin space $\vec{s}=(s,\theta,\phi)$; see Fig.~\ref{fig:SpinSph}. Each point on this sphere corresponds to a distinct orientation of the particle spin~\cite{PhysRevE.76.051104, PhysRevA.11.280}; see Fig.~\ref{fig:SpinSph}. Without loss of generality, the external field $\tilde{\vec{B}}$ is chosen to point along the $z$ axis. Further assuming that the Hamiltonian is axially symmetric about this direction and considering a spatially homogeneous but time varying magnetic field $\vec{B}$, the evolution equation reduces to~\cite{PhysRevE.76.051104, PhysRev.130.1677, Debye1929, KuboSPINFIRST, PhysRevA.11.280}
\begin{equation}\label{FokPM}
    \tau_s \frac{\partial P}{\partial \tau} = 
    \frac{1}{\sin \theta} \frac{\partial}{\partial \theta} \Bigg[ \sin \theta \Bigg(\frac{\partial}{\partial \theta} + \frac{\lambda}{D_{s}}\mu B(\tau) \sin \theta\, \Bigg) P\Bigg]\, , 
\end{equation}
where, $\tau_s\equiv 1/D_s$ represents spin relaxation time and $\mu$ is the magnitude of the magnetic moment of the particle. In analogy with ordinary translational Brownian motion, one expects the spin relaxation time to be connected to the dissipative coefficients appearing in spin hydrodynamics through an Einstein--Stokes-type relation~\cite{BitaghsirFadafan:2008adl}. However, even in the absence of a precise microscopic determination, a reasonable estimate of $\tau_s$ may be obtained from the characteristic energy and length scales governing the system~\cite{Hongo:2022izs, Hidaka:2023oze}.


\subsection{Solution of the Fokker-Planck equation} 

The Fokker–Planck equation in Eq.~\eqref{FokPM} may be expressed in the compact form as
\begin{align}\label{fokker_planck}
    \tau_s\, \partial_{\tau}P(\theta,\tau) = \mathcal{L}_{\theta}(\tau) \, P(\theta,\tau)\,,
\end{align}
where $\partial_\tau \equiv \frac{\partial}{\partial \tau}$, and $\mathcal{L}_\theta(\tau)$ denotes a time-dependent differential operator acting on $\theta$. The temporal evolution may then be represented in operator notation as
\begin{align}\label{fokpl_ope}
    \tau_s\,\partial_{\tau}\ket{P,\tau}=\widehat{\mathcal{L}}(\tau)\ket{P,\tau}.
\end{align}
One readily finds that the general solution takes the form
\begin{align}
    \ket{P,\tau}=\mathcal T\text{exp}\left[ \frac{1}{\tau_s}\int_{0}^{\tau}d\tau^\prime\, \widehat{\mathcal{L}}(\tau^\prime)\right] \ket{P,0} ,
\end{align}
where $\mathcal{T}$ represents the time-ordering operator. The operator $\widehat{\mathcal{L}}$ may be decomposed into a time-independent contribution, $\widehat{\mathcal{L}}^{0}$, and a time-dependent contribution, $\widehat{\mathcal{L}}^{\prime}(\tau)$. Assuming the magnetic field evolves according to $B(\tau)=B_{0}\phi(\tau)$, one finds~\cite{Dey:2025ail}
\begin{equation}
\widehat{\mathcal{L}}(\tau)=\widehat{\mathcal{L}}^{0}+\alpha \, \widehat{\mathcal{L}}^\prime(\tau),
\end{equation}
where $\alpha\equiv \mu B_0\lambda/D_s$. Moreover, by treating $\alpha$ as a perturbatively small quantity---an assumption that will subsequently be justified for heavy quarks propagating through the QGP---we may construct the solution through a Dyson expansion, organized in powers of $\alpha$~\cite{risken1996fokker, PhysRevResearch.3.043172}. To linear order in $\alpha$, one obtains
\begin{equation}\label{SolDyson} 
    \mathcal{P}\,(\theta,\tau;\theta_{0},0)\equiv \langle\theta|P,\tau\rangle = P^{0}(\theta,\tau;\theta_{0},0)
     +\alpha\!\int_\Omega \int_{0}^{\tau}\!\frac{d\tau^\prime}{\tau_s}\,d\Omega^\prime\,\mathcal{G}^{0}_{\tau-\tau^{\prime}}(\theta,\theta^\prime)\,\mathcal{L}^\prime_{\theta^\prime}(\tau^\prime)\,P^{0}(\theta^\prime\!,\tau^\prime;\theta_{0},0) \, , 
\end{equation}
where $d\Omega^\prime \equiv 2\pi \sin\theta^\prime\,d\theta^\prime$. In this expression, $P^{0}(\theta,\tau;\theta^\prime,0)$ denotes the solution of Eq.~\eqref{fokker_planck} in the limit $\alpha=0$, while $\mathcal{G}_{\tau}^{0}(\theta,\theta^\prime)$ is the associated Green's function for the same unperturbed equation.

When the magnetic field is switched off, the operator $\widehat{\mathcal{L}}$ appearing in Eq.~\eqref{fokpl_ope} reduces solely to its static component, $\widehat{\mathcal{L}}^{0}$, since no explicitly time-dependent contribution remains. The resulting operator is therefore given by
\begin{equation}
    \widehat{\mathcal{L}}^{0} = \frac{1}{\sin \theta} \frac{\partial}{\partial \theta} \bigg( \sin \theta \frac{\partial}{\partial \theta} \bigg).
\end{equation}
Under these conditions, Eq.~\eqref{fokker_planck} admits a solution that may be written in the form
\begin{align}
    P^{0}(\theta,\tau ;\theta^\prime,0)=\int d\Omega^\prime \, \mathcal{G}^{0}_{\tau}(\theta,\theta^\prime) \, P(\theta^\prime,0)\,,
\end{align}
where,
\begin{align}
    \mathcal{G}^{0}_{\tau}(\theta,\theta^\prime)=\sum_{n=0}^{\infty}\frac{(2n+1)}{4\pi}\exp\left[-n(n+1)\frac{\tau}{\tau_s}\right] 
    P_{n}(\cos\theta) \, P_{n}(\cos\theta^\prime).
\end{align}
Here, $P_n(x)$ denotes the Legendre polynomial of degree $n$ in the variable $x$. If it is assumed that, at the initial time, all heavy quarks are spin polarized along the direction of the magnetic field, corresponding to $\theta=\theta_0$, then the initial distribution is taken to be $P(\theta,0)=\frac{1}{2\pi}\delta(\cos\theta-\cos\theta_0)$. Under this choice of initial condition, one obtains
\begin{align}
    P^{0}(\theta,\tau ;\theta_{0},0) = \sum_{n=0}^{\infty}\frac{(2n+1)}{4\pi}\exp\left[-n(n+1)\frac{\tau}{\tau_s}\right] 
     P_{n}(\cos\theta)P_{n}(\cos\theta_{0}).
\end{align}
The solution obtained above for the case without a magnetic field serves as the starting point for the perturbative analysis in the presence of a time-dependent magnetic field. The dominant influence of the initially strong magnetic field is accounted for through the choice of initial condition, in which the spins of all heavy quarks are taken to be aligned along the magnetic-field direction.

We now turn to the case in which the magnetic field varies with time and is controlled by the small expansion parameter $\alpha$. The associated time-dependent contribution to the operator may then be written as
\begin{align}
    \mathcal{L}^\prime(\tau)=\frac{\phi(\tau)}{\sin\theta}\frac{\partial}{\partial\theta}\,\sin^{2}\theta\,.
\end{align}
Starting from the solution in Eq.~\eqref{SolDyson}, the moments $\langle \cos\theta \rangle$ and $\langle \cos^{2}\theta \rangle$ may be evaluated straightforwardly. These quantities characterize the vector and tensor polarizations, respectively~\cite{Leader_2001}. If it is assumed that the spin polarization of open heavy-flavor hadrons is primarily inherited from the polarization of the heavy quark, then $\langle \cos\theta \rangle$ is associated with the spin polarization of heavy-flavor baryons, whereas $\langle \cos^{2}\theta \rangle$ governs the spin alignment of heavy vector mesons. Such an interpretation naturally follows from a fragmentation picture in which polarized heavy quarks retain their spin information during hadronization. Employing Eq.~\eqref{SolDyson}, one finds
\begin{align}
    \langle \cos\theta \rangle = \cos\theta_{0} \, e^{- 2\tau/\tau_s} - \frac{2\alpha}{3\tau_s }\, e^{-2\tau/\tau_s}\int_{0}^{\tau} \! d\tau^\prime \, \phi(\tau^\prime) 
    \Big[e^{2\tau^\prime/\tau_s} - P_{2}(\cos\theta_{0}) e^{-4\tau^\prime/\tau_s}\Big]\,,\label{timPol}
\end{align}
and
\begin{align}
    \langle \cos^2\theta \rangle = \frac{1}{3} + \frac{2}{3} P_2( \cos\theta_{0}) e^{- 6\tau/\tau_s} + &\frac{2\alpha}{5\tau_s}\! \int_{0}^{\tau} \! d\tau^\prime \phi(\tau^\prime) 
    \bigg[P_1( \cos\theta_{0})e^{(4\tau^\prime-6\tau)/\tau_s} - P_3( \cos\theta_{0})e^{-6(\tau^\prime+\tau)/\tau_s}   \bigg]\,.\label{timPol2}
\end{align}
Our objective is to determine the heavy-quark spin polarization generated by the initially strong magnetic field. Owing to the coupling between the quark magnetic moment and the field, the heavy-quark spin is driven to orient either along or opposite to the magnetic-field direction, depending on the sign of the quark charge. As a result, the relevant initial angles are $\theta_0=0$ and $\theta_0=\pi$, for which $P_2(\cos\theta_0) = 1$.

In relativistic heavy-ion collisions, the magnetic field persists only for a very short duration and decreases rapidly as the spectator nuclei move away from the interaction region with ultrarelativistic speeds. Under these circumstances, the temporal evolution of the field may be modeled by an exponential decay, $\phi(\tau)=e^{-\tau/\tau_B}$, where $\tau_B$ characterizes the lifetime of the magnetic field. Substituting this profile into Eqs.~\eqref{timPol} and \eqref{timPol2}, the time integrals can be carried out analytically, yielding
\begin{align}
    \langle \cos\theta \rangle = \cos\theta_{0} \, e^{- 2\tau/\tau_s} -\frac{2\alpha\tau_{B}}{3} \, e^{- 2\tau/\tau_s} 
     \left[\frac{1-\exp(-\frac{(\tau_s-2\tau_{B})\tau}{\tau_s\tau_{B}})}{\tau_s-2\tau_{B}} -\frac{1-\exp(-\frac{(4\tau_{B}+\tau_s)\tau}{\tau_s\tau_{B}})}{4\tau_{B}+\tau_s} \right], \label{av_pol}
\end{align}
and
\begin{align}
    \langle \cos^2\theta \rangle = \frac{1}{3} + \frac{2}{3} \, e^{-6\tau/\tau_s} + &\frac{2\alpha\tau_{B}}{5} e^{-6\tau/\tau_s} \Bigg[\frac{1}{(\tau_s-4\tau_B)} 
    \bigg( 1-\exp\Big[-\frac{(\tau_s-4\tau_B)\tau}{\tau_s\tau_B} \Big] \bigg) \nonumber\\
    &-\frac{1}{(\tau_s+6\tau_B)}\bigg( 1-\exp\Big[-\frac{(\tau_s+6\tau_B)\tau}{\tau_s\tau_B} \Big] \bigg) \Bigg]\cos\theta_{0}, \label{av_pol2}
\end{align}
where the relations $P_{2n}(\cos\theta_{0})=1$ and $P_{2n+1}(\cos\theta_{0})=\cos\theta_{0}$ have been employed, taking into account that the allowed initial orientations are restricted to $\theta_{0}=0$ and $\theta_{0}=\pi$.

We now turn to heavy quarks characterized by a longitudinal rapidity $y$ and a transverse velocity defined as $v_T \equiv p_T/E$, where $p_T$ denotes the transverse momentum. The corresponding energy is $E = m_T \cosh y$, with the transverse mass given by $m_T = \sqrt{p_T^2 + m_Q^2}$, where $m_Q$ is the heavy-quark mass. Since the entire analysis has been carried out in the rest frame of the heavy quark, the magnetic field oriented along the $y$-direction in the laboratory frame, generated by the spectator charges, must be appropriately Lorentz transformed into this frame. The transformation of the initial electromagnetic field to the heavy-quark rest frame is then expressed as~\cite{Herbert:1997}
\begin{equation}\label{Lorentz_trans_B}
\vec{B} = \gamma_{v} \left( \vec{B}_{\rm Lab} - \vec{E}_{\rm Lab} \times \vec{v} \right) + \left( 1 - \gamma_{v} \right) \left( \frac{\vec{B}_{\rm Lab}\cdot\vec{v}}{|\vec{v}|^2} \right) \vec{v}.
\end{equation}
Here, $\gamma_v \equiv E/m_Q$ denotes the Lorentz factor associated with the heavy quark, while $\vec{v}$ represents its velocity. It follows from Eq.~\eqref{Lorentz_trans_B} that, for fixed transverse momentum $p_T$, the component of the magnetic field along the $y$-direction in the heavy-quark rest frame grows with increasing rapidity. Consequently, the polarization signal generated by the initial magnetic field is expected to be enhanced for heavy quarks at nonzero rapidity, as they experience a stronger effective magnetic field in their rest frame. In addition, the contributions involving $\vec{E}_{\rm Lab}\times\vec{v}$ and $\vec{B}_{\rm Lab}\cdot\vec{v}$ in Eq.~\eqref{Lorentz_trans_B} depend explicitly on the direction of the quark velocity. Owing to the random distribution of heavy-quark velocities, these terms are anticipated to average out and effectively cancel~\cite{Tuchin:2013ie}.

To assess the validity of treating the time-dependent magnetic field perturbatively, an estimate of the upper bound on the parameter $\alpha$ is required. In the laboratory frame, the magnetic field is oriented along the $y$-direction and, at the time of QGP formation, typically satisfies $eB_{\rm Lab} \lesssim 0.1\, m_{\pi}^{2}$. Based on this, the magnitude of the perturbative parameter $\alpha$ may be approximated as\footnote{In the absence of an Einstein--Stokes-type relation connecting the drag and diffusion coefficients for spin dynamics, an analogous relation is assumed on dimensional grounds, namely $D_{s} = \lambda T$.}
\begin{equation}
    \alpha \equiv \frac{\mu\, \gamma_v B_{\rm Lab}\lambda}{D_{s}}  = \frac{{\rm g}\, s\, q\, \gamma_v B_{\rm Lab}}{2\, m_QT} = \frac{(f\,e) \gamma_v B_{\rm Lab}}{2\, m_{Q}T}.
\end{equation}
In this estimate, the magnetic moment of a heavy quark is taken as $\mu = {\rm g}\, s\, q /(2 m_Q)$, with the gyromagnetic factor chosen as {\color{red} ${\rm g} \simeq 2$,} the spin $s = \hbar/2$, and the electric charge written as $q = f\,e$, where $f = 2/3$ for charm quarks and $f = -1/3$ for bottom quarks. Substituting these inputs and adopting a representative QGP temperature of $T \approx 300\,\mathrm{MeV}$, one finds that for heavy quarks carrying momentum of about $50\,\mathrm{GeV}$, the parameter $\alpha$ takes the values $\alpha \approx 0.034$ for charm and $\alpha \approx -0.007$ for bottom quarks. The smallness of these values supports the interpretation of $\alpha$ as a perturbative quantity and thereby justifies the expansion scheme employed above.

It is worth emphasizing that, owing to its transient nature, the initially generated magnetic field decays rapidly and becomes substantially weaker by the time the QGP is formed~\cite{Huang:2022qdn, Tuchin:2013ie, PhysRevC.105.054907, Huang:2017tsq, McLerran:2013hla}. In addition, theoretical estimates indicate that the spin relaxation time is much longer than the QGP formation timescale~\cite{Hongo:2022izs, Hidaka:2023oze, Kapusta:2019sad}. Consequently, in the regime where $\tau_s \gg \tau_B$, it is well justified to disregard the $\alpha$-dependent contributions in Eqs.~\eqref{av_pol} and \eqref{av_pol2}, which simplifies the expressions to
\begin{equation}\label{vec_ten_pol}
\langle \cos\theta \rangle = \cos\theta_{0} \, e^{- 2\tau/\tau_s} , \quad
\langle \cos^2\theta \rangle = \frac{1}{3} + \frac{2}{3} \, e^{-6\tau/\tau_s} . 
\end{equation}
In these expressions, $\tau$ represents the total time interval during which the heavy quark undergoes Brownian motion inside the QGP medium. It follows that an extended time spent within the plasma leads to a larger reduction in the resulting spin polarization. It is also evident that $\tau_s = 1/D_s$ in Eq.~\eqref{vec_ten_pol}, encodes the strength of the spin-spin interaction between the heavy quark and the surrounding medium. This spin relaxation timescale is sensitive to medium properties, such as the temperature of the QGP, and has been previously estimated in studies involving strange quarks~\cite{Kapusta:2019sad, Hongo:2022izs, Hidaka:2023oze}. However, it should be emphasized that the present framework does not incorporate the effects of heavy-quark energy loss. As a result, the identification of $\tau_s$ with a physical spin relaxation time should be interpreted as providing only an order-of-magnitude estimate rather than a precise determination.


\subsection{Connection with experimental observables}

In experimental analyses, the spin polarization of hadrons is not measured directly but is reconstructed from the angular distributions of their decay products. These distributions encode the correlation between the parent hadron’s spin orientation and the momentum of the emitted daughters. For baryons, weak decays are typically exploited to determine the polarization vector $\vec{P}$ of the parent particle \cite{STAR:2007ccu}. In contrast, for vector mesons, strong or electromagnetic decay channels are employed, where the spin alignment is commonly quantified through the tensor polarization parameter $\rho_{00}$ \cite{Mohanty:2021vbt}. From angular momentum conservation, the spin state of a produced hadron can be represented as a superposition of eigenstates of the angular momentum operator defined with respect to a chosen quantization axis $\widehat{n}$. The choice of this axis is conventional and is typically guided by the geometry of the collision system. In relativistic heavy-ion collisions, a natural choice is the direction perpendicular to the reaction plane, i.e., $\widehat{n} = \widehat{y}$ \cite{ALICE:2022byg, Faccioli:2010kd}, which coincides with the orientation of both the initial orbital angular momentum and the dominant magnetic field. An analogous choice has also been adopted in recent measurements of spin polarization of $D^{*+}$ mesons \cite{ALICE:2025cdf}. The momenta of the decay products are then analyzed with respect to this quantization axis, and the corresponding angular distribution can be written as
\begin{equation}
\frac{dN}{d\cos\theta^*}=
\frac{3}{4}
\left[
(1-\rho_{00})
+
(3\rho_{00}-1)\cos^2\theta^*
\right],
\label{eq:Ws_Messon}
\end{equation}
for  vector meson, and 
\begin{equation}
\frac{dN}{d\cos\theta^*}=
\frac{1}{2}
\left[
1
+\alpha\,|\vec{P}|\cos\theta^*
\right],
\label{eq:Ws_Baryon}
\end{equation}
for baryon decay. Here, $\theta^*$ is the angle between the momentum of the decay particle in the rest frame of the vector meson with respect to the quantisation axis along $\widehat{y}$ \cite{ALICE:2022byg, Faccioli:2010kd}, and $\alpha$ is the baryon decay parameter \footnote{In the previous discussion regarding $\Lambda$-hyperon, $\alpha$ has been indicated as $\alpha_{\Lambda}$ in the Eq.~\eqref{Eq1}. }. The extraction of spin polarisation in experiments is therefore carried out by evaluating the moments $\langle\cos\theta^*\rangle$ and $\langle\cos^{2}\theta^*\rangle$ over the anisotropic momentum distribution of the decay daughters, as given in Eqs.~\eqref{eq:Ws_Baryon} and \eqref{eq:Ws_Messon}. This allows one to determine the vector and tensor polarizations as
\begin{equation}
|\vec{P}| = \frac{3}{\alpha}\langle \cos\theta^* \rangle,
\qquad
\rho_{00} = \frac{5}{2}\langle \cos^{2}\theta^* \rangle - \frac{1}{2},
\end{equation}
respectively. Recently, the polarization of the $D^{*+}$ meson has been reported through measurements of $\rho_{00}$ extracted from its decay angular distribution~\cite{ALICE:2019aid}. 

In our calculation, we evaluated the expectation values of polynomials of $\cos\theta$ appearing in Eqs.~\eqref{av_pol}-\eqref{av_pol2}, where $\theta$ is the angle between the classical spin polarization vector and the magnetic field direction (see Fig.~\ref{fig:SpinSph}), which is along $\widehat{n}$ in the lab frame. Accordingly, the computed expectation values correspond to the axis along which the spin density matrix of the produced baryon or vector meson is diagonal, which is the direction of the heavy quark spin; for our reference we name it natural quantization axis. In general, this axis is different from the experimental quantization axis \footnote{It is worth emphasizing that, within a quantum-mechanical framework, one can always choose the polarization axis to coincide with the $z$-direction, so that the spin density matrix becomes diagonal in the basis of simultaneous eigenstates of $\widehat{J}^{2}$ and $\widehat{J}_{z}$ \cite{Sakurai, Leader_2001}. In contrast, in a classical description, the polarization vector itself naturally defines the preferred axis. This correspondence allows the expectation values appearing in Eqs.~\eqref{av_pol} and \eqref{av_pol2} to be interpreted as projections of a classical spin polarization onto the corresponding quantum-mechanical spin basis.} and the two are related by the angle $\theta$\,. Therefore, the vector and tensor polarizations, quantified by $\vec{P}$ and $\rho_{00}$ with respect to the experimental quantization axis $\widehat{y}$, can be related to those defined in the natural quantization axis via a Wigner rotation \footnote{The relation between $\rho_{00}$ ($\vec{P}$) and  $\rho'_{00}$ ($\vec{P}'$) can be shown follows directly from the rotation of the spin-$J$ density matrix. Under a rotation a rotation by $\theta$ between the axis of experimental quantization axis $\widehat{y}$ and the natural quantization axis \cite{Faccioli:2010kd, Leader_2001},
\begin{equation}
\rho_{J} = d^{(J)}(\theta)\,\rho'_{J}\,d^{(J)\dagger}(\theta),
\label{eq:rho_rotation_dmatrix}
\end{equation}
where $d^{(J)}(\theta)$ is the spin-1 Wigner small-$d$ matrix. Spin density matrix for $J=1$ and $J=1/2$ corresponds to vector meson and the baryon, and substituting proper representation of the Wigner d-matrix can helps to arrive at the relations \eqref{eq:rho00_rotation_final}-\eqref{eq:Pz_rotation_final}.  }
\begin{equation}
\rho_{00}
=
\rho'_{00}\cos^2\theta
+
\frac{1-\rho'_{00}}{2}\sin^2\theta,
\label{eq:rho00_rotation_final}
\end{equation}
and
\begin{equation}
\vec{P} = \vec{P}'\, P_{1}(\cos\theta).
\label{eq:Pz_rotation_final}
\end{equation}
Here, $\rho'_{00}$ and $\vec{P}'$ represent the tensor and vector polarizations defined with respect to the natural quantization axis. In the present framework, the spin polarized particle undergoes interactions with the surrounding medium, and these effects are incorporated by expressing the observables in terms of the expectation values $\langle\cos\theta\rangle$ and $\langle\cos^{2}\theta\rangle$, as given in Eqs.~\eqref{av_pol}–\eqref{av_pol2}. These expectation values are then to be substituted into Eqs.~\eqref{eq:rho00_rotation_final}–\eqref{eq:Pz_rotation_final}. Such quantities effectively quantify the degree to which the initial spin polarization persists after undergoing multiple, uncorrelated stochastic interactions with the medium. To account for these effects, an averaging procedure is carried out using the Fokker–Planck distribution function introduced in Eq.~\eqref{SolDyson}. Next, by combining Eqs.~\eqref{eq:rho00_rotation_final}–\eqref{eq:Pz_rotation_final} with Eq.~\eqref{vec_ten_pol}, the expressions can be recast in a compact form as
\begin{align}
\rho_{00}-\frac13 =
\left(\rho'_{00}-\frac13\right)\langle P_2(\cos\theta)\rangle = \left(\rho'_{00}-\frac13\right)e^{-6\tau/\tau_{s}} 
\implies \Delta \rho_{00} = \Delta \rho'_{00} e^{-6\tau/\tau_{s}} \,,\label{rhotwoparm}
\end{align}
and,
\begin{align}
\vec{P} = \vec{P}'\langle P_1(\cos\theta)\rangle \implies \vec{P}  = \vec{P}'\cos\theta_{0}e^{-2\tau/\tau_{s}}\,,\label{Ptwoparm}
\end{align}
where $\Delta \rho_{00}\equiv\rho_{00}-\frac13$ and $\Delta \rho'_{00}\equiv\rho'_{00}-\frac13$.

\begin{figure}[t]  
    \centering
    \includegraphics[width=0.4\textwidth]{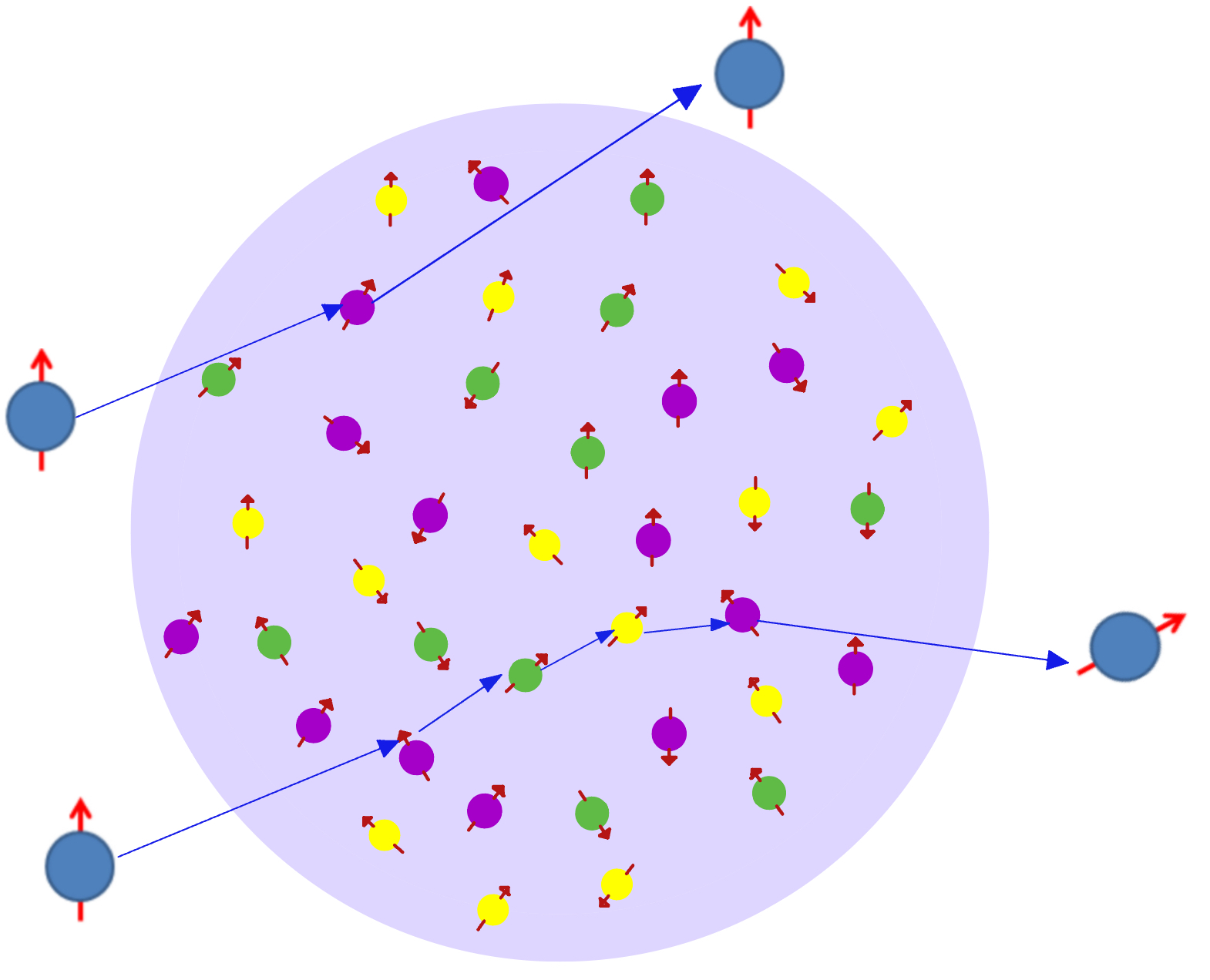}  
    \caption{ Schematic representation of heavy-quark spin dynamics in a quark–gluon plasma (QGP). Initially polarized heavy quarks, aligned by the early-time magnetic field (left), propagate through the later-formed QGP medium composed of light partons. During propagation, stochastic spin-dependent scatterings with medium constituents induce random torques, leading to partial depolarization. High-$p_T$ heavy quarks, spending shorter time in the medium, experience fewer interactions and thus retain a larger fraction of their initial polarization, whereas low-$p_T$ quarks undergo spin depolarization due to multiple scatterings. This mechanism qualitatively explains the observed increase of heavy-quark spin polarization with transverse momentum $p_T$. }
    \label{fig:HQSpin}
\end{figure}

Equation~\eqref{Ptwoparm} shows that the vector polarization, $\langle \cos\theta \rangle$, explicitly depends on the electric charge of the heavy quark, since particles with opposite charges acquire opposite spin orientations, corresponding to $\cos\theta_0=\pm 1$. In contrast, no such charge sensitivity appears in the tensor polarization $\langle \cos^{2}\theta \rangle$, as is evident from Eq.~\eqref{rhotwoparm}. This implies that open heavy-flavor baryons (e.g., $\Lambda_c$) and their corresponding anti-baryons should exhibit vector polarizations of opposite sign, whereas open heavy mesons and their antiparticles are expected to display tensor polarizations with the same sign. Furthermore, for heavy quarkonia, the net polarization induced by the magnetic field is expected to be suppressed, owing to the opposite spin alignment of the heavy quark and antiquark in the presence of the field~\cite{ALICE:2020iev}.  To estimate the space-time evolution, the average path length traversed by a heavy quark in the fireball is denoted by $L$, typically of order $L \simeq 10\,\mathrm{fm}$. Taking into account Lorentz contraction in the heavy-quark rest frame, the effective time spent undergoing Brownian motion in the QGP can be written as $\tau = L\, m_Q / |\vec{p}|$, where $|\vec{p}| = \sqrt{p_T^{2}\cosh^{2}y + m_Q^{2}\sinh^{2}y}$. As a result, the final expressions for polarization depend on both the transverse momentum $p_T$ and the rapidity $y$ of the heavy quark.

Note that Eqs.~\eqref{rhotwoparm} and \eqref{Ptwoparm}, can be written in a generic form for polarization function as
\begin{eqnarray}\label{pol_gen}
P \,(P_{T},y,\tau_{s}) = A\, e^{-a\,\tau/\tau_s},
\end{eqnarray}
where, $\vec{p}$ represents the heavy-quark momentum, $A=\Delta \rho'_{00}$, $a=6$ for mesons and $A=|\vec{P}'|\cos\theta_{0}$, $a=2$ for baryons. From the above equation, it is evident that both polarization and spin alignment increases with increasing transverse momentum. This behavior can be understood by noting that heavy quarks, whose spins are initially oriented by the strong magnetic field at early times, subsequently traverse the QGP medium formed at later stages and consisting predominantly of light partons. As they propagate, they undergo stochastic, spin-dependent interactions with the medium constituents, which generate random torques and gradually reduce their initial spin alignment. Heavy quarks with large transverse momentum ($p_T$) spend a relatively shorter time in the medium and therefore undergo fewer such interactions, allowing them to preserve a greater fraction of their original polarization. In contrast, low-$p_T$ heavy quarks remain in the medium for longer durations and experience repeated scatterings, resulting in a more pronounced depolarization; see Fig.~\ref{fig:HQSpin}.

\begin{figure}
    \centering
    \includegraphics[width=0.7\linewidth]{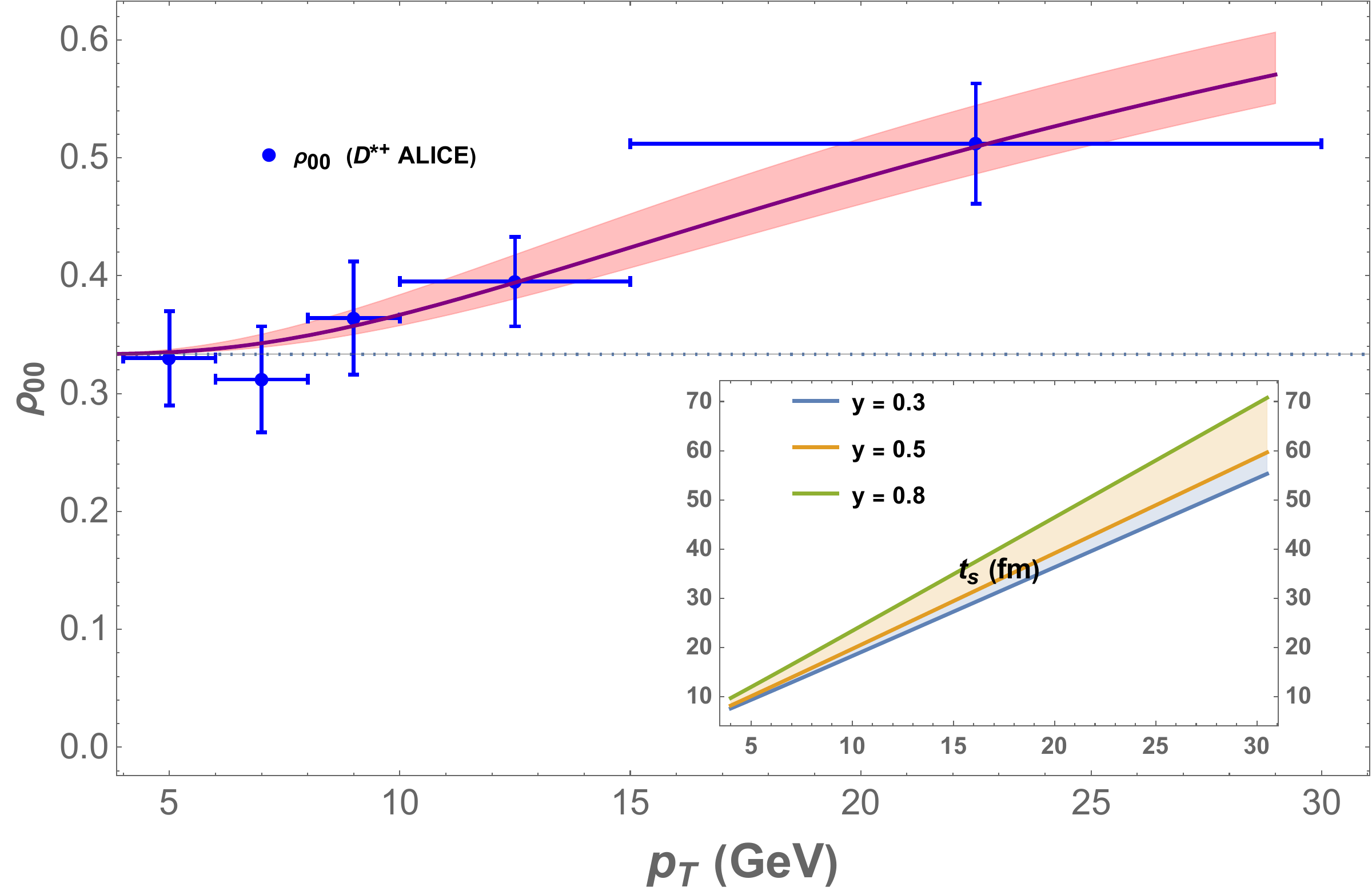}
    \caption{The ALICE Collaboration results for the spin-alignment observable $\rho_{00}$ of the $D^{*+}$ meson are shown as a function of transverse momentum~\cite{ALICE:2025cdf}. The corresponding best-fit curve, obtained from Eq.~\eqref{vec_ten_pol}, is overlaid for comparison. The inset displays the variation of the spin-relaxation time in the laboratory frame with $p_T$. The shaded band indicates the rapidity interval covered by the measurement, $0.3 < |y| < 0.8$.}
    \label{Fig}
\end{figure}

\begin{figure}
    \centering
    \includegraphics[width=0.7\linewidth]{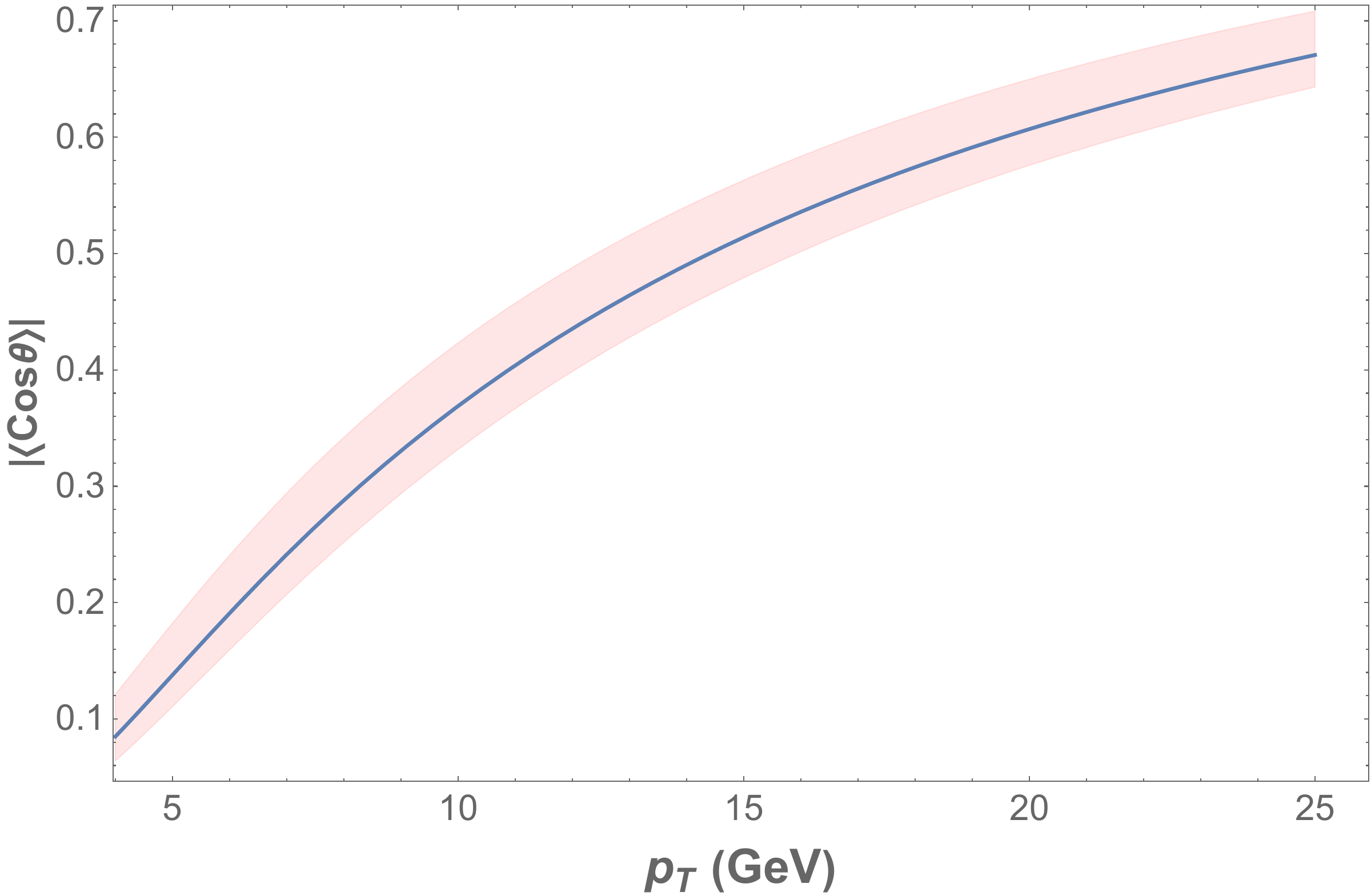}
    \caption{Predictions for polarization of open-charmed baryons, expressed as $|\langle \cos\theta \rangle|$, using the parameters obtained by fitting experimental results of ALICE collaboration for the spin alignment parameter $\rho_{00}$ for the $D^{*+}$~\cite{ALICE:2025cdf}.}
    \label{Fig2}
\end{figure}

For phenomenological comparisons, the expression for $\rho_{00}$ given in Eq.~\eqref{rhotwoparm}, involving the parameters $\Delta\rho'_{00}$ and $\tau_s$, can be fitted to experimental data to extract their values. A similar analysis may be carried out for baryons; however, since experimental measurements of heavy-baryon spin observables are currently unavailable, one may instead use the same value of $\tau_s$ to evaluate $\langle \cos\theta \rangle$, which is directly related to the vector polarization through Eq.~\eqref{Ptwoparm}. Figure~\ref{Fig} presents a comparison between the experimental measurements of $\rho_{00}$ for the $D^{*+}$ meson reported by the ALICE Collaboration~\cite{ALICE:2025cdf} (shown as blue points) in the rapidity interval $0.3 < |y| < 0.8$, and the corresponding theoretical fit. The fitting curve (blue line) is obtained using Eq.~\eqref{vec_ten_pol}, evaluated at a representative value $|y|=0.55$. The optimal agreement with the data is achieved for $\Delta\rho'_{00}=0.67$ and $\tau_s = 2.74\,\mathrm{fm}$, yielding a reduced $\chi^2$ of $0.21$. The inset displays the corresponding spin-relaxation time in the laboratory frame, $t_s = \gamma_v \tau_s$, which exhibits a monotonic increase with transverse momentum $p_T$.  In addition, Fig.~\ref{Fig2} shows the predicted behavior for open-charm baryons (magenta curve), obtained using the same fitted value of $\tau_s$. The observed rise of $\rho_{00}$ and $|\langle \cos\theta \rangle|$ with increasing $p_T$ can be understood as a consequence of kinematic effects: heavy quarks with larger momenta traverse the fireball more rapidly, thereby reducing their interaction time with the QCD medium and preserving a larger fraction of their initial spin polarization.

It is worth emphasizing that, if heavy-quark spin polarization originated solely from interactions with an already polarized medium, the resulting dependence on $p_T$ would be qualitatively different from that displayed in Fig.~\ref{Fig}. In such a scenario, heavy quarks with lower velocities would remain in the medium for longer times, allowing their spins to more effectively equilibrate with the medium’s polarization. Conversely, high-$p_T$ quarks would traverse the medium more rapidly, experiencing fewer interactions and consequently exhibiting weaker spin alignment (see Fig.~\ref{fig:HQSpin}). This mechanism would therefore lead to a decrease in spin polarization with increasing $p_T$. The experimentally observed rising trend thus indicates that polarization arising purely from medium-induced effects is insufficient to account for the data.


\subsection{Polarization harmonics for heavy quark polarization}

Heavy quarks are predominantly created in the initial stages of the collision through hard partonic processes. Owing to their early production and large masses, they traverse the evolving medium with minimal secondary production, making them effective probes of the initial conditions of the hot, deconfined matter formed in relativistic heavy-ion collisions. As discussed earlier, intense magnetic fields are generated in noncentral collisions, but they persist only for a very short duration and are therefore most relevant at these early times. Such rapidly varying electromagnetic fields can imprint a spin polarization on heavy quarks along the direction of the magnetic field. As these initially polarized heavy quarks propagate through the anisotropic quark–gluon plasma, their interactions with the surrounding medium gradually randomize their spin orientation, leading to a path-length–dependent depolarization~\cite{Dey:2025ail}. Consequently, measurements of heavy-quark spin polarization carry valuable information about the early-time dynamics and geometry of the fireball created in relativistic heavy-ion collisions.

Consider heavy quarks originating from randomly distributed production points in the transverse plane and moving along a fixed azimuthal angle $\phi$. The mean path length they traverse within the medium, denoted by $\langle L(\phi)\rangle$, reflects the underlying geometric anisotropy of the system. This angular dependence can be parametrized in a manner similar to the harmonic decomposition used for anisotropic flow observables. Accordingly, the azimuthal variation of the average path length may be represented as a Fourier expansion,
\begin{equation}\label{eq:Lphi_general}
\langle L(\phi)\rangle = L_0 \left[ 1 + \sum_{n=2}^{\infty} 2\,\ell_n \cos n\! \left( \phi-\Psi_n \right) \right],
\end{equation}
where $L_0 \equiv \langle\langle L(\phi)\rangle\rangle_{\phi}$ represents the path length averaged over all azimuthal directions, while the coefficients $\ell_n$ quantify the anisotropic components of the path length. The angles $\Psi_n$ correspond to the associated event-plane orientations for each harmonic. The terms with $n=2,3,4,\ldots$ capture the elliptic, triangular, quadrangular, and higher-order geometric anisotropies of the medium.

For a circular geometry of radius $R$, the average length of a chord can be expressed as $\rho R$, where the constant $\rho$ depends on the specific statistical prescription used to generate the ensemble of chords within the circle~\cite{Santalo_2004}. Since each chord corresponds to two possible directions of traversal, while a particle produced at a random location propagates only along one segment of that chord, the relevant mean path length is reduced by a factor of two, giving $\langle L\rangle = \rho R/2$. To incorporate deviations from perfect circular symmetry, we model the transverse profile as a weakly anisotropic geometry, with its boundary described by
\begin{equation}\label{eq:Rphi_app}
R(\varphi) = R_0 \left[ 1 + \sum_{n=2}^{\infty} a_n \cos n\! \left( \varphi-\Phi_n \right) \right]\!, \quad |a_n|\ll1 .
\end{equation}
Here, $R_0$ denotes the average radius of the system, while $a_n$ characterize the anisotropic deformations of the boundary, and $\Phi_n$ specify the corresponding symmetry-plane orientations, which are often taken to be approximately equal to the event-plane angles $\Psi_n$. Assuming that particles are produced uniformly throughout the medium and propagate along straight-line trajectories, the average path length for particles emitted at a given azimuthal angle $\phi$ is governed by the effective transverse size of the medium in that direction, such that $\langle L(\phi)\rangle \propto R(\phi)$. For convex geometries with small anisotropic distortions, as described in Eq.~\eqref{eq:Rphi_app}, this leads to
\begin{equation}
\frac{\delta\langle L(\phi)\rangle}{L_0} \simeq \frac{\delta R(\phi)}{R_0},
\end{equation}
which is in agreement with the mean chord-length arguments commonly employed in transport theory for convex systems~\cite{Santalo_2004}. In this notation, $\delta\langle L(\phi)\rangle \equiv \langle L(\phi)\rangle - L_0$ and $\delta R(\phi) \equiv R(\phi) - R_0$ represent the deviations from their respective azimuthally averaged values. A direct comparison between Eqs.~\eqref{eq:Lphi_general} and \eqref{eq:Rphi_app} then yields

\begin{equation}\label{2lnan}
2\,\ell_n=a_n,
\end{equation}
which is purely geometric in origin and remains valid for small deviations from isotropy, retaining accuracy to linear order in the anisotropy parameters.

The initial-state geometry in heavy-ion collisions is typically quantified through measures of spatial anisotropy, commonly expressed as~\cite{Bhalerao:2020ulk},
\begin{equation}\label{eq:epsn_def}
\epsilon_n \, e^{i n \Phi_n} \equiv - \frac{\langle r^n \, e^{i n\varphi}\rangle} {\langle r^n\rangle}, \qquad n \ge 2.
\end{equation}
Here, $\epsilon_n$ denote the spatial eccentricities, while $\Phi_n$ represent the associated participant-plane angles. The coordinates $(r,\varphi)$ correspond to polar coordinates in the transverse plane, and the averaging is performed over the initial matter density distribution. Since the participant-plane angles $\Phi_n$ are not directly measurable, it is standard practice to approximate them using the experimentally accessible event-plane angles $\Psi_n$. In the regime of small geometric deformations, the path-length anisotropy coefficients $\ell_n$ scale linearly with the corresponding eccentricities, i.e., $\ell_n \propto \epsilon_n$. At higher orders, however, additional nonlinear contributions can arise from combinations of lower-order eccentricities. Restricting to leading (linear) order in the anisotropies and assuming a smooth initial density profile, one obtains~\cite{Jaiswal:2026ixt}
\begin{equation}\label{eq:Lphi_small_eps}
\langle L(\phi)\rangle \simeq L_0 \left[ 1 - \sum_{n=2}^{\infty} \frac{\epsilon_n}{(n+2)} \, \cos n\! \left( \phi-\Psi_n \right) \right],
\end{equation}
where the negative sign indicates that propagation along directions with greater transverse size corresponds to an increased path length.

Taking into account the Lorentz contraction of the mean path length $\langle L(\phi) \rangle$ when viewed in the heavy-quark rest frame, the effective time interval over which the quark undergoes Brownian motion in the QGP can be expressed as $\tau = \big[\langle L(\phi) \rangle \, m_Q\big]/|\vec{p}|$. Inserting this relation for $\tau$ into Eq.~\eqref{pol_gen} then leads to
\begin{equation}\label{pol_mid}
P\, (p_T,\,\phi,\,y) = A \exp \left(-\frac{a\,m_Q\,\langle L(\phi) \rangle}{|\vec{p}|\,\tau_s} \right).
\end{equation}
Substituting the expression for $\langle L(\phi) \rangle$ from Eq.~\eqref{eq:Lphi_small_eps} and retaining terms only up to first order in the geometric anisotropy parameters, one obtains
\begin{align}\label{pol_full}
P\, (p_T,\,\phi,\,y) = A &\exp \left(-\frac{a\,m_Q\,L_0}{|\vec{p}|\,\tau_s} \right) \left[ 1 + \sum_{n=2}^{\infty} 2\,p_n \cos n \left( \phi-\Psi_n \right) \right].
\end{align}
Here, the expression for polarization harmonics are obtained as~\cite{Jaiswal:2026ixt} 
\begin{equation}\label{pol_har}
p_n\,(p_T,y) = \frac{a\,m_Q\,L_0\,\epsilon_n}{2\,(n+2)\,|\vec{p}|\,\tau_s}.
\end{equation}
This result shows that spatial anisotropies in the geometry translate directly into harmonic variations of path-length–dependent observables, yielding the proportionality $p_n \propto \epsilon_n$. It is worth noting that the mechanism proposed in Ref.~\cite{Dey:2025ail} to account for the spin alignment of the $D^{*+}$ meson corresponds to the leading term on the right-hand side of Eq.~\eqref{pol_full}. This framework thus leads to the introduction of a new class of observables, referred to as polarization harmonics, which is sensitive to initial fireball anisotropy. A quantum spin-density-matrix framework for heavy-quark spin dynamics has recently been developed which connects spin relaxation to observables such as $\rho_{00}$, baryon polarization, and polarization harmonics~\cite{Jaiswal:2026juz}. The resulting evolution of the heavy-quark polarization is found to be consistent with the leading-order behavior obtained in Ref.~\cite{Dey:2025ail} within a classical stochastic framework, as described in this section.


\section{Summary and outlook}
\label{sec6}

In this review, we have presented a detailed account of the theoretical and phenomenological developments in spin hydrodynamics and polarization phenomena in relativistic systems, with a primary focus on applications to relativistic heavy-ion collisions. The discovery of global spin polarization of $\Lambda$ hyperons has established that the quark--gluon plasma created in such collisions is an extremely vortical fluid, thereby opening a new frontier in the study of relativistic many-body systems~\cite{STAR:2007ccu}. This realization has elevated spin degrees of freedom to a central role, complementing traditional observables such as flow harmonics and particle spectra.

A key objective of this review has been to synthesize the various theoretical frameworks that have been developed to describe spin dynamics in relativistic media. These include relativistic spin hydrodynamics, kinetic theory approaches incorporating spin, and quantum statistical methods based on the Zubarev density operator. While each framework offers valuable insights, a unifying theme that emerges is the subtle interplay between macroscopic fluid properties and microscopic spin degrees of freedom. In particular, the coupling between spin and thermal vorticity provides a natural mechanism for generating polarization, yet its limitations, especially in describing local polarization observables, highlight the need for more comprehensive theoretical descriptions.

One of the central conceptual challenges discussed in this review is the issue of pseudo-gauge dependence. Although different pseudo-gauge choices lead to equivalent conservation laws, they can yield different expressions for spin polarization at the level of local observables. This ambiguity is not merely formal but has tangible consequences for theoretical predictions and their comparison with experimental data. Recent efforts toward constructing pseudo-gauge--invariant formulations of the local equilibrium density operator represent an important step forward, but a fully satisfactory resolution remains an open problem.

Another important aspect concerns the formulation of spin hydrodynamics itself. Extending conventional hydrodynamics to include spin requires not only the conservation of energy--momentum and charge but also the conservation of total angular momentum. This introduces additional dynamical variables, such as the spin chemical potential, whose proper treatment is still under active debate. As discussed in this review, different assumptions regarding the gradient ordering of the spin chemical potential lead to distinct hydrodynamic theories, each with different implications for entropy production, transport coefficients, and the structure of dissipative corrections. Moreover, first-order (Navier--Stokes) formulations of spin hydrodynamics are generically unstable and acausal, necessitating second-order (causal) extensions analogous to Israel--Stewart theory. The development and systematic analysis of such theories remain an important direction for future work.

From a phenomenological standpoint, we have examined how spin polarization observables arise from a combination of early-time electromagnetic fields, fluid vorticity, and subsequent interactions with the medium. While global polarization can be reasonably well described by spin--vorticity coupling, local polarization patterns, such as the azimuthal dependence of longitudinal polarization, pose significant challenges to existing models. Various mechanisms, including thermal shear, spin diffusion, and non-equilibrium effects, have been proposed to address these discrepancies. However, a unified description that simultaneously reproduces all available experimental data is still lacking.

The role of electromagnetic fields, particularly the strong but short-lived magnetic fields generated in noncentral collisions, has also been emphasized. These fields can induce spin polarization at early times and may leave observable imprints on heavy quarks and other probes that are sensitive to the initial stages of the collision~\cite{ALICE:2022byg}. In this context, heavy-flavor particles provide a particularly promising avenue for exploring spin dynamics, as their large masses and early production make them sensitive to both initial conditions and medium-induced effects.

Looking forward, several key challenges and opportunities can be identified:

\begin{itemize}
\item \textbf{Towards a unified and consistent theory:}  
A major goal is to develop a comprehensive framework that consistently incorporates spin dynamics across different scales, bridging hydrodynamics, kinetic theory, and quantum statistical approaches. Achieving pseudo-gauge invariance and resolving ambiguities in the definition of spin observables will be crucial in this endeavor.

\item \textbf{Determination of spin transport coefficients:}  
Quantitative predictions require knowledge of spin transport properties, such as spin diffusion and relaxation times. These quantities are currently poorly constrained and demand both theoretical calculations (e.g., from lattice QCD or effective field theories) and phenomenological extraction from experimental data.

\item \textbf{Causality and stability of spin hydrodynamics:}  
The construction of stable and causal second-order theories of spin hydrodynamics remains an open problem. Understanding the role of spin in the hydrodynamic gradient expansion and its impact on collective dynamics is essential for reliable modeling.

\item \textbf{Interplay of vorticity, electromagnetic fields, and shear:}  
A complete description of polarization phenomena must account for multiple sources of spin alignment, including vorticity, magnetic fields, and thermal gradients. Disentangling their relative contributions is both a theoretical and experimental challenge.

\item \textbf{Heavy-flavor and differential observables:}  
The study of spin polarization of heavy quarks, heavy-flavor hadrons, and quarkonia offers new opportunities to probe the early-time dynamics and transport properties of the QGP. Differential measurements in momentum, rapidity, and system size will provide stringent tests of theoretical models.

\item \textbf{Event-by-event fluctuations and correlations:}  
The role of fluctuations in the initial geometry and angular momentum of the system, and their impact on spin observables, is an important direction for future research~\cite{Giacalone:2025bgm}. This includes the study of polarization correlations and higher-order harmonic structures.

\item \textbf{Connections to other areas of physics:}  
The study of spin hydrodynamics and spin transport in relativistic systems has deep connections with condensed matter physics, chiral transport phenomena, and quantum information theory. Cross-disciplinary approaches may provide new insights and computational tools~\cite{Jaiswal:2024urq}.
\end{itemize}

In conclusion, the study of spin polarization in relativistic systems has rapidly evolved into a rich and multifaceted field at the intersection of nuclear physics, quantum field theory, and statistical mechanics. The interplay between theory and experiment continues to drive progress, with each new measurement posing fresh challenges and opportunities. As experimental capabilities improve and theoretical frameworks mature, spin observables are poised to become a powerful and indispensable tool for exploring the properties of strongly interacting matter under extreme conditions.


\section*{Acknowledgements}
S.D. and A.D. acknowledge the New Faculty Seed Grant (NFSG), NFSG/PIL/2024/P3825, provided by the Birla Institute of Technology and Science Pilani, Pilani Campus, India. A.J. gratefully acknowledges Department of Atomic Energy (DAE), India for financial support.  A.D. and A.J. acknowledge the Anusandhan National Research Foundation (ANRF), Advanced Research Grant (ARG), project number: ANRF/ARG/2025/000691/PS. 


\bibliography{ref.bib}{}
\bibliographystyle{utphys}

\end{document}